\documentclass[twocolumn,english,reprint,superscriptaddress]{revtex4-1}
\usepackage[T1]{fontenc}
\usepackage[latin9]{inputenc}
\usepackage[a4paper]{geometry}
\usepackage{xcolor}
\usepackage{pdfcolmk}
\usepackage{units}
\usepackage{amsmath}
\usepackage{amsthm}
\usepackage{amssymb}
\usepackage{graphicx}
\PassOptionsToPackage{normalem}{ulem}
\usepackage{ulem}

\makeatletter

\providecolor{lyxadded}{rgb}{0,0,1}
\providecolor{lyxdeleted}{rgb}{1,0,0}

\DeclareRobustCommand{\lyxsout}[1]{\ifx\\#1\else\sout{#1}\fi}

\theoremstyle{plain}
\newtheorem{thm}{\protect\theoremname}
\theoremstyle{definition}
\newtheorem{defn}[thm]{\protect\definitionname}
\theoremstyle{plain}
\newtheorem{lem}[thm]{\protect\lemmaname}
\theoremstyle{plain}
\newtheorem{cor}[thm]{\protect\corollaryname}

\usepackage[all]{xy}

\usepackage{multirow}

\def\frontmatter@abstractheading{}

\usepackage{babel}
\providecommand{\corollaryname}{Corollary}
  \providecommand{\definitionname}{Definition}
  \providecommand{\lemmaname}{Lemma}
  
\providecommand{\theoremname}{Theorem}

\makeatother

\usepackage{babel}
\providecommand{\corollaryname}{Corollary}
\providecommand{\definitionname}{Definition}
\providecommand{\lemmaname}{Lemma}
\providecommand{\theoremname}{Theorem}

\begin{document}
\title{Contextuality and noncontextuality measures and generalized Bell inequalities
for cyclic systems}
\author{Ehtibar N.\ Dzhafarov}
\email{To whom correspondence should be addressed. E-mail: ehtibar@purdue.edu}

\affiliation{Purdue University, USA}
\author{Janne V.\ Kujala}
\email{E-mail: jvk@iki.fi}

\affiliation{University of Turku, Finland}
\author{V\'ictor H. Cervantes}
\email{E-mail: cervantv@purdue.edu}

\affiliation{Purdue University, USA}
\begin{abstract}
Cyclic systems of dichotomous random variables have played a prominent
role in contextuality research, describing such experimental paradigms
as the Klyachko-Can-Binicio\u{g}lu-Shumovsky, Einstein-Podolsky-Rosen-Bell,
and Leggett-Garg ones in physics, as well as conjoint binary choices
in human decision making. Here, we understand contextuality within
the framework of the Contextuality-by-Default (CbD) theory, based
on the notion of probabilistic couplings satisfying certain constraints.
CbD allows us to drop the commonly made assumption that systems of
random variables are consistently connected (i.e., it allows for all
possible forms of ``disturbance'' or ``signaling'' in them). Consistently
connected systems constitute a special case in which CbD essentially
reduces to the conventional understanding of contextuality. We present
a theoretical analysis of the degree of contextuality in cyclic systems
(if they are contextual) and the degree of noncontextuality in them
(if they are not). By contrast, all previously proposed measures of
contextuality are confined to consistently connected systems, and
most of them cannot be extended to measures of noncontextuality. Our
measures of (non)contextuality are defined by the $L_{1}$-distance
between a point representing a cyclic system and the surface of the
polytope representing all possible noncontextual cyclic systems with
the same single-variable marginals. We completely characterize this
polytope, as well as the polytope of all possible probabilistic couplings
for cyclic systems with given single-variable marginals. We establish
that, in relation to the maximally tight Bell-type CbD inequality
for (generally, inconsistently connected) cyclic systems, the measure
of contextuality is proportional to the absolute value of the difference
between its two sides. For noncontextual cyclic systems, the measure
of noncontextuality is shown to be proportional to the smaller of
the same difference and the $L_{1}$-distance to the surface of the
box circumscribing the noncontextuality polytope. These simple relations,
however, do not generally hold beyond the class of cyclic systems,
and noncontextuality of a system does not follow from noncontextuality
of its cyclic subsystems.
\end{abstract}
\maketitle

\section{Introduction}

A \emph{cyclic system of rank} $n=2,3,\ldots$, is a system
\begin{equation}
\mathcal{R}=\left\{ \left\{ R_{i}^{i},R_{i\oplus1}^{i}\right\} :i=1,\ldots,n\right\} ,\label{eq:cyclic_n}
\end{equation}
where $i\oplus1=i+1$ for $i<n$, and $n\oplus1=1$; $R_{j}^{i}$
denotes a Bernoulli ($0/1$) random variable measuring \emph{content}
$q_{j}$ in \emph{context} $c_{i}$ ($j=i,i\oplus1$). A content is
any property that can be present or absent (e.g., spin of a half-spin
particle in a given direction), a context here is defined by which
two contents are measured together (simultaneously or in a specific
order). A cyclic system of rank $n$ has $n$ contexts containing
two \emph{jointly distributed} random variables each, $\left\{ R_{i}^{i},R_{i\oplus1}^{i}\right\} $.
Each of such pairs is referred to as a \emph{bunch} (of random variables).
The system also has $n$ \emph{connections} $\left\{ R_{i}^{i\ominus1},R_{i}^{i}\right\} $
(where $i\ominus1=i-1$ for $i>1$, and $1\ominus1=n$), each of which
contains two \emph{stochastically unrelated} (i.e., possessing no
joint distribution) random variables measuring the same content in
two different contexts.

Cyclic systems have played a prominent role in contextuality studies
\citep{Araujoetal2013,KujDzhLar2015}. The matrices below represent
cyclic systems of rank 5 (describing, e.g., the Klyachko-Can-Binicio\u{g}lu-Shumovsky
experiment \citep{KCBS2008,Lapkiewicz2015}), rank 4 (describing,
e.g., Bell's ``Alice-Bob'' experiments \citep{Bell1964,Bell1966,Fine1982,CHSH1969}),
rank 3 (describing, e.g., the Leggett-Garg experiments \citep{LeggGarg1985,Bacciagaluppi2015,KoflerBrukner2013,SuppesZanotti1981}),
and rank 2 (of primary interest outside quantum physics, e.g., describing
the question-order experiment in human decision making \citep{DzhZhaKuj2016,Wang}).
\begin{equation}
\begin{array}{cc}
\begin{array}{|c|c|c|c|c||c|}
\hline R_{1}^{1} & R_{2}^{1} &  &  &  & c^{1}\\
\hline  & R_{2}^{2} & R_{3}^{2} &  &  & c^{2}\\
\hline  &  & R_{3}^{3} & R_{4}^{3} &  & c^{3}\\
\hline  &  &  & R_{4}^{4} & R_{5}^{4} & c^{4}\\
\hline R_{1}^{5} &  &  &  & R_{5}^{5} & c^{5}\\
\hline\hline q_{1} & q_{2} & q_{3} & q_{4} & q_{5} & \mathcal{R}_{5}
\\\hline \end{array} & \begin{array}{|c|c|c|c||c|}
\hline R_{1}^{1} & R_{2}^{1} &  &  & c^{1}\\
\hline  & R_{2}^{2} & R_{3}^{2} &  & c^{2}\\
\hline  &  & R_{3}^{3} & R_{4}^{3} & c^{3}\\
\hline R_{1}^{4} &  &  & R_{4}^{4} & c^{4}\\
\hline\hline q_{1} & q_{2} & q_{3} & q_{4} & \mathcal{R}_{4}
\\\hline \end{array}\\
\\
\begin{array}{|c|c|c||c|}
\hline R_{1}^{1} & R_{2}^{1} &  & c^{1}\\
\hline  & R_{2}^{2} & R_{3}^{2} & c^{2}\\
\hline R_{1}^{3} &  & R_{3}^{3} & c^{3}\\
\hline\hline q_{1} & q_{2} & q_{3} & \mathcal{R}_{3}
\\\hline \end{array} & \begin{array}{|c|c||c|}
\hline R_{1}^{1} & R_{2}^{1} & c^{1}\\
\hline R_{1}^{2} & R_{2}^{2} & c^{2}\\
\hline\hline q_{1} & q_{2} & \mathcal{R}_{2}
\\\hline \end{array}
\end{array}\label{eq: R5-R2}
\end{equation}

A cyclic system is \emph{consistently connected} (satisfies the ``no-disturbance''
or ``no-signaling'' condition) if $R_{i}^{i}$ and $R_{i}^{i\ominus1}$
are identically distributed for $i=1,\ldots,n$. This assumption is
commonly made in quantum physical applications. The present paper,
however, is based on the Contextuality-by-Default (CbD) theory \citep{DzhCerKuj2017,KujDzhMeasures,DzhKujFoundations2017},
which is not predicated on this assumption, that is, the systems of
random variables we consider are generally \emph{inconsistently connected}.
Cyclic systems have been intensively analyzed within the framework
of CbD \citep{Bacciagaluppi2015,DzhKujLar2015,DzhZhaKuj2016,KujDzhLar2015,KujDzhProof2016}.
In this paper they are studied in relation to the measures of contextuality
and noncontextuality considered in Ref. \citep{KujDzhMeasures}.

The familiarity of the reader with CbD (e.g., Refs. \citep{DzhCerKuj2017,KujDzhMeasures})
for understanding this paper is not necessary, even if desirable.
We recapitulate here all relevant definitions and results, although
they are presented in the form specialized to cyclic systems rather
than in complete generality, so the broader motivation behind the
constructs may not always be apparent. In particular, we take it for
granted in this paper that it is important not to be constrained by
the confines of consistent connectedness \citep{KujDzhLar2015,DzhKujLar2015}.
The simplest reason for this is that if a consistently connected system
is contextual or noncontextual by one's definition, then it is reasonable
to require from this definition that the system's contextuality status
should not change under sufficiently small perturbations rendering
it inconsistently connected. Another reason is that inconsistent connectedness
is ubiquitous. Thus, in accordance with the quantum-mechanical laws,
consistent connectedness does not generally hold for sequential measurements,
e.g., for the Leggett-Garg system \citep{KoflerBrukner2013,Bacciagaluppi2015}.
In other experimental paradigms it is often violated due to unavoidable
or inadvertent design biases \citep{Lapkiewicz2015}. In all such
cases, use of CbD to analyze data has proved to be useful \citep{Malinowskietal.,Ariasetal.2015,Crespietal.2017,Fluhmannetal2018,Zhanetal.2017,KujDzhLar2015,Bacciagaluppi2015}.
At the same time, all contextuality measures proposed outside CbD
are confined to consistent connectedness \citep{Grudkaetal2014,Kleinmanetal2011,AbramBarbMans2017,Brunneretal2014,AmaralCunhaCabello2015}.

We also take for granted in this paper that it is desirable to seek
principled and unified ways of measuring both contextuality and noncontextuality
\citep{KujDzhMeasures}. Degree of contextuality has been related
to such concepts as quantum advantage in computation and communication
complexity \citep{Bermejoetal2017,Howardetal.2014,Brukneretal.2004},
and generally is viewed as a measure of nonclassicality of a system.
Moreover, it is intrinsically interesting to compare different contextual
systems in terms of which of them can be more easily rendered noncontextual
by perturbing its random variables (see Ref. \citep{Brunneretal2014}
for an overview). Intrinsic interest in measures of noncontextuality
can be justified similarly. It is too uninformative to simply view
noncontextual systems as having zero contextuality: some of them would
be easier than others to render contextual by perturbing their random
variables. Remarkably, there seem to be no measures of noncontextuality
proposed in the literature prior to Ref. \citep{KujDzhMeasures},
and most of the proposed measures of contextuality (e.g., the Contextual
Fraction measure proposed in Ref. \citep{AbramBarbMans2017} and generalized
to inconsistently connected systems in Ref. \citep{KujDzhMeasures})
do not naturally extend to measures of noncontextuality. By ``natural
extension'' we mean the extension to noncontextual systems using
the same principles as in constructing a contextuality measure being
extended.

Note that the term ``degree of noncontextuality'' in this paper
always applies to noncontextual systems only, in the same way as ``degree
of contextuality'' only applies to contextual systems. This is useful
to mention because ``degree of noncontextuality'' has been used
in the literature in a different meaning: as a measure complementary
to the degree of contextuality in contextual systems. Thus, the Noncontextual
Fraction measure in Ref. \citep{AbramBarbMans2017} is unity minus
Contextual Fraction measure. Both are defined for contextual systems,
while noncontextual ones all have Noncontextual Fraction equal to
unity.

Of the several measures of contextuality considered in Ref. \citep{KujDzhMeasures}
we focus here on two, labeled $\textnormal{CNT}_{1}$ and $\textnormal{CNT}_{2}$.
The former is the oldest measure introduced within the framework of
CbD \citep{DzhKujLar2015,KujDzhLar2015,KujDzhProof2016}, whereas
$\textnormal{CNT}_{2}$ is the newest one, discussed in Ref. \citep{KujDzhMeasures}.
A detailed description of these measures will have to wait until we
have introduced the necessary definitions and results. In a nutshell,
however, a cyclic system (with the distribution of each of the random
variables $R_{j}^{i}$ being fixed) is represented in CbD by two vectors
of product expectations, $\mathbf{p_{b}}$ and $\mathbf{p_{c}}$,
conventionally referred to as vectors of ``correlations.'' (We will
only use this term, strictly speaking, incorrect, in this informal
introduction, due to its familiarity in the contextuality literature.)
The subscripts $\mathbf{b}$ and $\mathbf{c}$ stand for the just-defined
CbD terms ``bunch'' and ``connection.'' The vector $\mathbf{p_{b}}$
encodes the correlations within the bunches $\left\{ R_{i}^{i},R_{i\oplus1}^{i}\right\} $,
$i=1,\ldots,n$. The vector $\mathbf{p_{c}}$ encodes the correlations
imposed on the within-connection pairs $\left\{ R_{i}^{i\ominus1},R_{i}^{i}\right\} $,
$i=1,\ldots,n$, defining thereby so-called \emph{couplings} of the
connections (recall that the connections themselves do not possess
joint distributions). A cyclic system whose (non)contextuality we
measure is represented by vectors $\mathbf{p_{b}^{*}},\mathbf{p_{c}^{*}}$,
where $\mathbf{p_{b}^{*}}$ consists of the observed bunch correlations,
and $\mathbf{p_{c}^{*}}$ consists of the correlations computed for
the connections in a special way (the \emph{maximal coupling}s of
the connections). In the case of $\textnormal{CNT}_{1}$, the $L_{1}$-distance
is measured between $\mathbf{\mathbf{p_{c}^{*}}}$ and the \emph{feasibility}
\emph{polytope} $\mathbb{P}_{\mathbf{c}}$ comprising all possible
$\mathbf{p_{c}}$-vectors compatible with $\mathbf{p_{b}^{*}}$. In
the case of $\textnormal{CNT}_{2}$, $L_{1}$-distance is computed
between $\mathbf{p_{b}^{*}}$ and the \emph{noncontextuality} \emph{polytope}
$\mathbb{P}_{\mathbf{b}}$ comprising all $\mathbf{p_{b}}$-vectors
compatible with $\mathbf{p_{c}^{*}}$. The two measures therefore
are, in a well-defined sense, mirror images of each other.
\begin{center}
\begin{figure}[h]
\begin{centering}
\includegraphics[scale=0.25]{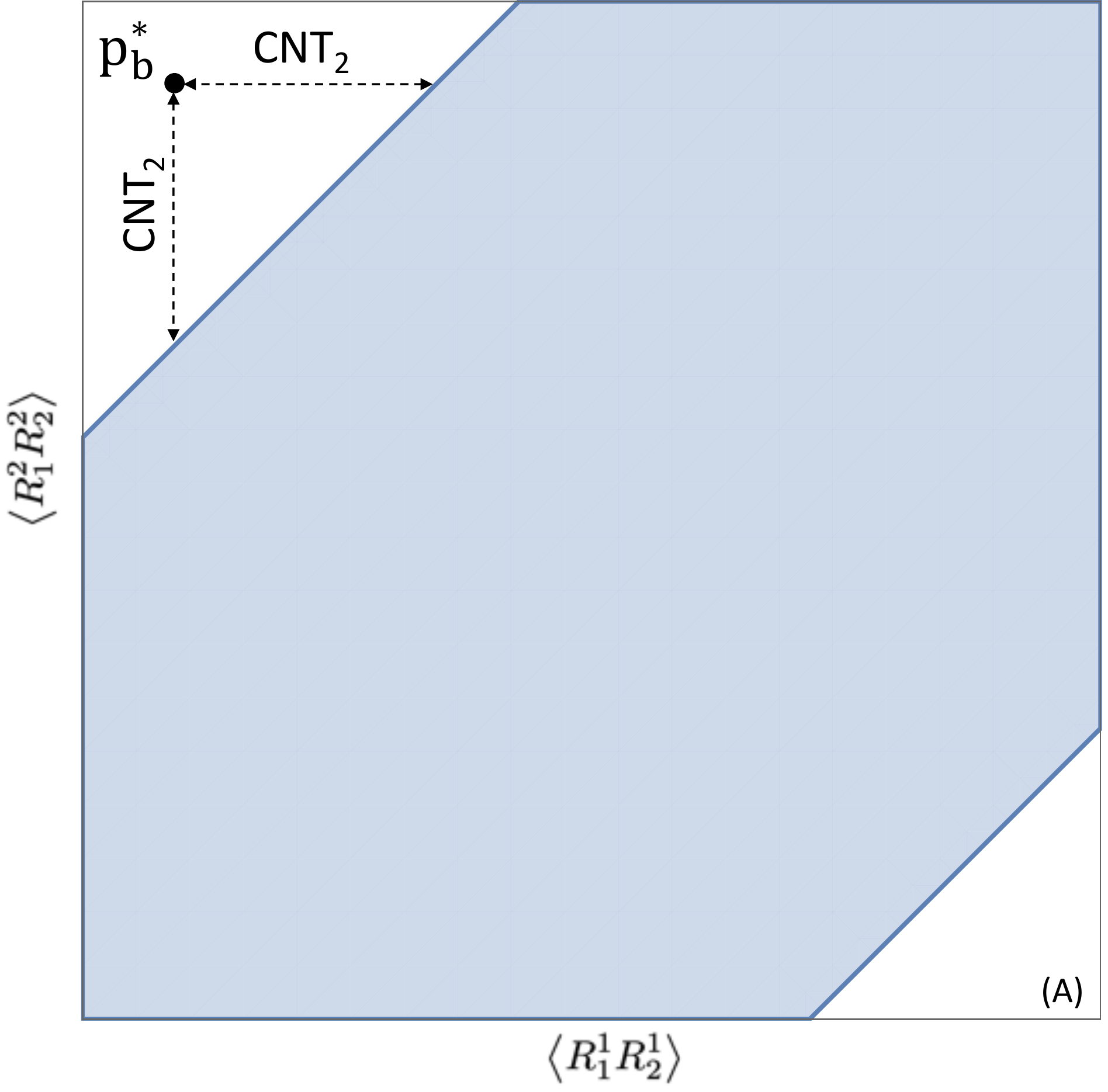}
\par\end{centering}
\begin{centering}
\includegraphics[scale=0.25]{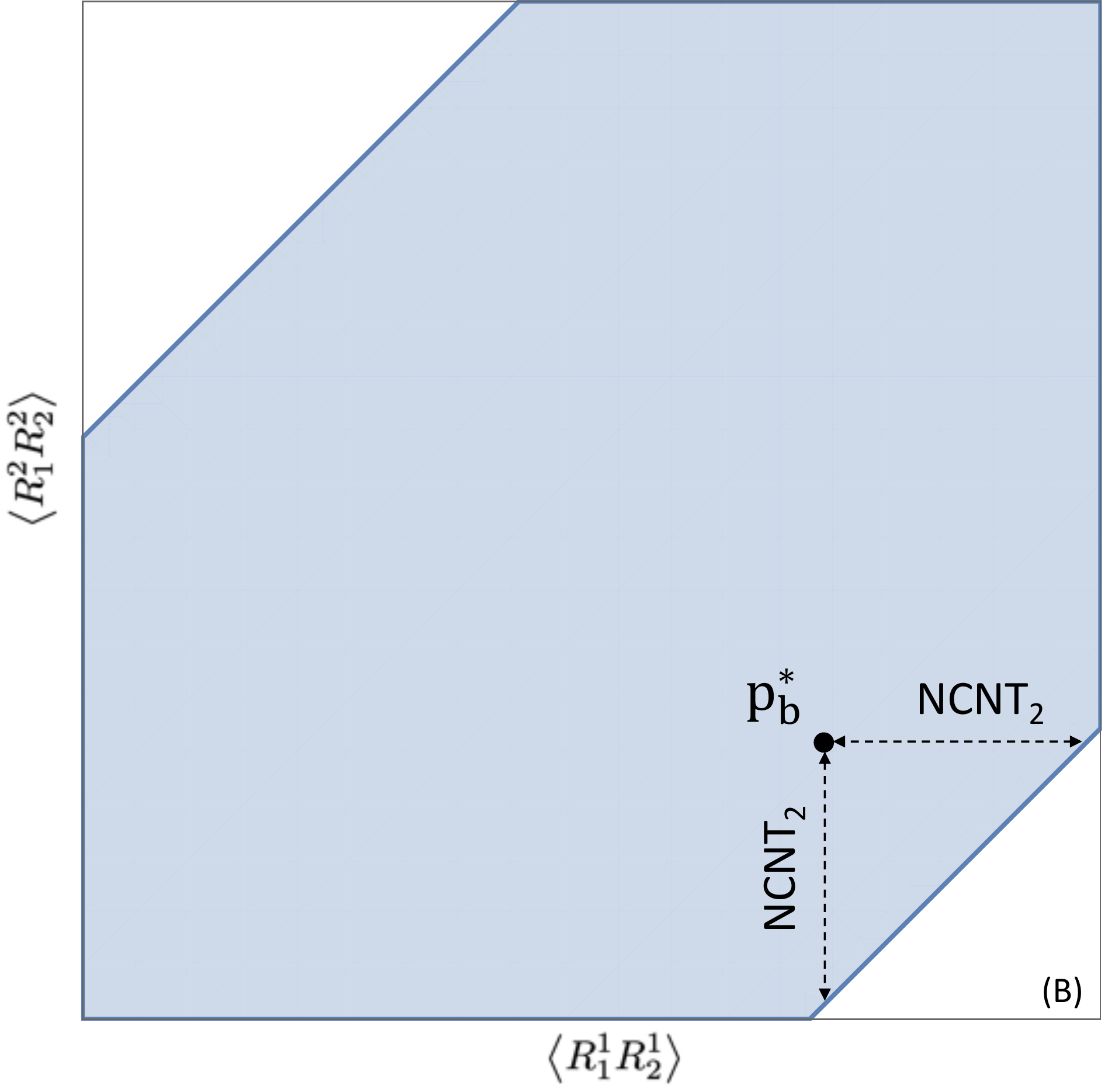}
\par\end{centering}
\begin{centering}
\includegraphics[scale=0.25]{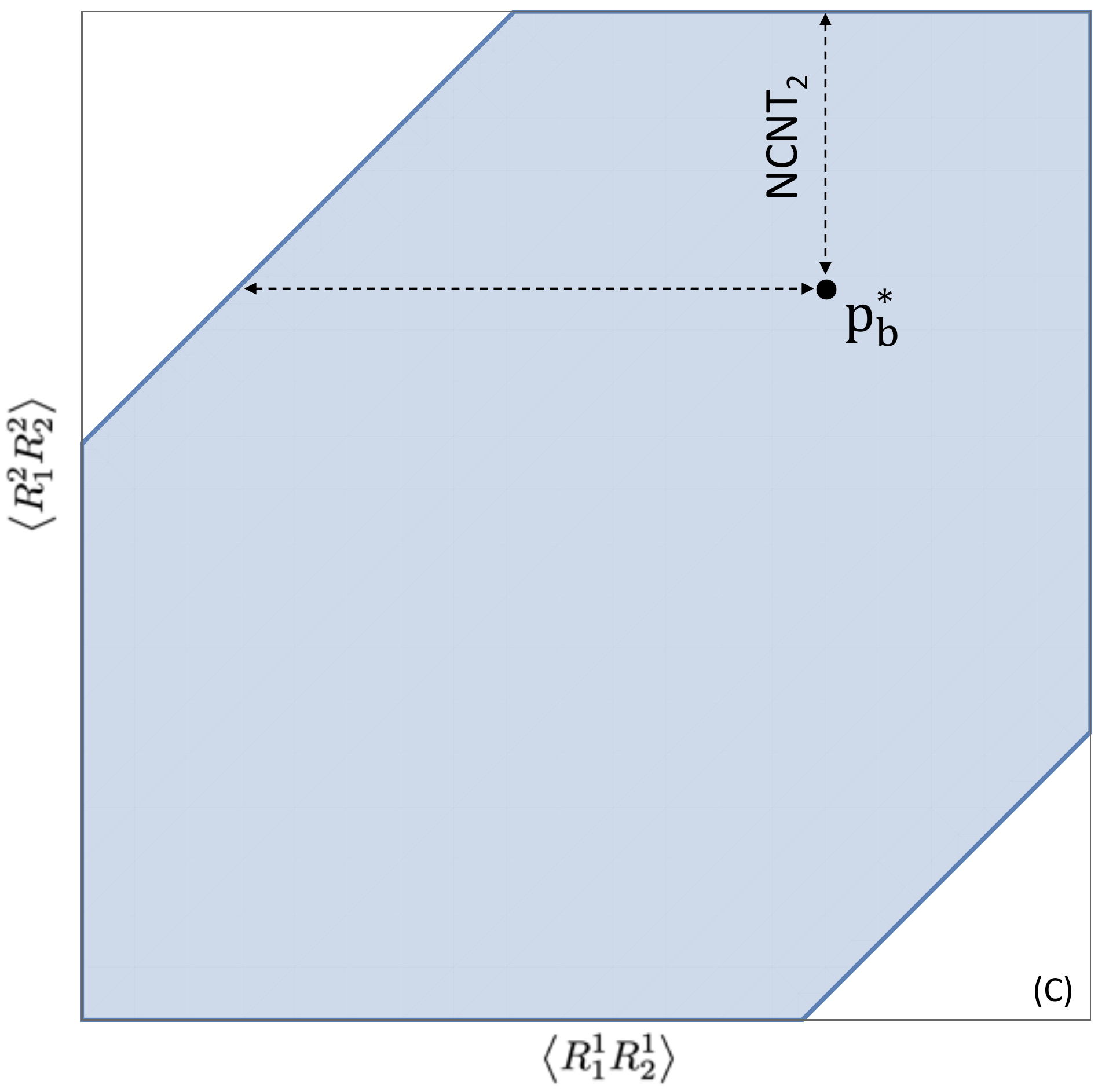}
\par\end{centering}
\caption{\label{fig:points}The relationship between a noncontextuality polytope
$\mathbb{P}_{\mathbf{b}}$ (light-colored hexagon), the circumscribing
box $\mathbb{R}_{\mathbf{b}}$ (rectangle), and a point $\mathbf{p_{b}^{*}}$
representing a system (small circle). The illustration uses cyclic
systems of rank 2, shown as $\mathcal{R}_{2}$ in (\ref{eq: R5-R2}).
The point $\mathbf{p_{b}^{*}}$ in (A) represents a contextual system,
and its $L_{1}$-distance to the polytope is $\textnormal{CNT}_{2}$;
it can be measured along any of the (here, two) coordinates. The point
$\mathbf{p_{b}^{*}}$ in (B) represents a noncontextual system, and
since its $L_{1}$-distance to an internal face of the polytope (here,
edge of the hexagon) is smaller than that to the surface of $\mathbb{R}_{\mathbf{b}}$,
this distance is $\textnormal{NCNT}_{2}$, and it can be measured
along any single coordinate. The point $\mathbf{p_{b}^{*}}$ in (C)
represents a noncontextual system which is $L_{1}$-closer to the
surface of $\mathbb{R}_{\mathbf{b}}$, and this (also single-coordinate)
distance from the surface is $\textnormal{NCNT}_{2}$.}
\end{figure}
\par\end{center}

In this paper, we provide a complete characterization of the noncontextuality
polytope, and show that the $L_{1}$-distance between this polytope
and the observed vector $\mathbf{p_{b}^{*}}$ is a single-coordinate
distance, i.e. it can be computed along a single coordinate of $\mathbf{p_{b}}$.
Moreover, when $\mathbf{p_{b}^{*}}$ is outside this polytope, this
distance is the same along all coordinates of $\mathbf{p_{b}}$ (see
Fig. \ref{fig:points}A), and it is proportional to the amount of
violation of the generalized Bell criterion derived in Ref. \citep{KujDzhProof2016}
for noncontextuality of (generally inconsistently connected) cyclic
systems.\citep{footnote2} In other words, if we schematically present
the Bell criterion as stating that a system is noncontextual if and
only if some expression $E$ does not exceed a constant $k$, then
$\textnormal{CNT}_{2}$ is proportional to $E-k$ when this value
is positive. Since precisely the same is true for $\textnormal{CNT}_{1}$
\citep{KujDzhProof2016}, with the same proportionality coefficient,
we have
\begin{equation}
\textnormal{CNT}_{2}=\textnormal{CNT}_{1}.
\end{equation}
To understand why this is the case, we characterize the polytope $\mathbb{P}$
of all possible vectors $\left(\mathbf{p_{b}},\mathbf{p_{c}}\right)$,
and show that its $L_{1}$-distance from the vector $\left(\mathbf{p_{b}^{*}},\mathbf{p_{c}^{*}}\right)$
representing the observed contextual cyclic system has the same properties
as above: it is a single-coordinate distance, the same along any of
the coordinates of $\left(\mathbf{p_{b}},\mathbf{p_{c}}\right)$.
The equality of the two measures follows from this immediately.

Despite the fact that $\textnormal{CNT}_{1}$ and $\textnormal{CNT}_{2}$
are ``mirror images'' of each other, only one of them, $\textnormal{CNT}_{2}$,
was shown in Ref. \citep{KujDzhMeasures} to be naturally extendable
to a measure of the degree of noncontextuality in noncontextual systems,
$\textnormal{NCNT}_{2}$. Geometrically, this measure is the $L_{1}$-distance
between a point $\mathbf{p_{b}^{*}}$ inside the noncontextuality
polytope $\mathbb{P}_{\mathbf{b}}$ and the polytope's surface. It
is, too, a single-coordinate distance (as is the case for any internal
point of any convex region \citep{Tuenter2006}), but its properties
are somewhat more complicated due to the structure of $\mathbb{P}_{\mathbf{b}}$.
The polytope $\mathbb{P}_{\mathbf{b}}$ is circumscribed by an $n$-box
$\mathbb{R}_{\mathbf{b}}$, so that some of the faces of $\mathbb{P}_{\mathbf{b}}$
lie within the box's interior, while others lie within its surface.
If the point $\mathbf{p_{b}^{*}}$ is $L_{1}$-closer to an internal
face of $\mathbb{P}_{\mathbf{b}}$ than to the surface of the box,
$\textnormal{NCNT}_{2}$ can be measured along any single coordinate
of $\mathbf{p_{b}}$ (see Fig. \ref{fig:points}B), and it is proportional
to the amount of compliance of the system with the generalized Bell
criteria of noncontextuality \citep{KujDzhProof2016}. In other words,
in this case $\textnormal{NCNT}_{2}$ is proportional to $k-E$ if
the criterion is written as $E\leq k$. However, $\textnormal{NCNT}_{2}$
becomes the $L_{1}$-distance between $\mathbf{p_{b}^{*}}$ and the
surface of the box $\mathbb{R}_{\mathbf{b}}$ when this distance is
smaller than that to any internal face of $\mathbb{P}_{\mathbf{b}}$
(Fig. \ref{fig:points}C). In this case, $\textnormal{NCNT}_{2}$
is not related to the Bell inequalities.

One might wonder why we could not simply define the degree of contextuality
by the amount of violation of the appropriate Bell criterion (and,
by extension, define the degree of noncontextuality by the amount
of compliance with it). Brunner and coauthors address this approach
in Ref. \citep{Brunneretal2014}, where they discuss contextuality
in the special form of nonlocality. They call this approach ``a common
choice for quantifying nonlocality,'' and correctly point out that
it is untenable, because there can be a potential infinity of the
alternatives $E'\leq k'$ to $E\leq k$ such that the two inequalities
are equivalent but $E-k$ and $E'-k'$ are grossly different. Our
approach is to define contextuality and noncontextuality as certain
distances in the space of points representing cyclic systems, and
then to see how these distances are related to specific forms of the
generalized Bell criteria of noncontextuality.

The choice of $L_{1}$-distances is natural and convenient when dealing
with probabilities, because of their additivity. However, due to the
special structure of the noncontextuality polytope, any $L_{p}$-distance
($p\geq1$), including the Euclidean ($L_{2}$) and supremal ($L_{\infty}$)
ones, are simply scaled versions of $L_{1}$:
\begin{equation}
L_{p}\equiv n^{\frac{1-p}{p}}L_{1},
\end{equation}
where $n$ is the rank of the cyclic system. The consequences of replacing
$L_{1}$-distances with other $L_{p}$-distances in our measures of
contextuality and noncontextuality are discussed in Sec. \ref{sec:Main-theorems}.

In the concluding section we consider the question of whether the
regularities established in this paper for cyclic systems extend to
noncyclic systems as well. We answer this question in the negative:
in particular, $\textnormal{CNT}_{1}$ and $\textnormal{CNT}_{2}$
are not generally equal, nor is one of them any function of the other.

\section{Cyclic systems}

\begin{figure}[h]
\begin{centering}
\includegraphics[scale=0.3]{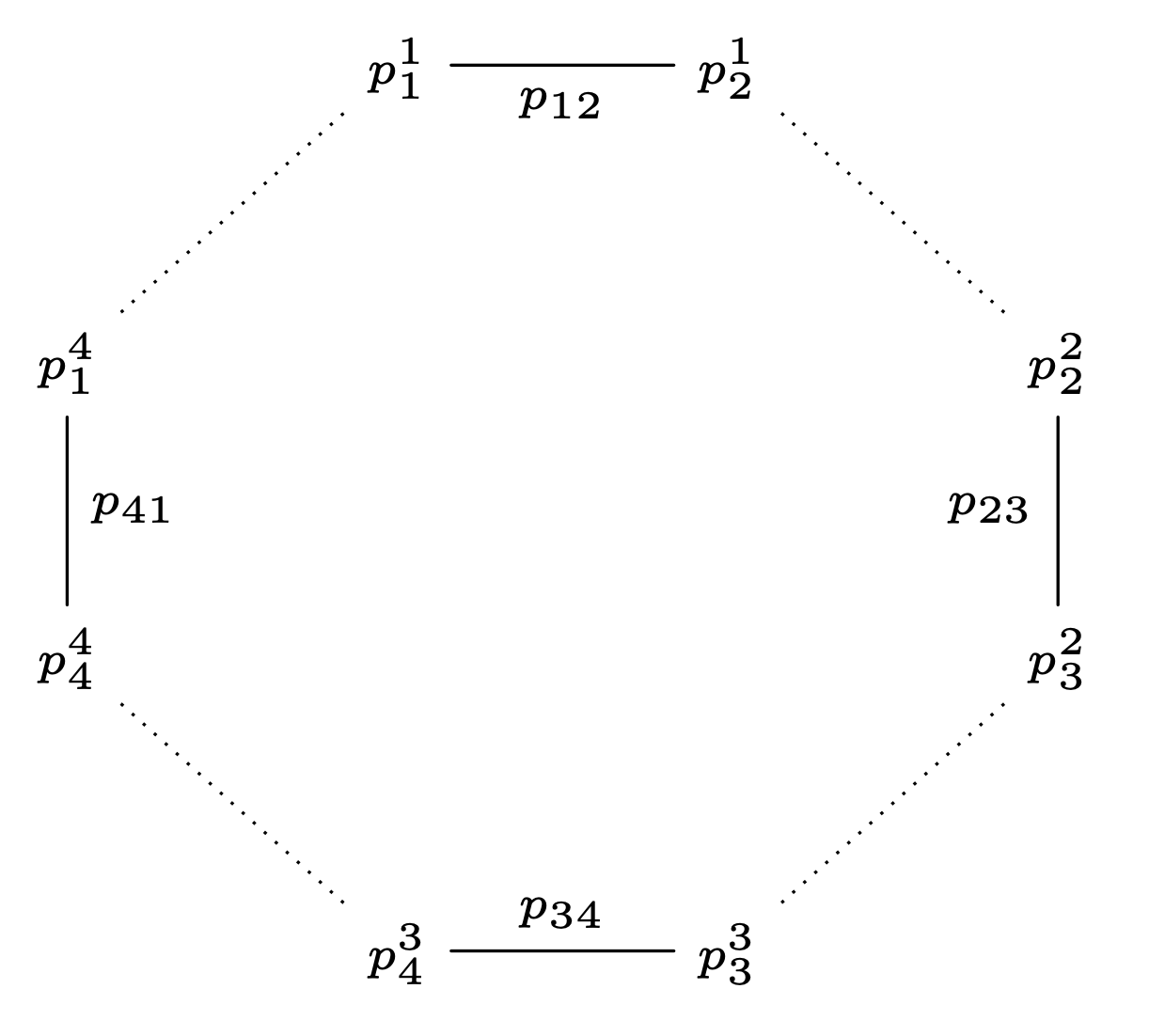}
\par\end{centering}
\caption{\label{fig:Diagram-4}Diagram of all bunch distributions for a rank-4
cyclic system (e.g., the distribution in context $c_{1}$ is described
by $p_{1}^{1}=\Pr\left[R_{1}^{1}=1\right]$, $p_{2}^{1}=\Pr\left[R_{2}^{1}=1\right]$,
and $p_{12}=\Pr\left[R_{1}^{1}=R_{2}^{1}=1\right]$.)}
\end{figure}

In each context $i=1,\ldots,n$ of the cyclic system (\ref{eq:cyclic_n}),
the joint distribution of the bunch $\left\{ R_{i}^{i},R_{i\oplus1}^{i}\right\} $
is described by three numbers,
\begin{equation}
\begin{array}{c}
\left\langle R_{i}^{i}\right\rangle =p_{i}^{i}=\Pr\left[R_{i}^{i}=1\right],\\
\left\langle R_{i\oplus1}^{i}\right\rangle =p_{i\oplus1}^{i}=\Pr\left[R_{i\oplus1}^{i}=1\right],\\
\left\langle R_{i}^{i}R_{i\oplus1}^{i}\right\rangle =p_{i,i\oplus1}=\Pr\left[R_{i}^{i}=R_{i\oplus1}^{i}=1\right].
\end{array}
\end{equation}
(One does not need a superscript for the product expectation because
the context is uniquely determined by the two contents measured in
this context.) For instance, a cyclic system of rank 4 has all bunch
distributions in it described as shown in Fig. \ref{fig:Diagram-4}.

A cyclic system therefore can be represented by two column vectors:
\begin{equation}
\mathbf{p_{l}}=\left(1,p_{1}^{1},p_{2}^{1}\ldots,p_{n}^{n},p_{1}^{n}\right)^{\intercal},
\end{equation}
which is the vector of single-variable expectations preceded by $\left\langle \right\rangle =1$
(the index $\mathbf{l}$ stands for ``low-level marginals''), and
\begin{equation}
\mathbf{p_{b}}=\left(p_{12},p_{23},\ldots,p_{n-1,n},p_{n1}\right)^{\intercal},
\end{equation}
the vector of all bunch \emph{product expectations}.

\begin{figure}
\begin{centering}
\includegraphics[scale=0.3]{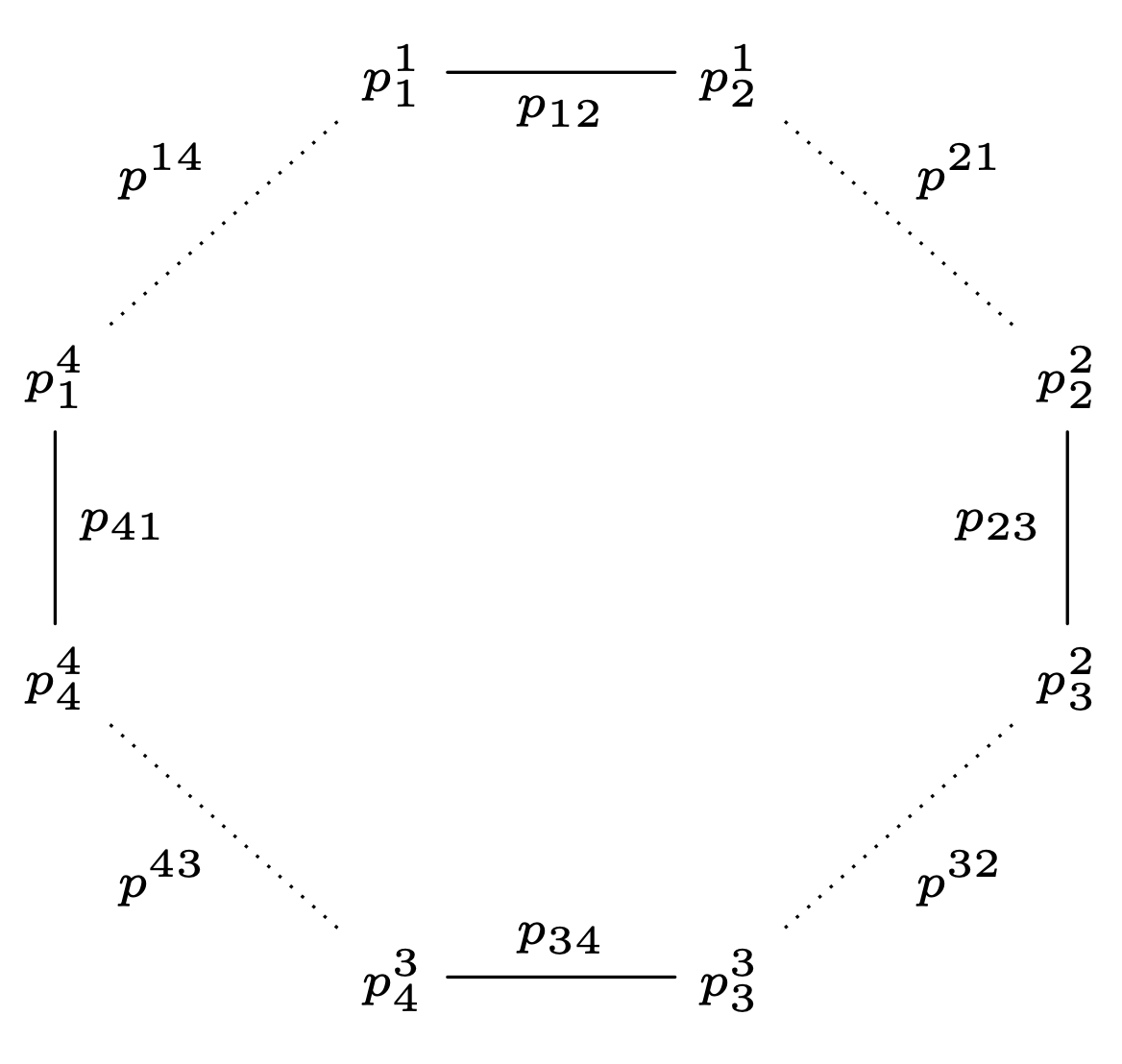}
\par\end{centering}
\caption{\label{fig:Diagram-4 with conenctions}The same diagram as in Fig.
\ref{fig:Diagram-4}, complemented with coupling distributions for
the connections (e.g., for the coupling of the connection corresponding
to $q_{1}$ the distribution is described by $p_{1}^{1}=\Pr\left[T_{1}^{1}=1\right]=\Pr\left[R_{1}^{1}=1\right]$,
$p_{1}^{4}=\Pr\left[T_{1}^{4}=1\right]=\Pr\left[R_{1}^{4}=1\right]$,
and $p^{14}=\Pr\left[T_{1}^{1}=T_{1}^{4}=1\right]$).}
\end{figure}

A \emph{coupling of a connection} $\left\{ R_{i}^{i},R_{i}^{i\ominus1}\right\} $
is a pair of \emph{jointly distributed} random variables $\left\{ T_{i}^{i},T_{i}^{i\ominus1}\right\} $
with the same 1-marginals:
\begin{equation}
\begin{array}{c}
\left\langle T_{i}^{i}\right\rangle =\left\langle R_{i}^{i}\right\rangle =p_{i}^{i},\\
\left\langle T_{i}^{i\ominus1}\right\rangle =\left\langle R_{i}^{i\ominus1}\right\rangle =p_{i}^{i\ominus1}.
\end{array}\label{eq: connect coupling}
\end{equation}
In other words, a coupling adds to each pair $p_{i}^{i},p_{i}^{i\ominus1}$
describing the connection a product expectation 
\begin{equation}
\left\langle T_{i}^{i}T_{i}^{i\ominus1}\right\rangle =p^{i,i\ominus1}=\Pr\left[T_{i}^{i}=T_{i}^{i\ominus1}=1\right],
\end{equation}
as it is shown in Fig. \ref{fig:Diagram-4 with conenctions}. This
can generally be done in an infinity of ways, constrained only by
\begin{equation}
\max\left(0,p_{i}^{i}+p_{i}^{i\ominus1}-1\right)\leq p^{i,i\ominus1}\leq\min\left(p_{i}^{i},p_{i}^{i\ominus1}\right).
\end{equation}
If couplings are constructed for all connections, they are represented
by a vector of \emph{connection product expectations},
\begin{equation}
\mathbf{p_{c}}=\left(p^{1n},p^{21},p^{32},\ldots,p^{n,n-1}\right)^{\intercal}.
\end{equation}

An (overall) \emph{coupling of the entire system} $\mathcal{R}$ is
a set 
\begin{equation}
\mathcal{S}=\left\{ S_{j}^{i}:j=i,i\oplus1;i=1,\ldots,n\right\} 
\end{equation}
of jointly distributed random variables such that, for $i=1,\ldots,n$,
\begin{equation}
\begin{array}{c}
\left\langle S_{i}^{i}\right\rangle =\left\langle R_{i}^{i}\right\rangle =p_{i}^{i},\\
\left\langle S_{i\oplus1}^{i}\right\rangle =\left\langle R_{i\oplus1}^{i}\right\rangle =p_{i\oplus1}^{i},\\
\left\langle S_{i}^{i}S_{i\oplus1}^{i}\right\rangle =\left\langle R_{i}^{i}R_{i\oplus1}^{i}\right\rangle =p_{i,i\oplus1}.
\end{array}\label{eq: coupling}
\end{equation}
In other words, a coupling $\mathcal{S}$ induces as its 1-marginals
and 2-marginals the same vectors $\mathbf{p_{l}},\mathbf{p_{b}}$
as those representing $\mathcal{R}$. An overall coupling also induces
couplings of all connections as its 2-marginals $\left(S_{i}^{i},S_{i}^{i\ominus1}\right)$,
which means that it induces a vector $\mathbf{p_{c}}$ of connection
product expectations.

\section{(Non)contextuality}

In the following it is convenient to speak of cyclic systems as \emph{represented}
by vectors
\begin{equation}
\mathbf{p}=\left(\begin{array}{c}
\mathbf{p_{l}}\\
\mathbf{p_{b}}\\
\mathbf{p_{c}}
\end{array}\right),
\end{equation}
even though $\mathbf{p_{c}}$ is computed and added to a given system.
Since this can be done in multiple ways, one and the same system is
represented by multiple vectors $\mathbf{p}$.

If in a vector $\mathbf{p_{c}}$, 
\begin{equation}
p^{i,i\ominus1}=\min\left(p_{i}^{i},p_{i}^{i\ominus1}\right),i=1,\ldots,n,
\end{equation}
then the values of $p^{i,i\ominus1}$ are maximal possible ones, and
the couplings of the connections used to compute these product expectations
are called \emph{maximal couplings}. In particular, if the system
is consistently connected, i.e., 
\begin{equation}
p_{i}^{i}=p_{i}^{i\ominus1}=p_{i},i=1,\ldots,n,
\end{equation}
then the joint and marginal probabilities in the maximal coupling
are as shown,
\begin{equation}
\begin{array}{c||c|c||c}
\textnormal{probability of} & T_{i}^{i}=1 & T_{i}^{i}=0\\
\hline\hline T_{i}^{i\ominus1}=1 & p_{i} & 0 & p_{i}\\
\hline T_{i}^{i\ominus1}=0 & 0 & 1-p_{i} & 1-p_{i}\\
\hline\hline  & p_{i} & 1-p_{i}
\end{array}\:,
\end{equation}
whence
\begin{equation}
\Pr\left[T_{i}^{i}\not=T_{i}^{i\ominus1}\right]=0,i=1,\ldots,n.\label{eq:identity}
\end{equation}
In other words, in a consistently connected system the random variables
in each connection are treated as if they were essentially the same
random variable. In the general case, with $p_{i}^{i}$ and $p_{i}^{i\ominus1}$
not necessarily equal, it is easy to show that
\begin{equation}
\Pr\left[T_{i}^{i}\not=T_{i}^{i\ominus1}\right]=\left|p_{i}^{i}-p_{i}^{i\ominus1}\right|,i=1,\ldots,n.
\end{equation}
That is, the maximal coupling $\left\{ T_{i}^{i},T_{i}^{i\ominus1}\right\} $
of $\left\{ R_{i}^{i},R_{i}^{i\ominus1}\right\} $ provides a natural
measure of difference between the two variables (in fact, it is the\emph{
total variation distance} between them). The intuitive meaning of
contextuality can be presented in the form of the following counterfactual:
if all the random variables in the system containing pairs $R_{i}^{i}$
and $R_{i}^{i\ominus1}$ were jointly distributed, it would force
some of these pairs (measuring ``the same thing'' in different contexts)
to be more dissimilar than they are in isolation.

Let us agree that an \emph{observed}, or \emph{target} system $\mathcal{R}$
(one being investigated) is represented by the vector\textbf{
\begin{equation}
\mathbf{p^{*}}=\left(\begin{array}{c}
\mathbf{p_{l}^{*}}\\
\mathbf{p_{b}^{*}}\\
\mathbf{p_{c}^{*}}
\end{array}\right),
\end{equation}
}where $\mathbf{p_{l}^{*}}$ and $\mathbf{p_{b}^{*}}$ are as they
are observed, and $\mathbf{p_{c}^{*}}$ is the vector of the maximal
connection product expectations.
\begin{defn}
\label{def:contextual}A target system $\mathcal{R}$ represented
by vector $\left(\mathbf{p_{l}^{*}},\mathbf{p_{b}^{*}},\mathbf{p_{c}^{*}}\right)^{\intercal}$
is \emph{noncontextual} if it has a coupling $\mathcal{S}$ that induces
as its marginals the vector $\mathbf{p_{c}^{*}}$ (of maximal connection
product expectations). If no such coupling exists, the system is contextual.
\end{defn}

In other words, if a system is noncontextual it has an overall coupling
that (by definition) satisfies (\ref{eq: coupling}), and also
\begin{equation}
\left\langle S_{i}^{i}S_{i}^{i\ominus1}\right\rangle =p^{i,i\ominus1}=\min\left(p_{i}^{i},p_{i}^{i\ominus1}\right),i=1,\ldots,n.
\end{equation}
In the case of consistent connectedness, CbD essentially reduces to
the conventional contextuality analysis (see Refs. \citep{DzhKujFoundations2017}
and \citep{Dzh2019} for logical ramifications of this reduction).
As an example, for a consistently connected cyclic system of rank
3,
\begin{equation}
\begin{array}{|c|c|c||c|}
\hline R_{1}^{1} & R_{2}^{1} &  & c^{1}\\
\hline  & R_{2}^{2} & R_{3}^{2} & c^{2}\\
\hline R_{1}^{3} &  & R_{3}^{3} & c^{3}\\
\hline\hline q_{1} & q_{2} & q_{3} & \mathcal{R}_{3}
\\\hline \end{array},
\end{equation}
if it is noncontextual, its coupling satisfying Definition \ref{def:contextual},
due to (\ref{eq:identity}), can be presented as
\begin{equation}
\begin{array}{|c|c|c||c|}
\hline S_{1} & S_{2} &  & c^{1}\\
\hline  & S_{2} & S_{3} & c^{2}\\
\hline S_{1} &  & S_{3} & c^{3}\\
\hline\hline q_{1} & q_{2} & q_{3} & \mathcal{S}_{3}
\\\hline \end{array},
\end{equation}
involving just three random variables recorded two at a time.

Let 
\begin{equation}
\mathbf{M=\left(\begin{array}{c}
\mathbf{M_{l}}\\
\mathbf{M_{b}}\\
\mathbf{M_{c}}
\end{array}\right)}
\end{equation}
 be a Boolean (incidence) matrix with $0/1$ cells. The $2^{2n}$
columns of $\mathbf{M}$ are indexed by events 
\begin{equation}
S_{1}^{1}=r_{1}^{1},S_{2}^{1}=r_{2}^{1},\ldots,S_{n}^{n}=r_{n}^{n},S_{1}^{n}=r_{1}^{n},\label{eq:events}
\end{equation}
while its rows are indexed by the elements of $\mathbf{p}$ (with
$\mathbf{M_{l}}$ corresponding to $\mathbf{p_{l}}$, $\mathbf{M_{b}}$
to $\mathbf{p_{b}}$, and $\mathbf{M_{c}}$ to $\mathbf{p_{c}}$).
A cell $\left(l,m\right)$ of $\mathbf{M}$ is filled with 1 if the
following is satisfied: for each random variable $S_{j}^{i}$ entering
the expectation that indexes the $l$th row of $\mathbf{M}$, the
value of $S_{j}^{i}$ in the event indexing the $m$th column of $\mathbf{M}$
is equal to 1. Otherwise the cell is filled with zero. For instance,
if the $l$th row of $\mathbf{M}$ corresponds to the expectation
$\left\langle S_{1}^{1}S_{1}^{2}\right\rangle $ in $\mathbf{p}$,
we put 1 in the cell $\left(l,m\right)$ if both $r_{1}^{1}$ and
$r_{1}^{2}$ in the event (\ref{eq:events}) corresponding to the
$m$th column of $\mathbf{M}$ are 1; otherwise the cell is filled
with zero.

Once $\mathbf{p^{*}}$and $\mathbf{M}$ are defined, one can reformulate
the definition of (non)contextuality as follows.
\begin{defn}[equivalent to Definition \ref{def:contextual}]
A target system $\mathcal{R}$ represented by vector $\mathbf{p^{*}=\left(\mathbf{p_{l}^{*}},\mathbf{p_{b}^{*}},\mathbf{p_{c}^{*}}\right)^{\intercal}}$
is noncontextual if and only if there is a vector $\mathbf{h}\ge0$
(componentwise) such that 
\begin{equation}
\mathbf{M}\mathbf{\mathbf{h}}=\mathbf{p^{*}}.
\end{equation}
Otherwise the system is contextual.
\end{defn}

It is easy to show that if such a vector $\mathbf{h}$ exists, then
it can always be interpreted as the column-vector of probabilities
\[
\Pr\left[S_{1}^{1}=r_{1}^{1},S_{2}^{1}=r_{2}^{1},\ldots,S_{n}^{n}=r_{n}^{n},S_{1}^{n}=r_{1}^{n}\right]
\]
for some overall coupling $\mathcal{S}$ of a system, across all $2^{2n}$
combinations of\textbf{ $r_{j}^{i}=0/1$}. In particular, the elements
of $\mathbf{h}$ sum to 1, because the first row of $\mathbf{M}$
and the first element of $\mathbf{p^{*}}$ consist of 1's only.

\section{Relabeling from $0/1$ to $\pm1$}

For many aspects of cyclic systems it is more convenient to label
the values of the random variables $\pm1$ rather than consider them
Bernoulli, $0/1$. This amounts to switching from $R_{j}^{i}$ variables
to $A_{j}^{i}=2R_{j}^{i}-1$. In the case of the connection couplings
(\ref{eq: connect coupling}), this means switching from $T_{j}^{i}$
to $U_{j}^{i}=2T_{j}^{i}-1$. A cyclic system $\mathcal{R}$ with
Bernoulli variables will then be renamed into a cyclic system $\mathcal{A}$
with $\pm1$-variables. We have, for $i=1,\ldots,n$,
\begin{equation}
\begin{array}{c}
\left\langle A_{j}^{i}\right\rangle =e_{j}^{i}=2p_{j}^{i}-1,j=i,i\oplus1,\\
\left\langle A_{i}^{i}A_{i\oplus1}^{i}\right\rangle =e_{i,i\oplus1}=4p_{i,i\oplus1}-2p_{i}^{i}-2p_{i\oplus1}^{i}+1,\\
\left\langle U_{i}^{i}U_{i}^{i\ominus1}\right\rangle =e^{i,i\ominus1}=4p^{i,i\ominus1}-2p_{i}^{i}-2p_{i}^{i\ominus1}+1,
\end{array}
\end{equation}
and this defines componentwise the transformation of the expectation
vectors
\begin{equation}
\left(\begin{array}{c}
\mathbf{e_{l}}\\
\mathbf{e_{b}}\\
\mathbf{e_{c}}
\end{array}\right)=\phi\left(\begin{array}{c}
\mathbf{p_{l}}\\
\mathbf{p_{b}}\\
\mathbf{p_{c}}
\end{array}\right).\label{eq:phi-transformation}
\end{equation}
The relabeling in question is useful in the formulation of the \emph{Bell-type
criterion of noncontextuality}. Let us denote

\begin{equation}
s_{1}\left(\mathbf{e_{b}}\right)=\max_{\begin{array}{c}
\lambda_{i}=\pm1,i=1,\ldots,n\\
\prod_{i=1}^{n}\lambda_{i}=-1
\end{array}}\sum\lambda_{i}e_{i,i\oplus1},\label{eq:s1}
\end{equation}
\begin{equation}
\delta\left(\mathbf{e_{l}}\right)=\sum_{i=1}^{n}\left|e_{i}^{i}-e_{i}^{i\ominus1}\right|.\label{eq:delta}
\end{equation}
and 
\begin{equation}
\Delta\left(\mathbf{e_{l}}\right)=\min\left(n-2+\delta\left(\mathbf{e_{l}}\right),n\right).\label{eq:Delta}
\end{equation}
Note that $\delta$ and $\Delta$ depend on $\mathbf{e_{l}}$, but
since this vector is fixed, we may (and will henceforth) consider
$\delta$ and $\Delta$ as constants \citep{footnote3}.
\begin{thm}[Kujala-Dzhafarov \citep{KujDzhProof2016}]
\label{thm:KD2015}A cyclic system $\mathcal{A}$ represented by
vector $\left(\mathbf{e_{l}^{*}},\mathbf{e_{b}^{*}},\mathbf{e_{c}^{*}}\right)^{\intercal}$
is noncontextual if and only if
\begin{equation}
s_{1}\left(\mathbf{e_{b}^{*}}\right)-\Delta\leq0.
\end{equation}
\end{thm}

This result generalizes the criterion derived in Ref. \citep{Araujoetal2013}
for consistently connected cyclic systems (those with $\delta=0$).

\section{\label{sec:Measures-of-contextuality}Measures of contextuality and
a measure of noncontextuality}

The idea of the two measures of contextuality considered in Ref. \citep{KujDzhMeasures},
$\textnormal{CNT}_{1}$ and $\textnormal{CNT}_{2}$, is as follows.
First we think of the space of all $\mathbf{p}=\left(\mathbf{p_{l}},\mathbf{p_{b}},\mathbf{p_{c}}\right)^{\intercal}$
obtainable as $\mathbf{p}=\mathbf{Mh}$ with $\mathbf{h}\geq0$. In
this space, we fix the 1-marginals $\mathbf{p_{l}}$ at $\mathbf{p_{l}^{*}}$
(observed values), and define the polytope
\begin{equation}
\mathbb{P}=\left\{ \mathbf{\left.\left(\begin{array}{c}
\mathbf{p_{b}}\\
\mathbf{p_{c}}
\end{array}\right)\right|}\exists\mathbf{h}\geq0:\left(\begin{array}{c}
\mathbf{p_{l}^{*}}\\
\mathbf{p_{b}}\\
\mathbf{p_{c}}
\end{array}\right)=\left(\begin{array}{c}
\mathbf{M_{l}}\\
\mathbf{M_{b}}\\
\mathbf{M_{c}}
\end{array}\right)\mathbf{h}\right\} .\label{all-couplings}
\end{equation}
This polytope describes all possible couplings of all systems with
low-marginals $\mathbf{p_{l}^{*}}$. Then we do one of the two: either
we fix $\mathbf{M_{b}h}=\mathbf{p_{b}^{*}}$ and see how close $\mathbf{p_{c}=M_{c}h}$
can be made to $\mathbf{p_{c}^{*}}$ by changing $\mathbf{h}$; or
we fix $\mathbf{M_{c}h}=\mathbf{p_{c}^{*}}$ and see how close $\mathbf{p_{b}=M_{b}h}$
can be made to $\mathbf{p_{b}^{*}}$. These two procedures define
two polytopes that we use to define $\textnormal{CNT}_{1}$ and $\textnormal{CNT}_{2}$,
respectively.
\begin{defn}
If a system $\mathcal{R}$ represented by vector $\mathbf{p}^{*}$
is contextual,
\begin{equation}
\textnormal{CNT}_{1}=L_{1}\left(\mathbf{p_{c}^{*}},\mathbb{P}_{\mathbf{c}}\right)
\end{equation}
the $L_{1}$-distance between $\mathbf{p_{c}^{*}}$ and the \emph{feasibility
polytope}
\begin{equation}
\mathbb{P}_{\mathbf{c}}=\left\{ \left.\begin{array}{c}
\\
\\
\end{array}\mathbf{p_{c}}\right|\exists\mathbf{h}\geq0:\left(\begin{array}{c}
\mathbf{p_{l}^{*}}\\
\mathbf{p_{b}^{*}}\\
\mathbf{p_{c}}
\end{array}\right)=\left(\begin{array}{c}
\mathbf{M_{l}}\\
\mathbf{M_{b}}\\
\mathbf{M_{c}}
\end{array}\right)\mathbf{h}\right\} .
\end{equation}
\end{defn}

Written in extenso, 
\begin{equation}
\textnormal{CNT}_{1}=\min_{\mathbf{p_{c}}\in\mathbb{P}_{\mathbf{c}}}\left\Vert \mathbf{p_{c}^{*}}-\mathbf{p_{c}}\right\Vert _{1}=\mathbf{1}\cdot\mathbf{p_{c}^{*}}-\max_{\mathbf{p_{c}}\in\mathbb{P}_{\mathbf{c}}}\left(\mathbf{1}\cdot\mathbf{p_{c}}\right).
\end{equation}
Because $\mathbf{p_{l}}$ is fixed at $\mathbf{p_{l}^{*}}$, the transformation
$\phi$ in (\ref{eq:phi-transformation}) has the form $4p^{i,i\ominus1}+\textnormal{const}$
for each component of $\mathbf{p_{c}}$, and we have
\begin{equation}
\left\Vert \mathbf{p_{c}^{*}}-\mathbf{p_{c}}\right\Vert _{1}=\frac{\left\Vert \mathbf{e_{c}^{*}}-\mathbf{e_{c}}\right\Vert _{1}}{4}.
\end{equation}
This allows us to redefine the measure in the way more convenient
for our purposes,
\begin{equation}
\textnormal{CNT}_{1}=\frac{1}{4}L_{1}\left(\mathbf{e_{c}^{*}},\mathbb{E}_{\mathbf{c}}\right),
\end{equation}
where (pointwise)
\begin{equation}
\mathbb{E}_{\mathbf{c}}=\phi\left(\mathbb{P}_{\mathbf{c}}\right).
\end{equation}

\begin{defn}
If a system $\mathcal{R}$ represented by vector $\mathbf{p}^{*}$
is contextual,
\begin{equation}
\textnormal{CNT}_{2}=L_{1}\left(\mathbf{p_{b}^{*}},\mathbb{P}_{\mathbf{b}}\right),
\end{equation}
the $L_{1}$-distance between $\mathbf{p_{b}^{*}}$ and the \emph{noncontextuality
polytope}
\begin{equation}
\mathbb{P}_{\mathbf{b}}=\left\{ \left.\begin{array}{c}
\\
\\
\end{array}\mathbf{p_{b}}\right|\exists\mathbf{h}\geq0:\left(\begin{array}{c}
\mathbf{p_{l}^{*}}\\
\mathbf{p_{b}}\\
\mathbf{p_{c}^{*}}
\end{array}\right)=\left(\begin{array}{c}
\mathbf{M_{l}}\\
\mathbf{M_{b}}\\
\mathbf{M_{c}}
\end{array}\right)\mathbf{h}\right\} .
\end{equation}
\end{defn}

Here, 
\begin{equation}
\textnormal{\ensuremath{\textnormal{CNT}_{2}}}=\min_{\mathbf{p_{b}}\in\mathbb{P}_{\mathbf{b}}}\left\Vert \mathbf{p_{b}^{*}}-\mathbf{p_{b}}\right\Vert _{1}.
\end{equation}
For the same reason as above, the transformation $\phi$ in (\ref{eq:phi-transformation})
has the form $4p_{i,i\oplus1}+\textnormal{const}$ for each component
of $\mathbf{p_{b}}$. We have therefore
\begin{equation}
\textnormal{CNT}_{2}=\frac{1}{4}L_{1}\left(\mathbf{e_{b}^{*}},\mathbb{E}_{\mathbf{b}}\right),
\end{equation}
the $L_{1}$-distance between $\mathbf{e_{b}^{*}=\phi\left(p_{b}^{*}\right)}$
and the polytope
\begin{equation}
\mathbb{E}_{\mathbf{b}}=\phi\left(\mathbb{P}_{\mathbf{b}}\right).\label{eq:E_b}
\end{equation}
For convenience, we will use the same term, ``feasibility polytope,''
for both $\mathbb{P}_{\mathbf{c}}$ and $\mathbb{E}_{\mathbf{c}}$.
Analogously, both $\mathbb{P}_{\mathbf{b}}$ and $\mathbb{E}_{\mathbf{b}}$
can be referred to as ``noncontextuality polytope.''

As for any two $\pm1$-random variables, we have
\begin{equation}
\left|e_{i}^{i}+e_{i\oplus1}^{i}\right|-1\leq e_{i,i\oplus1}\leq1-\left|e_{i}^{i}-e_{i\oplus1}^{i}\right|,i=1,\ldots,n.
\end{equation}
Therefore the convex polytope $\mathbb{E}_{\mathbf{b}}$ is circumscribed
by the $n$-box
\begin{equation}
\mathbb{R_{\mathbf{b}}}=\prod_{i=1}^{n}\left[\left|e_{i}^{i}+e_{i\oplus1}^{i}\right|-1,1-\left|e_{i}^{i}-e_{i\oplus1}^{i}\right|\right].\label{eq: rectangle}
\end{equation}
We can analogously define the $n$-box circumscribing $\mathbb{E}_{c}$,
but we do not need this notion.

The idea of the noncontextuality measure $\textnormal{NCNT}_{2}$
extending $\textnormal{CNT}_{2}$ to noncontextual systems is as follows.
\begin{defn}
If a system $\mathcal{R}$ represented by vector $\mathbf{p}^{*}$
is noncontextual,
\begin{equation}
\textnormal{NCNT}_{2}=L_{1}\left(\mathbf{p_{b}^{*}},\partial\mathbb{P}_{\mathbf{b}}\right)=\frac{1}{4}L_{1}\left(\mathbf{e_{b}^{*}},\partial\mathbb{E}_{\mathbf{b}}\right),
\end{equation}
the $L_{1}$ distance between $\mathbf{p_{b}^{*}}$ and the surface
$\partial\mathbb{P}_{\mathbf{b}}$ of the noncontextuality polytope
$\mathbb{P}_{\mathbf{b}}$.
\end{defn}

Note that $\textnormal{CNT}_{2}$, too, could be defined as the distance
from a point to $\partial\mathbb{P}_{\mathbf{b}}$, so the definition
is the same for both $\textnormal{CNT}_{2}$ and $\textnormal{NCNT}_{2}$,
only the position of the $\mathbf{p_{b}^{*}}$ changes from the outside
to the inside of the polytope. In extenso,
\begin{equation}
\textnormal{\ensuremath{\textnormal{NCNT}_{2}}}=\frac{1}{4}\min_{\mathbf{e_{b}}\in\partial\mathbb{E}_{\mathbf{b}}}\left\Vert \mathbf{e_{b}^{*}}-\mathbf{e_{b}}\right\Vert _{1}=\frac{1}{4}\inf_{\mathbf{x}\in\mathbf{R}^{n}-\mathbb{E}_{\mathbf{b}}}\left\Vert \mathbf{e_{b}^{*}}-\mathbf{x}\right\Vert _{1},
\end{equation}
where $\mathbf{R}$ is the set of reals. As shown in Ref. \citep{KujDzhMeasures},
no such extension to a noncontextuality measure exists for $\ensuremath{\textnormal{CNT}_{1}}$
(see Sec. \ref{sec:Polytope-of-all} for the argument by which this
is established).

\section{Additional terminology and conventions\label{sec:Additional-terminology-and}}

To focus now on $\textnormal{\ensuremath{\textnormal{CNT}_{2}}}$
and $\textnormal{\ensuremath{\textnormal{NCNT}_{2}}}$, we need a
few additional terms and conventions. We confine our consideration
to the space of all possible points $\mathbf{e_{b}}$, which is the
n-cube
\begin{equation}
\mathbb{C}_{\mathbf{b}}=\left[-1,1\right]^{n}.
\end{equation}
Given an arbitrary $n$-box
\begin{equation}
\mathbb{X}=\prod_{i=1}^{n}\left[\min x_{i},\max x_{i}\right]\subseteq\mathbb{C}_{\mathbf{b}},
\end{equation}
a vertex $V$ of $\mathbb{\mathbb{X}}$ is called \emph{odd} if its
coordinates contain an odd number of $\min x_{i}$'s; otherwise the
vertex is \emph{even}. A hyperplane is said to be \emph{pocket-forming}
at vertex $V$ if it cuts each of the $n$ edges emanating from $V$,
i.e., if it intersects each of them between $V$ and the edge's other
end. The region within $\mathbb{\mathbb{X}}$ strictly above the pocket-forming
hyperplane at $V$ is called a \emph{pocket} at $V$. This pocket
is said to be \emph{regular} if the pocket-forming hyperplane cuts
all $n$ edges emanating from $V$ at an equal distance from $V$.
We apply this terminology to two special $n$-boxes: the $n$-box
$\mathbb{R}_{\mathbf{b}}$ circumscribing the noncontextuality polytope
(\ref{eq: rectangle}), and the ambient n-cube $\mathbb{C}_{\mathbf{b}}$
itself.

We will assume in the following that no context in the system contains
a deterministic variable. If such a context exists, the $n$-box $\mathbb{R}_{\mathbf{b}}$
is degenerate (has lower dimensionality than $n$), and 
\[
\mathbb{E}_{\mathbf{b}}=\mathbb{R_{\mathbf{b}}},
\]
making the system trivially noncontextual. Indeed, assume, e.g., that
$A_{1}^{1}$ is a deterministic variable. We know that any deterministic
variable can be removed from a system without affecting its (non)contextuality
\citep{Dzh2017Nothing}. The system therefore can be presented as
a noncyclic chain
\[
A_{2}^{1},A_{2}^{2},A_{3}^{2},\ldots,A_{n}^{n},A_{1}^{n}.
\]
Whatever the joint distributions of adjacent pairs in such a chain,
there is always a global joint distribution that agrees with these
pairwise distributions as its marginals: for any assignment of values
to the links of the chain, the coupling probability is obtained as
the product of the chained conditional probabilities.

A cyclic system $\mathcal{A}$ is called a \emph{variant} of a cyclic
system $\mathcal{B}$ of the same rank if
\begin{equation}
\left\{ A_{i}^{i},A_{i}^{i\ominus1}\right\} =\pm1\cdot\left\{ B_{i}^{i},B_{i}^{i\ominus1}\right\} ,
\end{equation}
for $i=1,\ldots,n$.
\begin{lem}[Kujala-Dzhafarov \citep{KujDzhProof2016}]
\label{lem:variants}All variants of a system have the same values
of $s_{1}$$\left(\mathbf{e_{b}}\right)$ and $\left|e_{i}^{i}-e_{i}^{i\ominus1}\right|$,
$i=1,\ldots,n$ (hence also they have the same value of $\Delta$).
\end{lem}

\begin{lem}[Kujala-Dzhafarov \citep{KujDzhProof2016}]
\label{lem:Canonical form}Among the $2^{n}$ variants of a cyclic
system there is one, called canonical, in which (following a circular
permutation of indices)
\begin{equation}
\left|e_{n1}\right|\leq e_{i,i+1},i=1,\ldots,n-1.
\end{equation}
\end{lem}

Clearly, a canonical variant of a system is a canonical variant of
any variant of the system, including itself. In a canonical variant
of a system,

\begin{equation}
s_{1}\left(\mathbf{e_{b}}\right)=\sum_{i=1}^{n-1}e_{i,i+1}-e_{n1}.\label{eq: canonical s1}
\end{equation}

\section{Properties of the noncontextuality polytope\label{sec:Properties-of-the}}

In this section we present a series of lemmas establishing the remarkably
simple structure of the noncontextuality polytope. The proofs of these
results are relegated to the Appendix.
\begin{lem}
\textup{\emph{\label{lem:1 pocket}For each odd vertex $V=\left\{ \lambda_{i}:i=1,\ldots,n\right\} $
of $\mathbb{C_{\mathbf{b}}}$, the inequality $\sum\lambda_{i}e_{i,i\oplus1}>\Delta$
describes a regular pocket at $V$. The distance at which the hyperplane
segment $\sum\lambda_{i}e_{i,i\oplus1}=\Delta$ cuts each of the edges
of the cube emanating from $V$ is $n-\Delta$. (See Fig. \ref{fig:1 pocket}.)}}
\end{lem}

\begin{center}
\begin{figure}[h]
\begin{centering}
\includegraphics[scale=0.35]{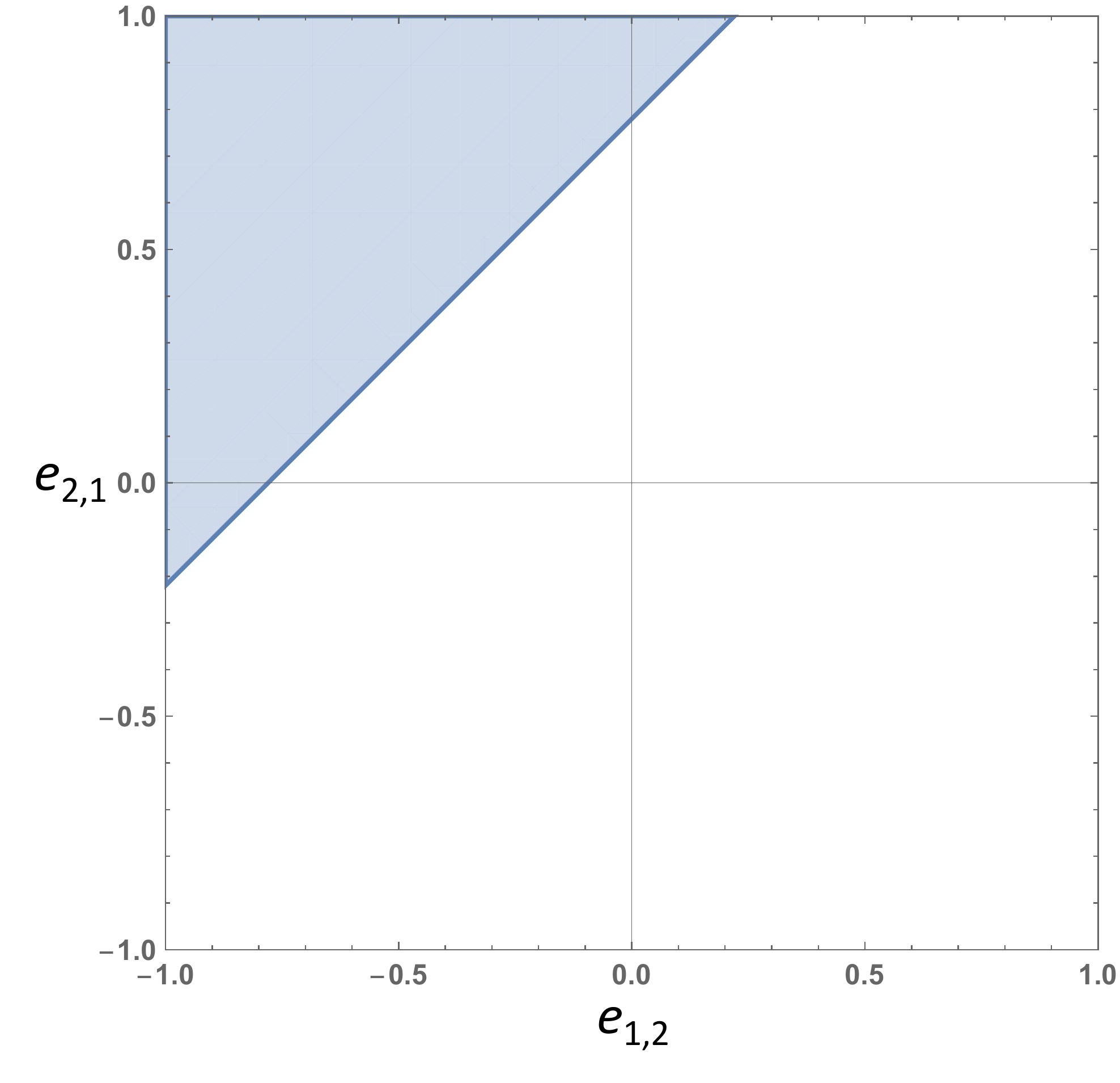}
\par\end{centering}
\begin{centering}
\includegraphics[scale=0.4]{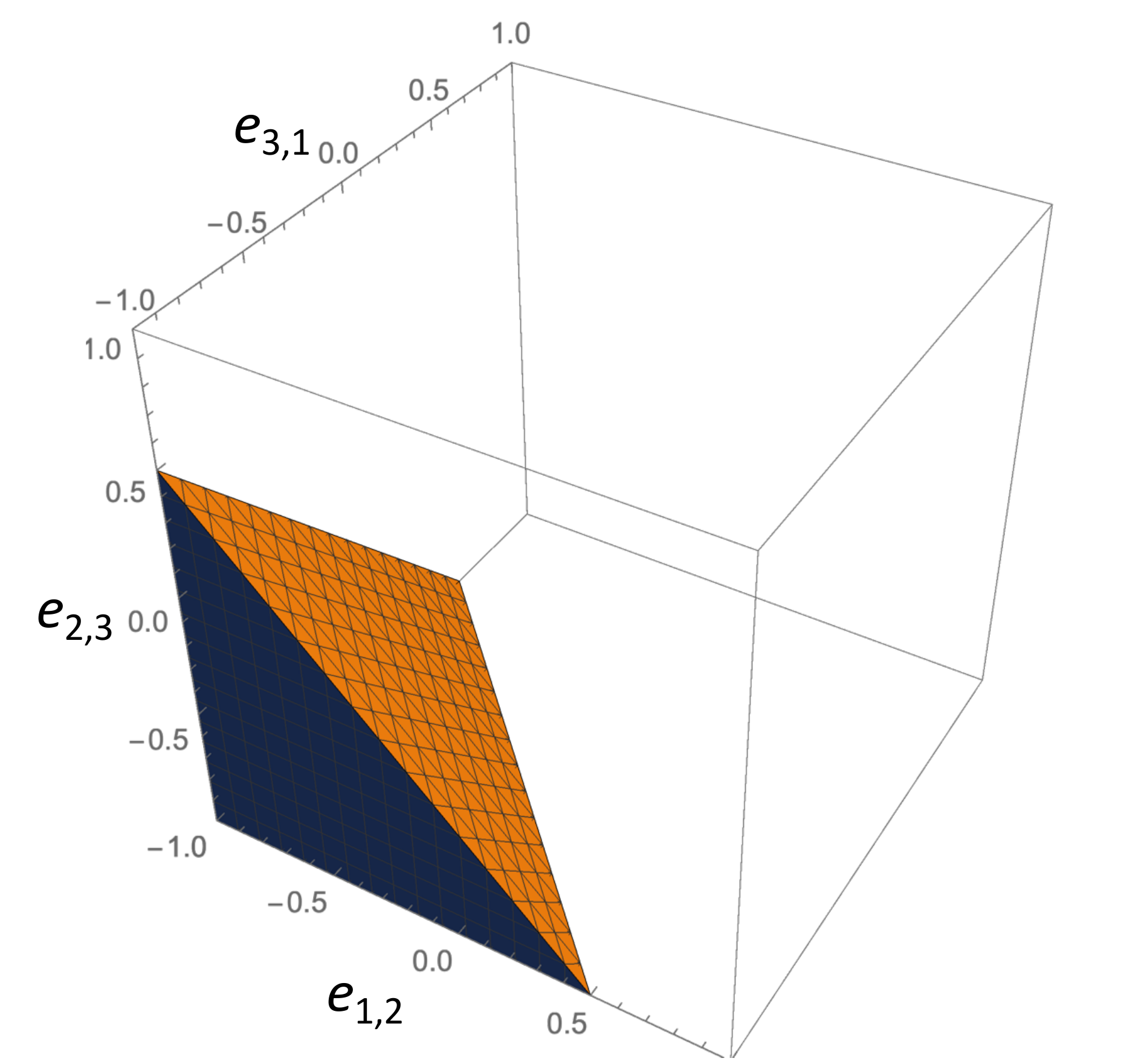}
\par\end{centering}
\caption{\label{fig:1 pocket}Illustration for Lemma \ref{lem:1 pocket}, $n=2$
and $n=3$.}
\end{figure}
\par\end{center}
\begin{lem}
\label{lem:pockets}For a given $\Delta$, no two pockets \textup{\emph{$\sum\lambda_{i}e_{i,i\oplus1}>\Delta$}}
and \textup{\emph{$\sum\lambda'_{i}e_{i,i\oplus1}>\Delta$}} formed
by the hyperplanes at different odd vertices of $\mathbb{C_{\mathbf{b}}}$
intersect. The pocket-forming hyperplanes at the odd vertices are
also disjoint within $\mathbb{C}_{\mathbf{b}}$ unless $\Delta=n-2$.
(See Figs. \ref{fig:pockets} and \ref{fig:pockets-1}.)
\end{lem}

\begin{center}
\begin{figure}[h]
\begin{centering}
\includegraphics[scale=0.35]{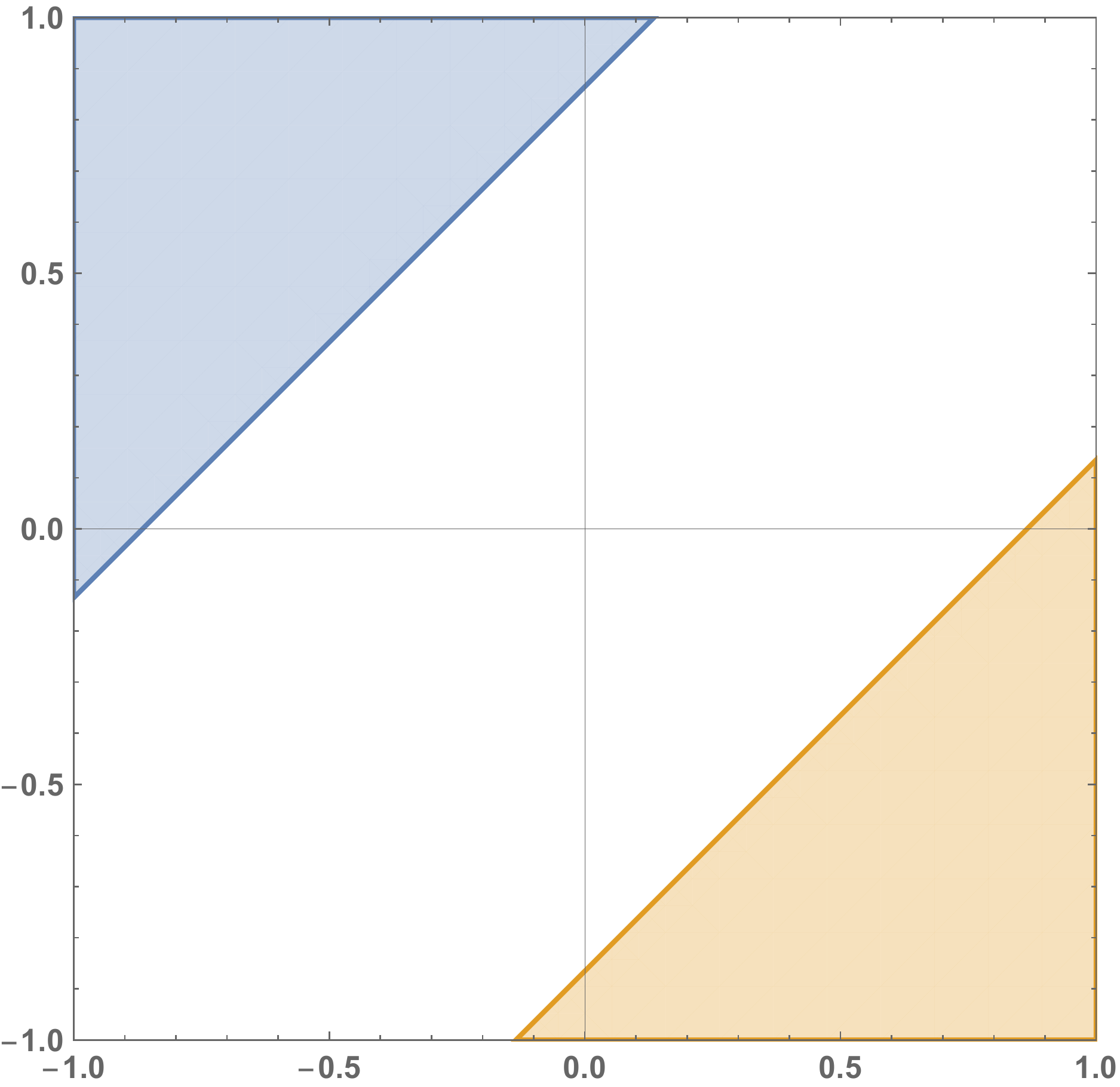}
\par\end{centering}
\begin{centering}
\includegraphics[scale=0.4]{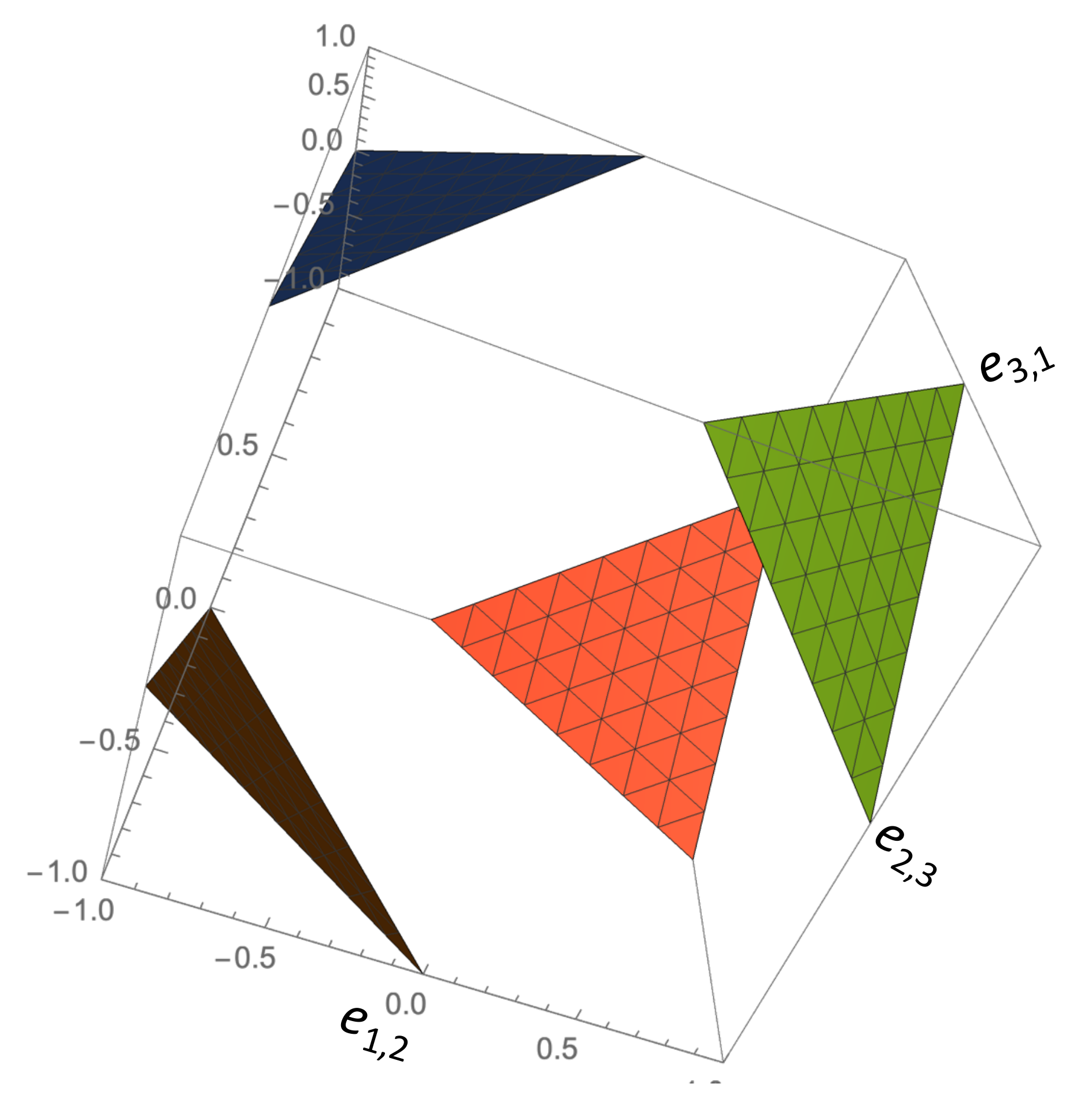}
\par\end{centering}
\caption{\label{fig:pockets}Illustration for Lemma \ref{lem:pockets}, $n=2$
and $n=3$. The pocket-forming hyperplanes at different odd vertices
of $\mathbb{C_{\mathbf{b}}}$ do not touch within $\mathbb{C}_{\mathbf{b}}$
when $\Delta>n-2$.}
\end{figure}
\par\end{center}

\begin{center}
\begin{figure}[h]
\begin{centering}
\includegraphics[scale=0.35]{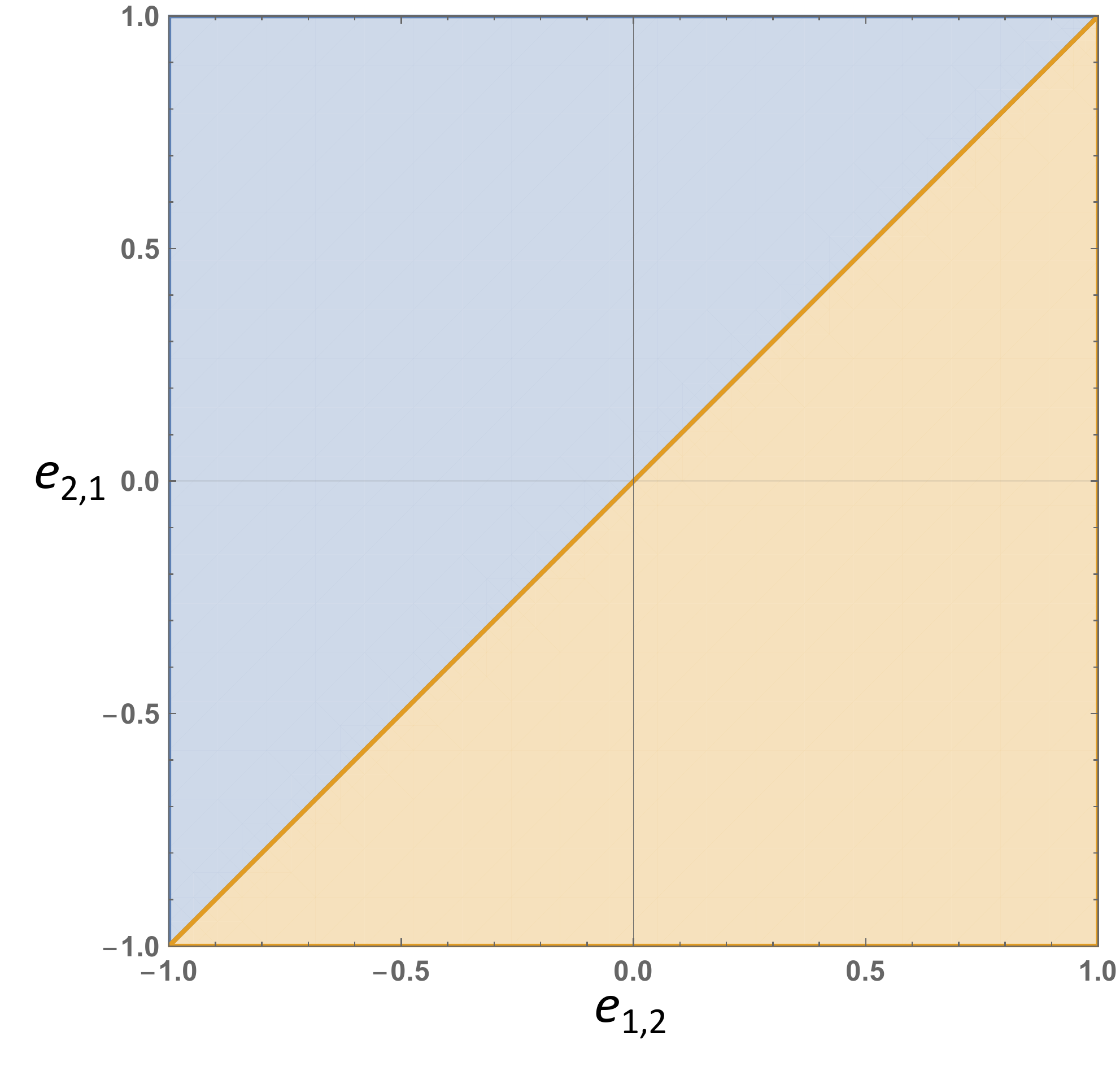}
\par\end{centering}
\begin{centering}
\includegraphics[scale=0.4]{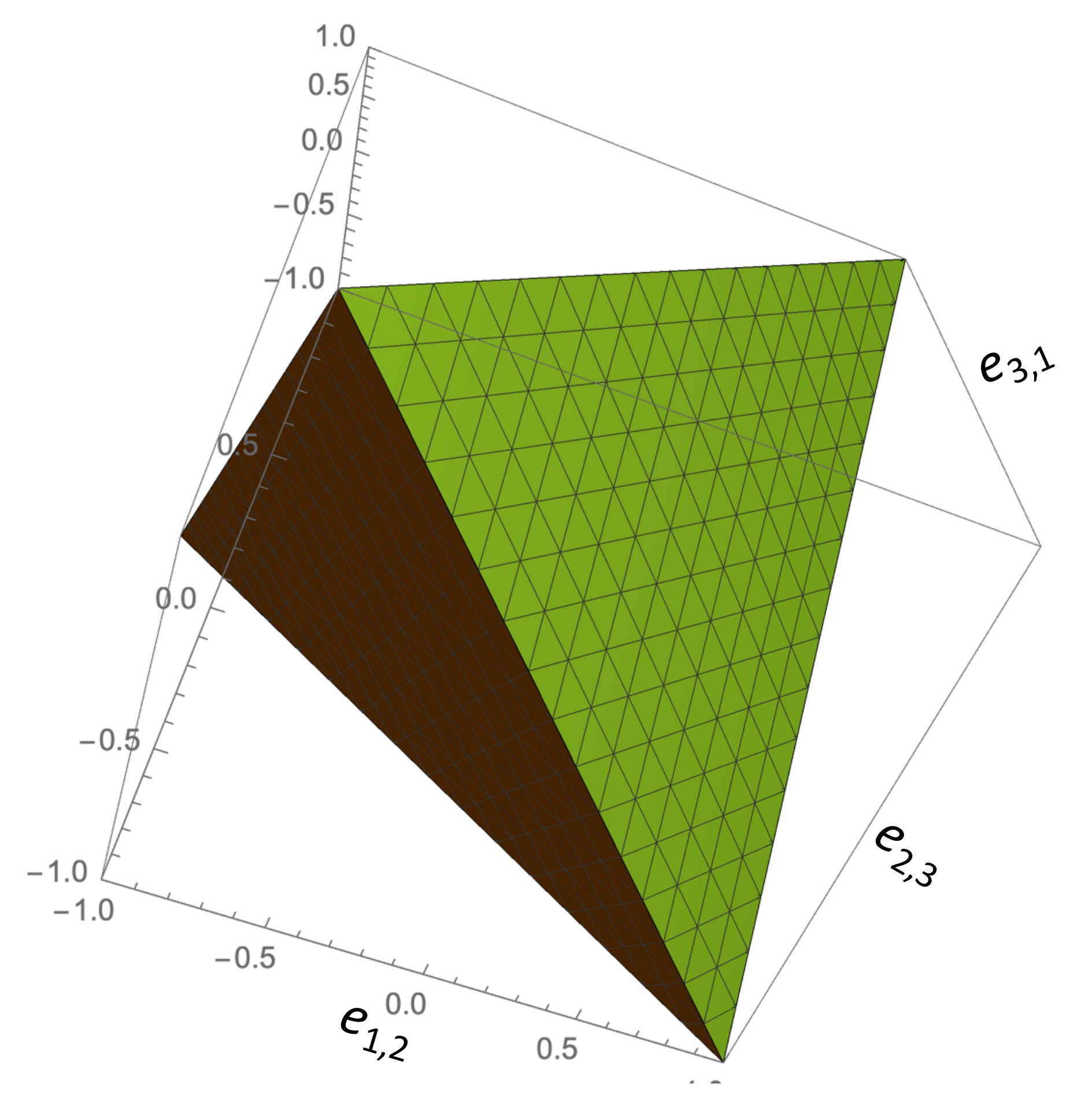}
\par\end{centering}
\caption{\label{fig:pockets-1}Illustration for Lemma \ref{lem:pockets} (continued),
$n=2$ and $n=3$. When $\Delta=n-2$, the pocket-forming hyperplanes
at different odd vertices of $\mathbb{C_{\mathbf{b}}}$ form an $n$-demicube,
the convex hull of the $2^{n-1}$ even vertices of $\mathbb{C_{\mathbf{b}}}$.}
\end{figure}
\par\end{center}
\begin{lem}
\label{lem:point outside}If a point $\mathbf{x}$ is within the pocket
formed at an odd vertex \textup{$V=\left\{ \lambda_{i}:i=1,\ldots,n\right\} $
of }$\mathbb{C}_{\mathbf{b}}$ by a hyperplane segment \textup{$\sum\lambda_{i}e_{i,i\oplus1}=\Delta$,
}\textup{\emph{then 
\[
s_{1}\left(\mathbf{x}\right)=\sum\lambda_{i}x_{i,i\oplus1}=\Delta_{\mathbf{x}}>\Delta,
\]
and $s_{1}\left(\mathbf{x}\right)-\Delta$ is the distance between
the points at which the two hyperplane segments $\sum\lambda_{i}e_{i,i\oplus1}=\Delta_{\mathbf{x}}$
and }}\textup{$\sum\lambda_{i}e_{i,i\oplus1}=\Delta$}\textup{\emph{
cut any of the edges emanating from $V$. (See Fig. \ref{fig:point outside}.)}}
\end{lem}

\begin{center}
\begin{figure}[h]
\begin{centering}
\includegraphics[scale=0.4]{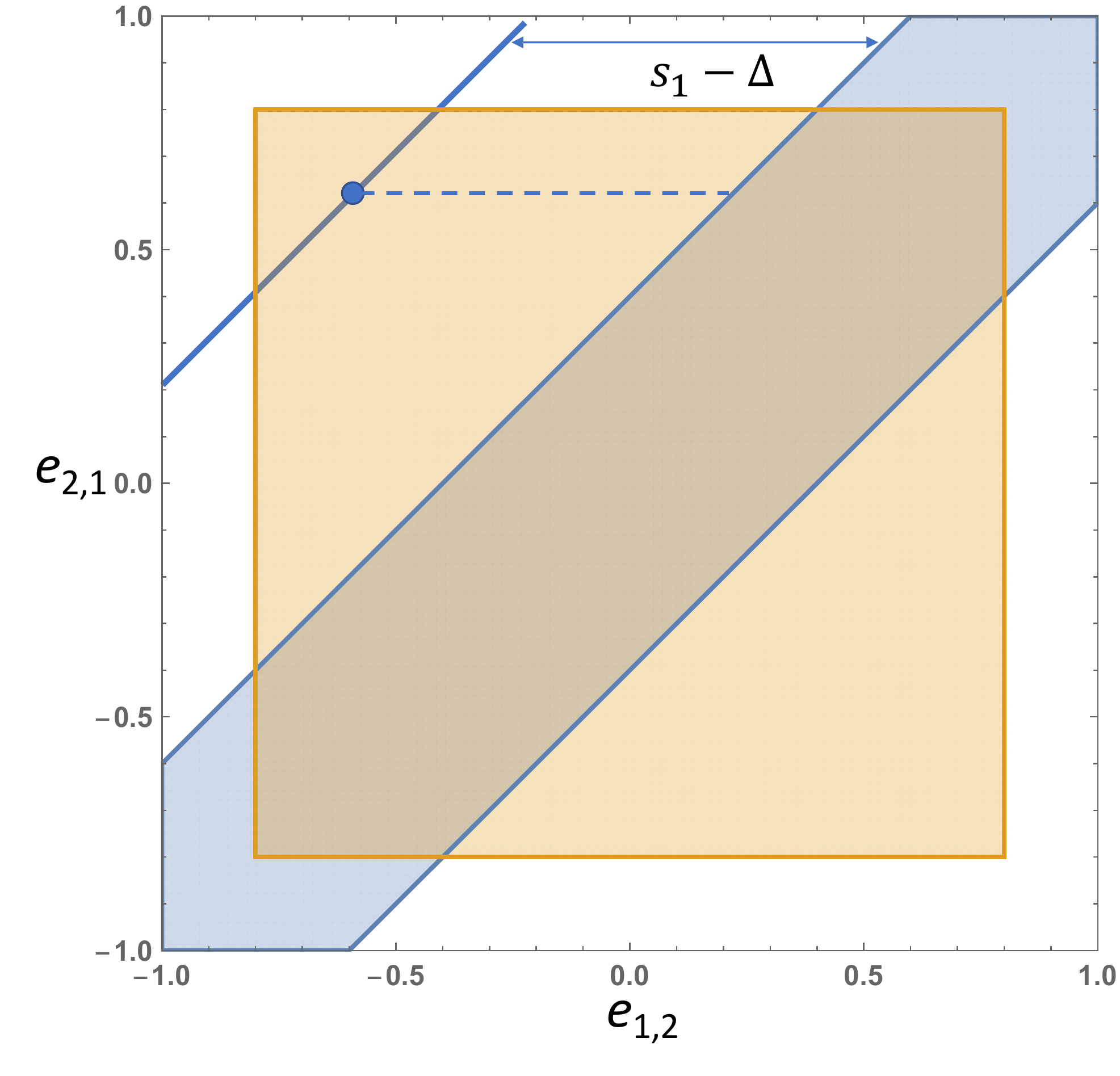}
\par\end{centering}
\caption{\label{fig:point outside}Illustration for Lemma \ref{lem:point outside},
$n=2$, and for subsequent development. The hyperplane segment $\sum\lambda_{i}e_{i,i\oplus1}=\Delta$
is shown as the left boundary of the extended noncontextuality polytope
(here, hexagon) \emph{$\mathbb{N}_{\mathbf{b}}$}. The smaller internal
rectangle represents $\mathbb{R_{\mathbf{b}}}$.}
\end{figure}
\par\end{center}

The \emph{extended noncontextuality polytope} $\mathbb{N_{\mathbf{b}}\subseteq\mathbb{C}_{\mathbf{b}}}$
is defined by $2^{n-1}$ half-space inequalities
\begin{equation}
\sum_{i=1}^{n}\lambda_{i}e_{i,i\oplus1}\leq\Delta,i=1,\ldots,n,\label{eq:halfplanes}
\end{equation}
where $n-2\leq\Delta\leq n$ and $\left\{ \lambda:i=1,\ldots,n\right\} $
are odd vertices of \emph{$\mathbb{C}_{\mathbf{b}}$ .} Therefore
we can identify $\mathbb{N}_{\mathbf{b}}$ by the value of $\Delta$,
and write $\mathbb{N}_{\mathbf{b}}=\mathbb{N}_{\mathbf{b}}\left(\Delta\right)$.
(See Fig. \ref{fig:polytope N_b}.)
\begin{center}
\begin{figure}[h]
\begin{centering}
\includegraphics[scale=0.4]{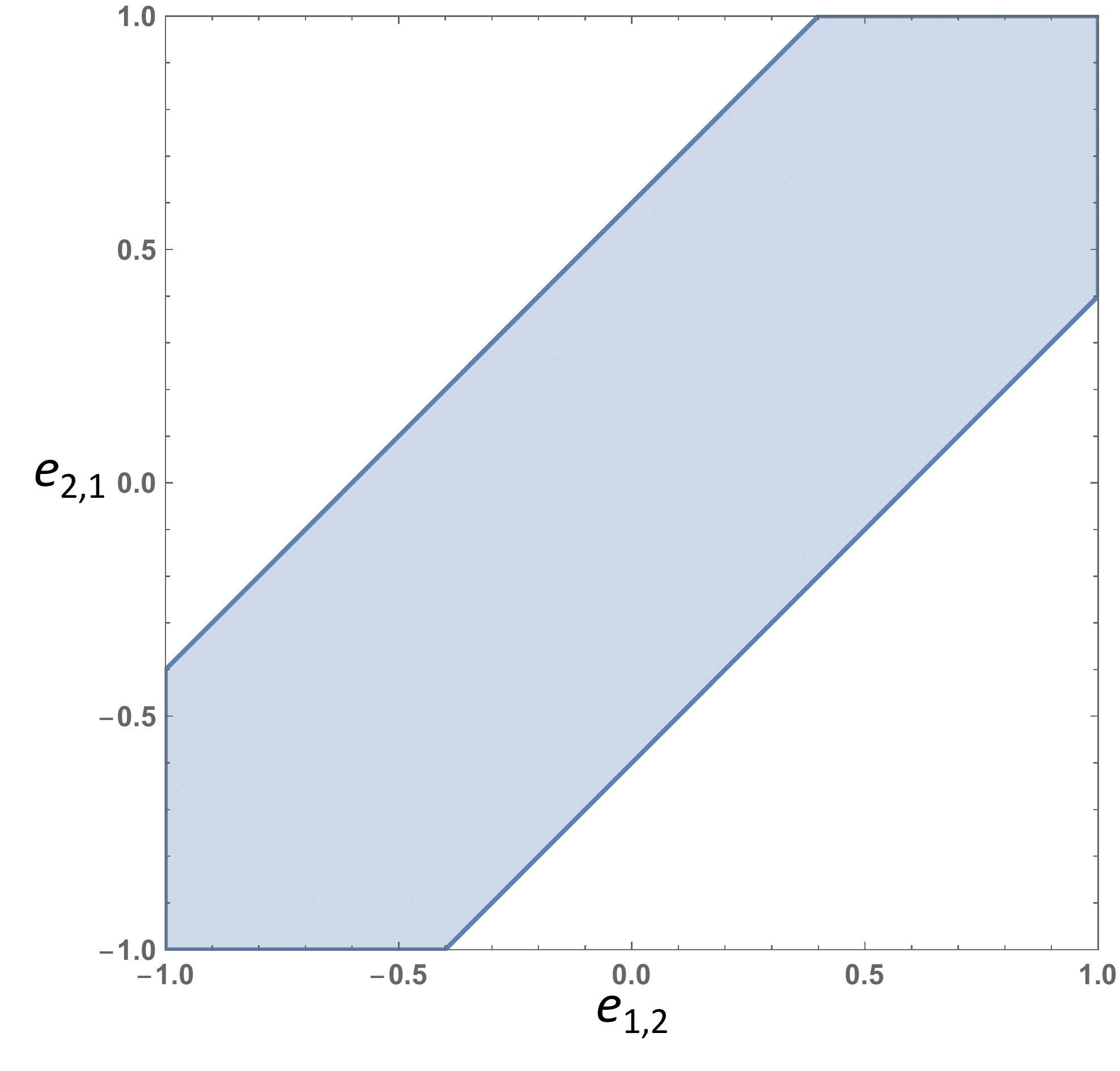}
\par\end{centering}
\begin{centering}
\includegraphics[scale=0.4]{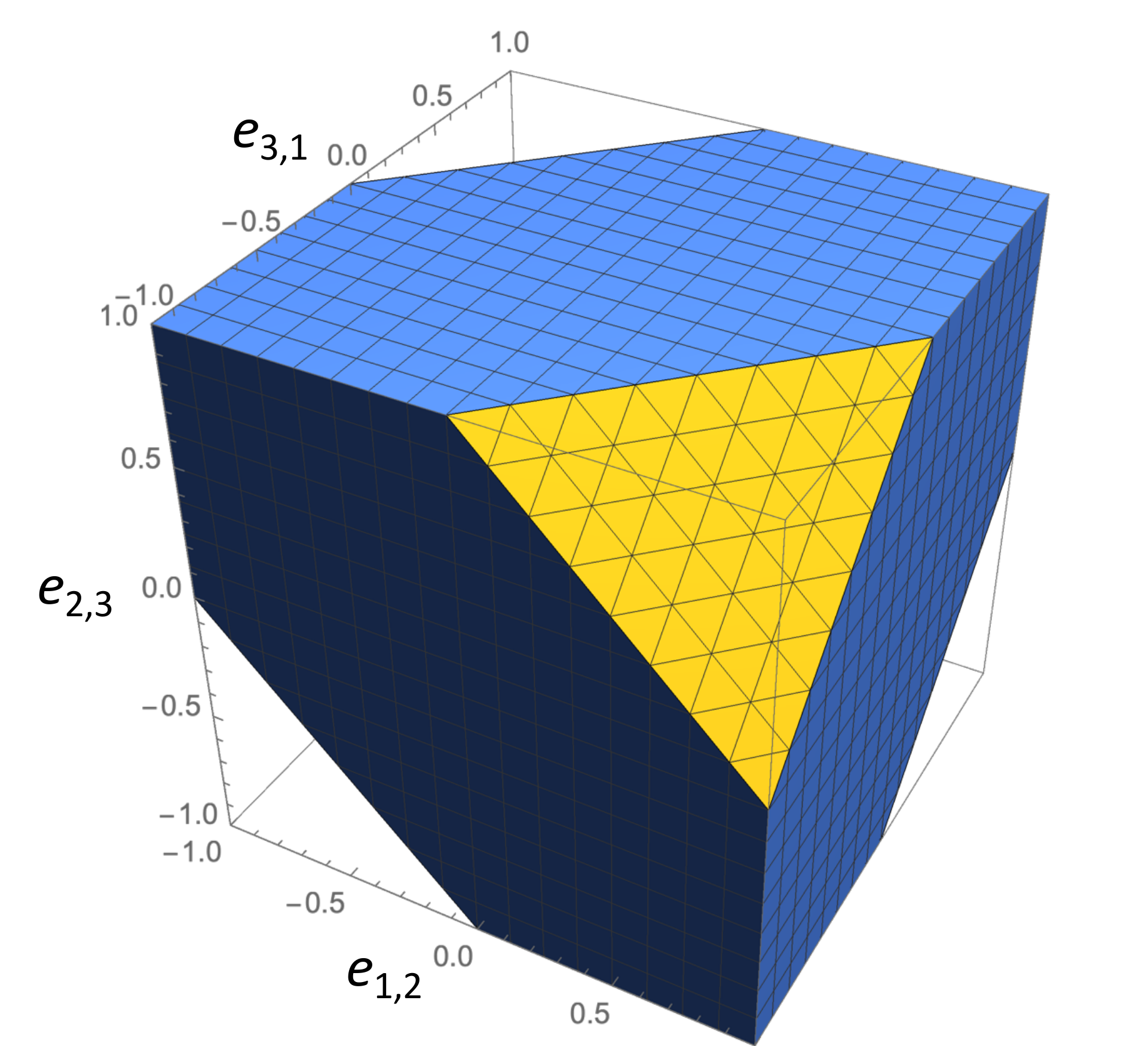}
\par\end{centering}
\caption{\label{fig:polytope N_b}Examples of polytope $\mathbb{N}_{\mathbf{b}}$
within cube $\mathbb{C_{\mathbf{b}}},$ for $n=2$ and $n=3$.}
\end{figure}
\par\end{center}
\begin{lem}
\label{lem:point within}If a point $\mathbf{x}$ is within the extended
noncontextuality polytope\emph{ $\mathbb{N}_{\mathbf{b}}\left(\Delta\right)$}\textup{\emph{,
then
\[
s_{1}\left(\mathbf{x}\right)=\sum\lambda_{i}x_{i,i\oplus1}=\Delta_{\mathbf{x}}\leq\Delta,
\]
where $V=\left\{ \lambda:i=1,\ldots,n\right\} $ is an odd vertex
of $\mathbb{C}_{\mathbf{b}}$ (unique if $\Delta>n-2$) at which the
hyperplane segment $\sum\lambda{}_{i}e_{i,i\oplus1}=\Delta_{\mathbf{x}}$
forms a pocket. The difference $\Delta-\Delta_{\mathbf{x}}$ is the
distance between the points at which the two hyperplanes cut any of
the edges emanating from $V$. (See Fig. \ref{fig:point within}.)}}
\end{lem}

The proof of Lemma \ref{lem:point within} is obvious, in view of
the previous results.
\begin{center}
\begin{figure}[h]
\begin{centering}
\includegraphics[scale=0.4]{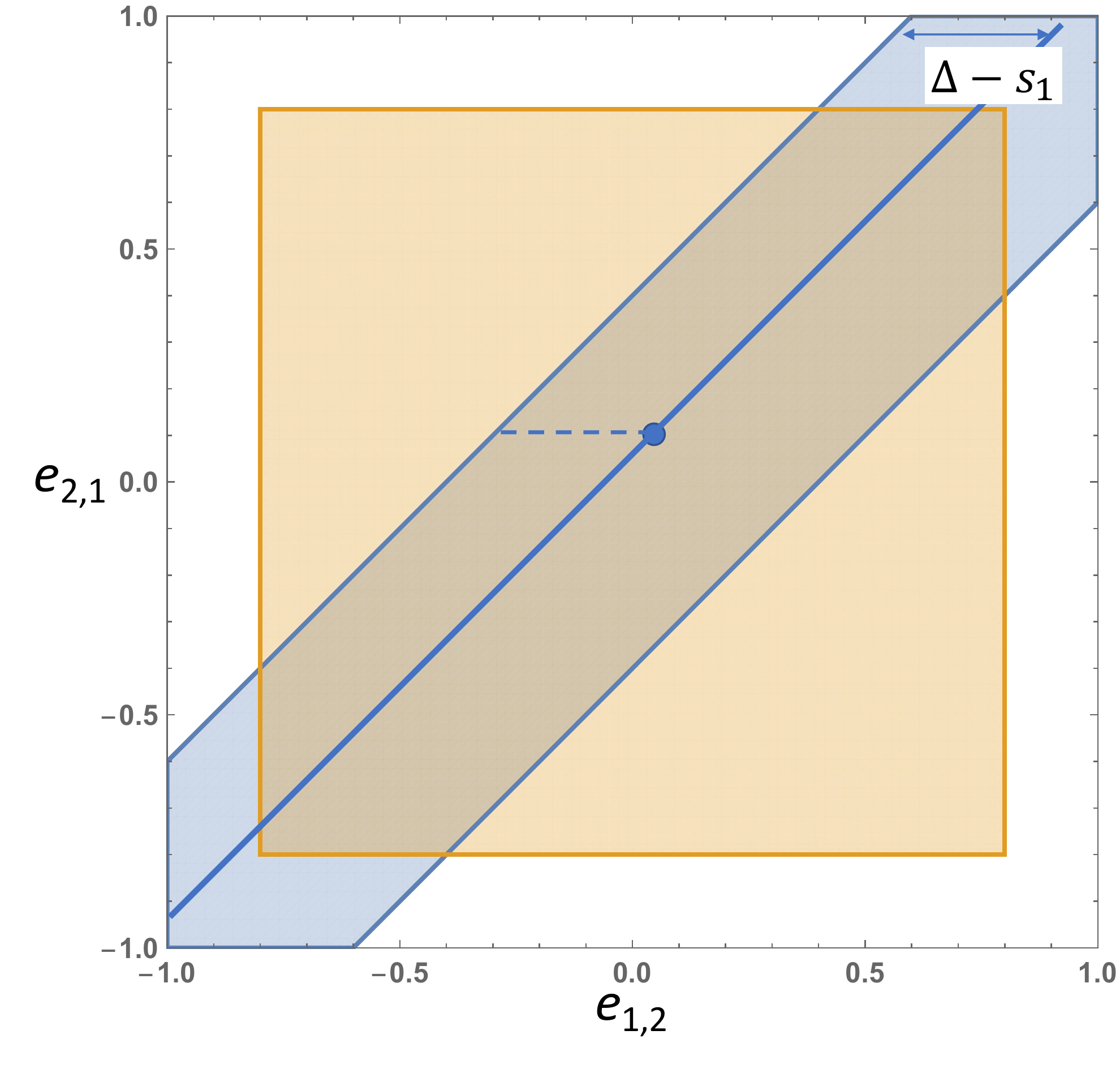}
\par\end{centering}
\caption{\label{fig:point within}Illustration for Lemma \ref{lem:point within},
$n=2$. The polytope and the box $\mathbb{R_{\mathbf{b}}}$ are as
in Fig. \ref{fig:point outside}.}
\end{figure}
\par\end{center}

We know that $\mathbb{E}_{\mathbf{b}}$ is the intersection of $\mathbb{R_{\mathbf{b}}}$
and the polytope $\mathbb{N}_{\mathbf{b}}\left(\Delta\right)$. The
following lemma stipulates an important property of this intersection.
\begin{lem}
\label{lem:even vertices}All even vertices of $\mathbb{R_{\mathbf{b}}}$
are within $\mathbb{E}_{\mathbf{b}}$. (See Fig. \ref{fig:even vertices}.)
\end{lem}

\begin{center}
\begin{figure}[h]
\begin{centering}
\includegraphics[scale=0.4]{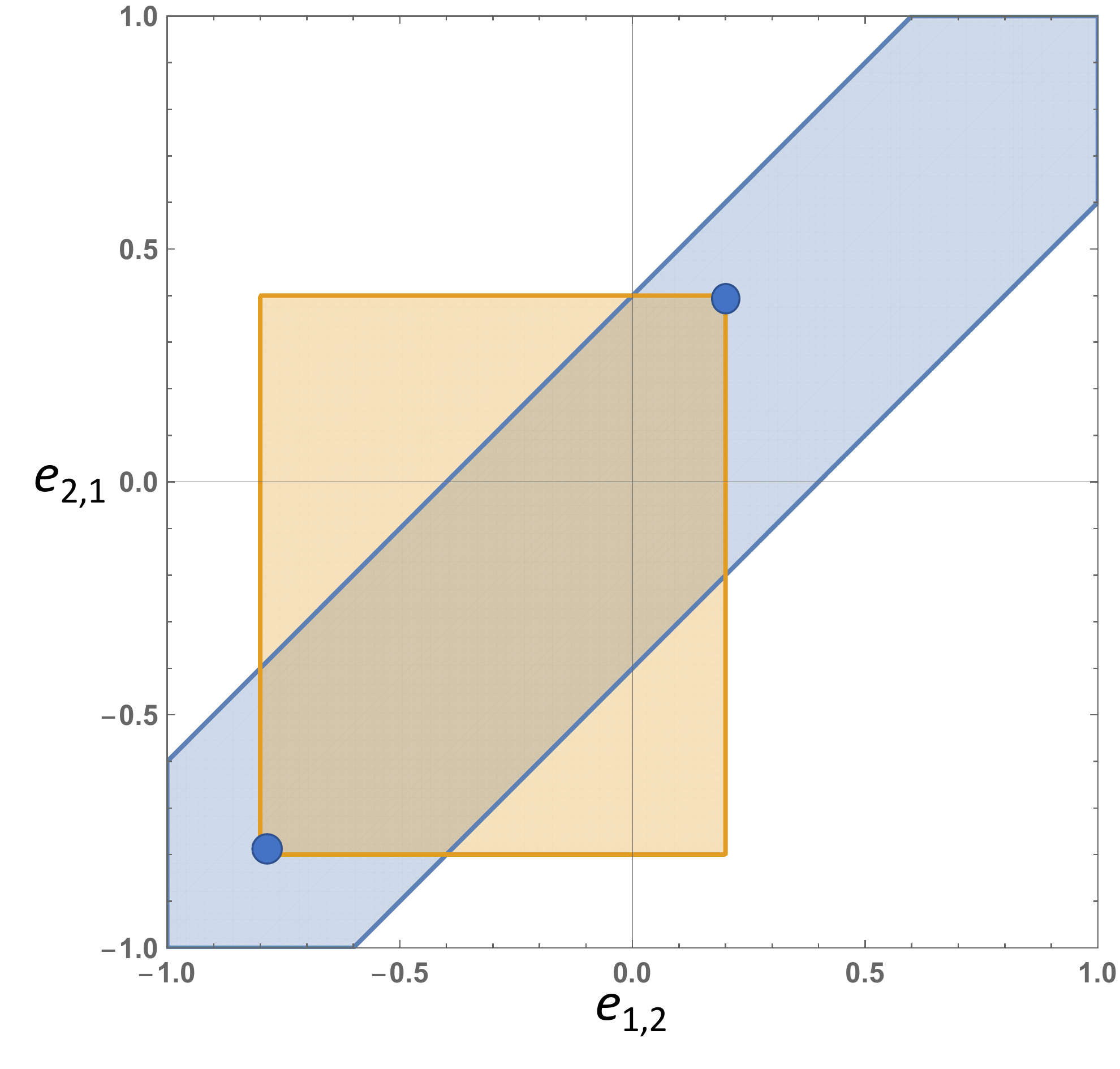}
\par\end{centering}
\begin{centering}
\includegraphics[scale=0.4]{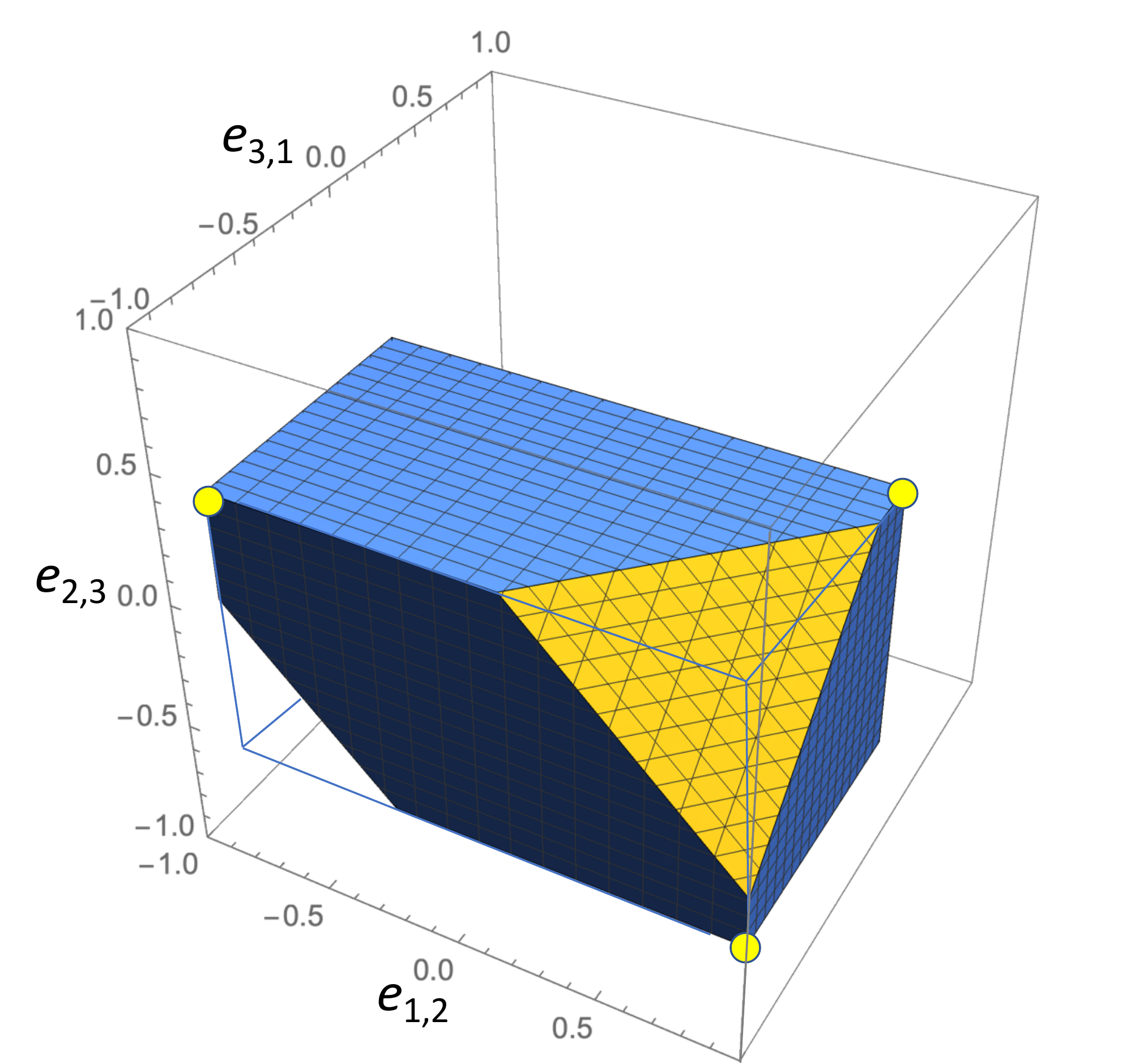}
\par\end{centering}
\caption{\label{fig:even vertices}Illustration for Lemma \ref{lem:even vertices},
$n=2$ and $n=3$: even vertices are shown by small circles.}
\end{figure}
\par\end{center}
\begin{cor}
\label{cor:k pockets}A point $\mathbf{x}$ represents a contextual
system if and only if it belongs to a pocket formed by a pocket-forming
hyperplane segment \textup{\emph{$\sum\lambda_{i}e_{i,i\oplus1}=\Delta$}}
at an odd vertex \textup{\emph{$V=\left\{ \lambda_{i}:i=1,\ldots,n\right\} $}}
of $\mathbb{R}_{\mathbf{b}}$. These pockets are regular and their
number is $0\leq k\leq2^{n-1}$. (See Fig. \ref{fig:k pockets}.)
\end{cor}

\begin{center}
\begin{figure}[h]
\begin{centering}
\includegraphics[scale=0.25]{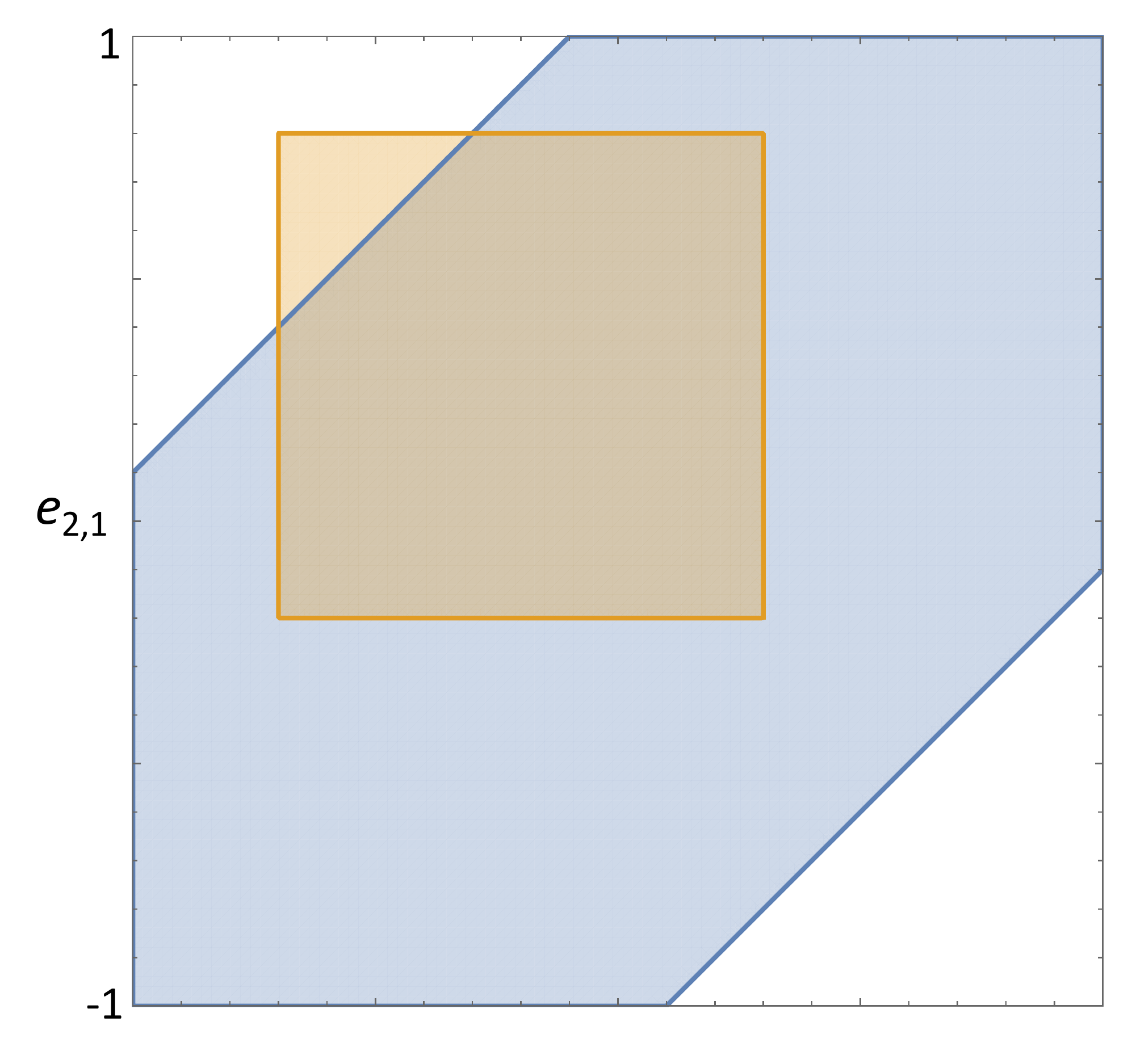}\includegraphics[scale=0.25]{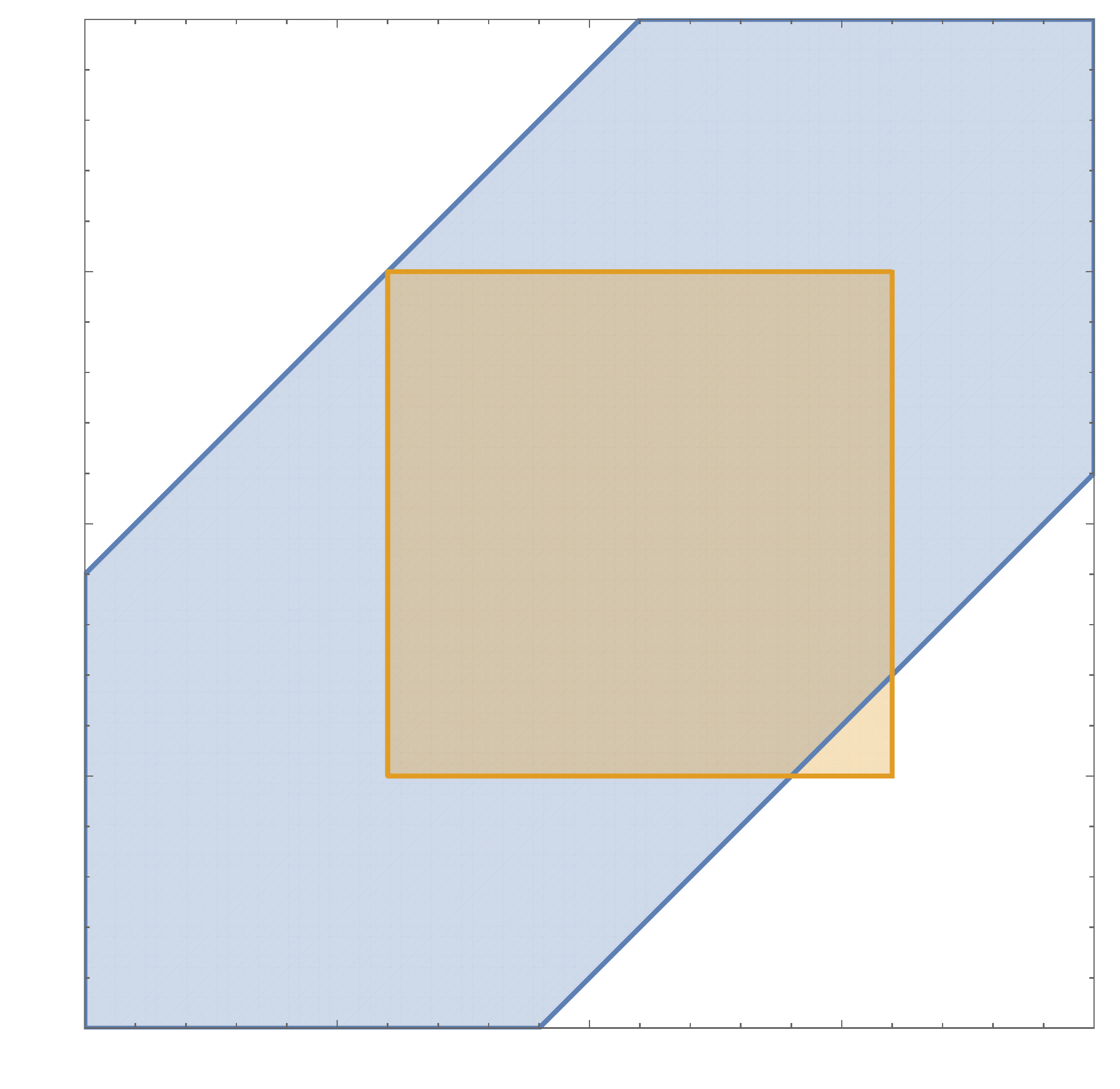}
\par\end{centering}
\begin{centering}
\includegraphics[scale=0.25]{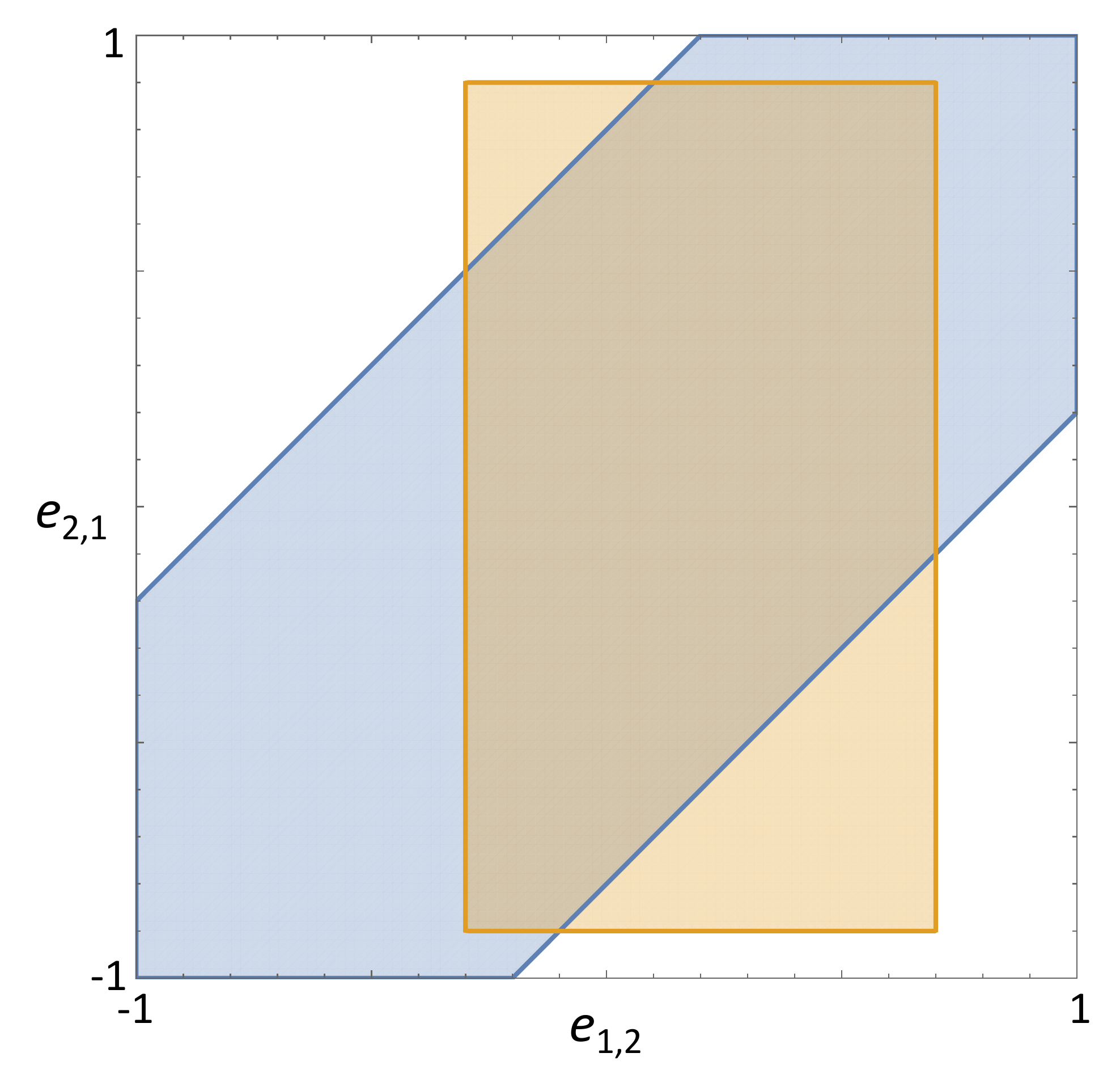}\includegraphics[scale=0.25]{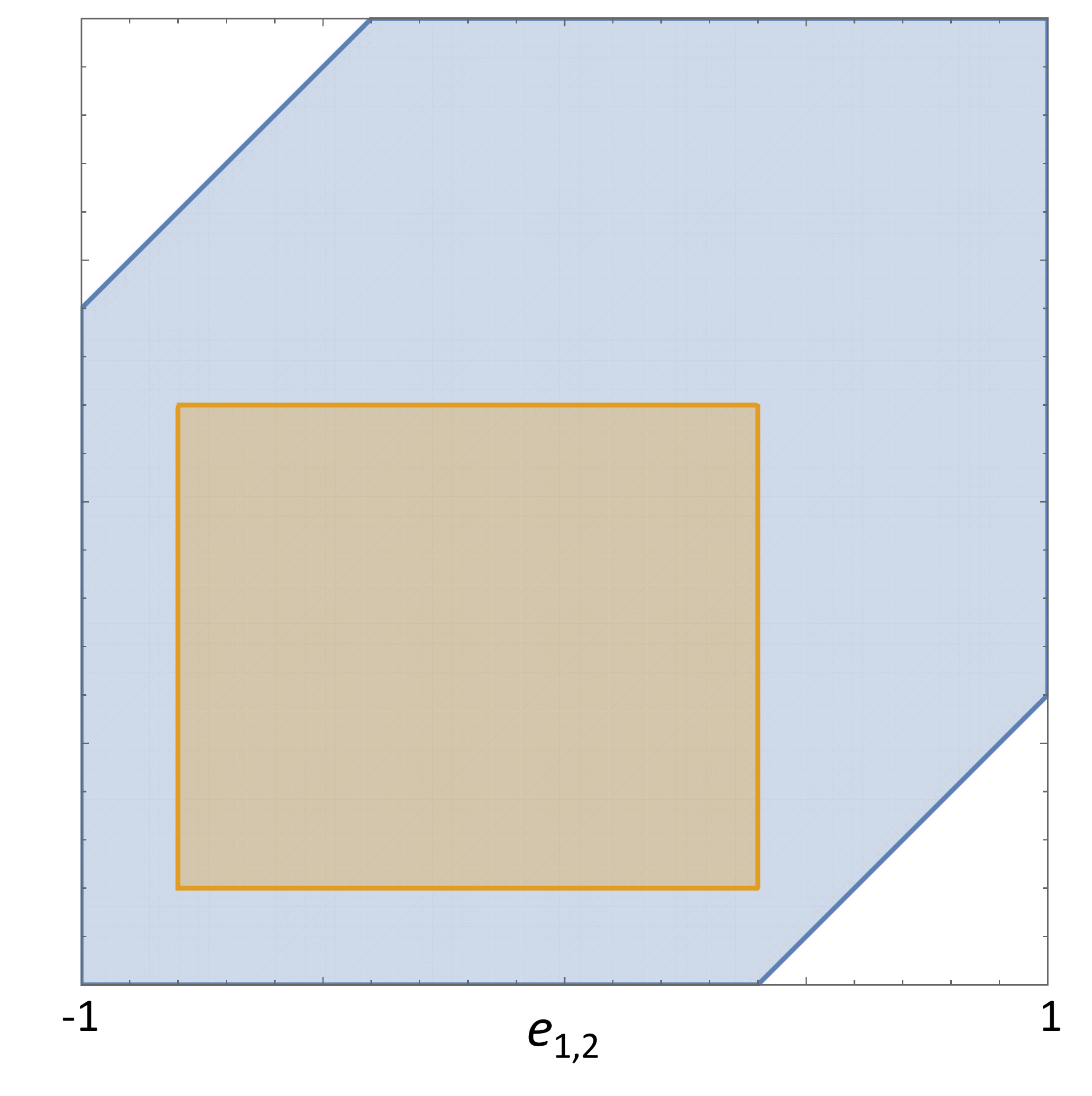}
\par\end{centering}
\caption{\label{fig:k pockets}Illustration for Corollary \ref{cor:k pockets},
$n=2$. The number of regular pockets formed at odd vertices of box
$\mathbb{R}_{\mathbf{b}}$ can be 1 (upper panels), 2 (left lower),
or zero (right lower).}
\end{figure}
\par\end{center}

\section{Main theorems\label{sec:Main-theorems}}

The following two theorems now are simple corollaries of the previous
results. Consider a noncontextuality polytope 
\begin{equation}
\mathbb{E}_{\mathbf{b}}=\mathbb{R}_{\mathbf{b}}\cap\mathbb{N}_{\mathbf{b}}\left(\Delta\right).
\end{equation}

\begin{thm}
\label{thm:Main contextual}The $L_{1}$-distance between $\mathbb{E}_{\mathbf{b}}$
and a point $\mathbf{e_{b}^{*}}$ representing a contextual system\emph{
}\textup{\emph{is a single-coordinate distance, equal to $s_{1}\left(\mathbf{e_{b}^{*}}\right)-\Delta$
for all coordinates. This is the value of }}$\textnormal{4\ensuremath{\cdot}\ensuremath{\ensuremath{\textnormal{CNT}_{2}}}}$.
(See Fig. \ref{fig:CNT2}.)
\end{thm}

\begin{center}
\begin{figure}[h]
\begin{centering}
\includegraphics[scale=0.4]{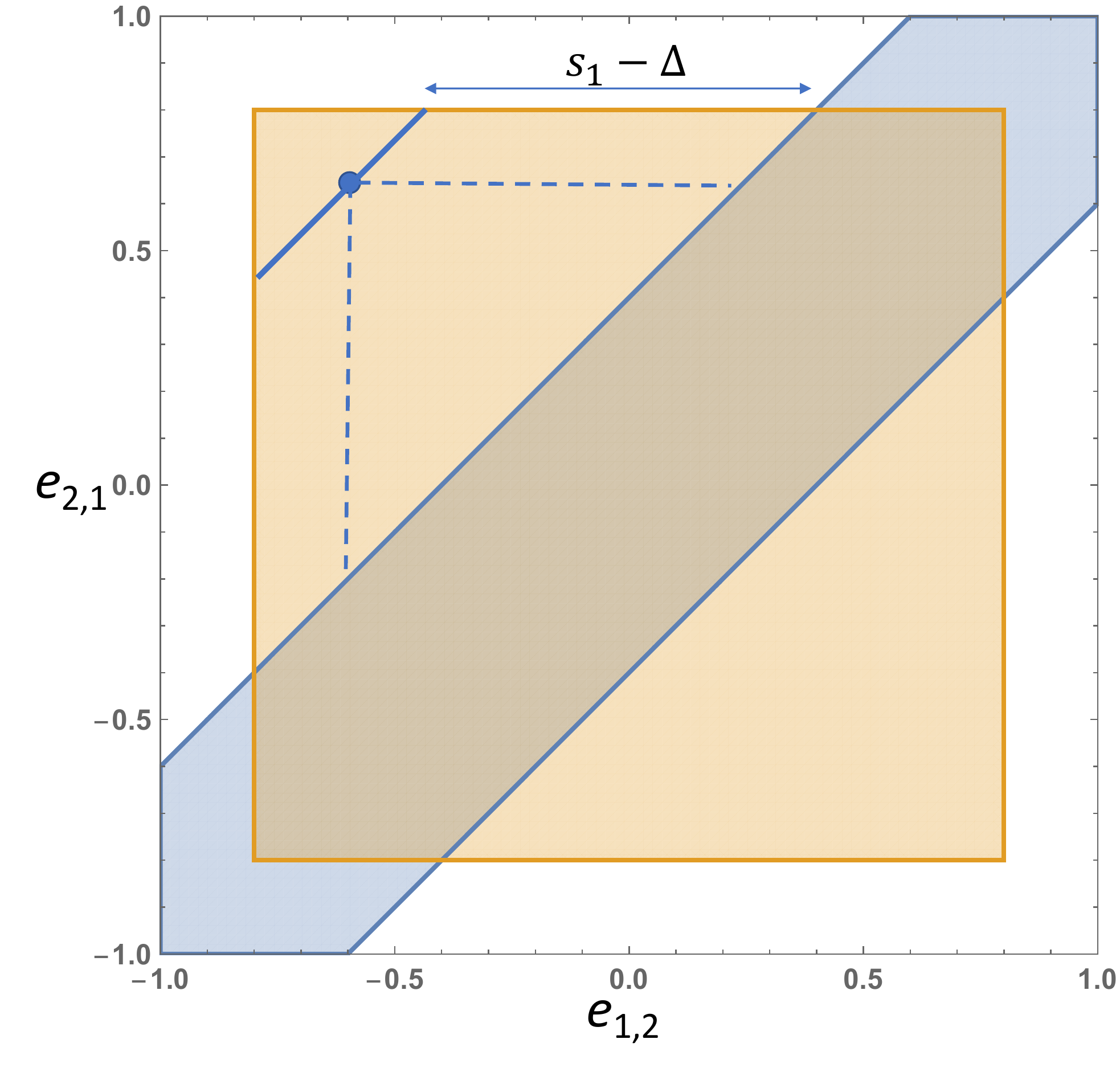}
\par\end{centering}
\caption{\label{fig:CNT2}Illustration for Theorem \ref{thm:Main contextual},
$n=2$, a detailed analog of Fig. \ref{fig:points}A.}
\end{figure}
\par\end{center}

It is easy to show that for any $p\geq1$, the $\frac{1}{4}$ of the
$L_{p}$-distance between $\mathbf{e_{b}^{*}}$ and $\mathbb{E}_{\mathbf{b}}$
(let us call it $\textnormal{CNT}_{2}^{\left(p\right)}$) is simply
\begin{equation}
\textnormal{CNT}_{2}^{\left(p\right)}=n^{\frac{1-p}{p}}\textnormal{CNT}_{2},
\end{equation}
where $n$ is the rank of the cyclic system. This means that in the
case of contextual cyclic systems $L_{1}$-distance can be, if one
so wishes, replaced by any $L_{p}$-distance with no nontrivial changes
in the theory. However, this may not be possible for noncyclic systems,
where the faces of the noncontextuality polytope need not have the
simple structure of $\mathbb{E}_{\mathbf{b}}$.

Let us define a new measure now, the $L_{1}$-distance between the
box $\mathbb{R}_{\mathbf{b}}$ and a point $\mathbf{e_{b}}$ within
the box: 
\begin{equation}
m\left(\mathcal{\mathbf{\mathbf{e_{b}}}}\right)=\min_{i=1,\ldots,n}\left(\min\left(\begin{array}{c}
e_{i,i\oplus1}-\left|e_{i}+e_{i\oplus1}\right|+1,\\
1-\left|e_{i}^{i}-e_{i\oplus1}^{i}\right|-e_{i,i\oplus1}
\end{array}\right)\right).
\end{equation}

\begin{thm}
\label{thm:NCNT2}The $L_{1}$-distance between the surface of $\mathbb{E}_{\mathbf{b}}$
and a point $\mathbf{\mathbf{e_{b}^{*}}}$ representing a noncontextual
system \textup{\emph{is a single-coordinate distance, equal to $\min\left(\Delta-s_{1}\left(\mathbf{\mathbf{\mathbf{e_{b}^{*}}}}\right),m\left(\mathbf{\mathbf{e_{b}^{*}}}\right)\right)$.
This is the value of }}$4\cdot\textnormal{\ensuremath{\textnormal{NCNT}_{2}}}$.\textup{\emph{
If this value equals $s_{1}\left(\mathbf{e^{*}}\right)-\Delta,$ it
is the same for all coordinates. (See Fig. \ref{fig:NCNT2}.)}}
\end{thm}

\begin{center}
\begin{figure}[h]
\begin{centering}
\includegraphics[scale=0.35]{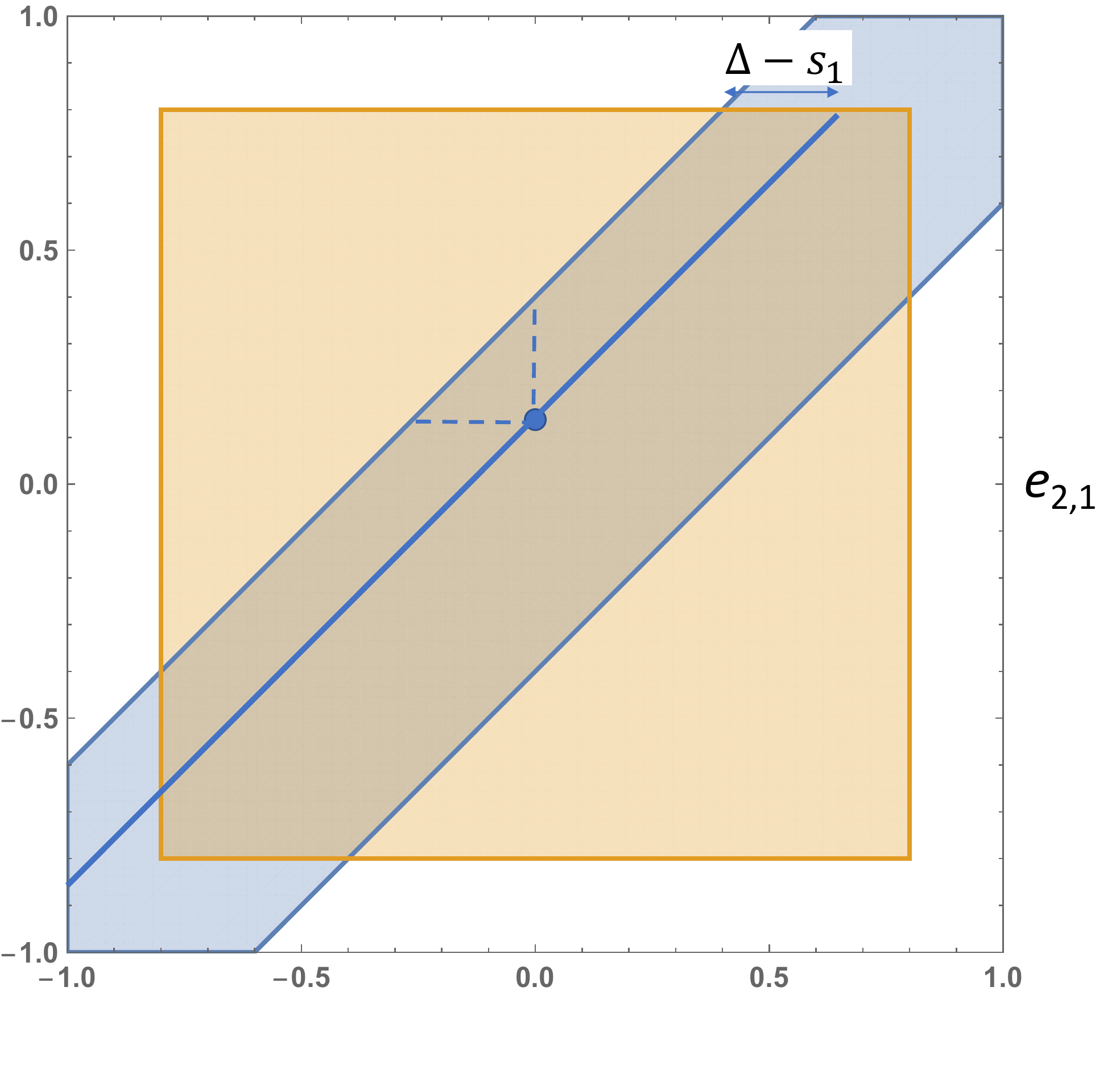}
\par\end{centering}
\begin{centering}
\includegraphics[scale=0.35]{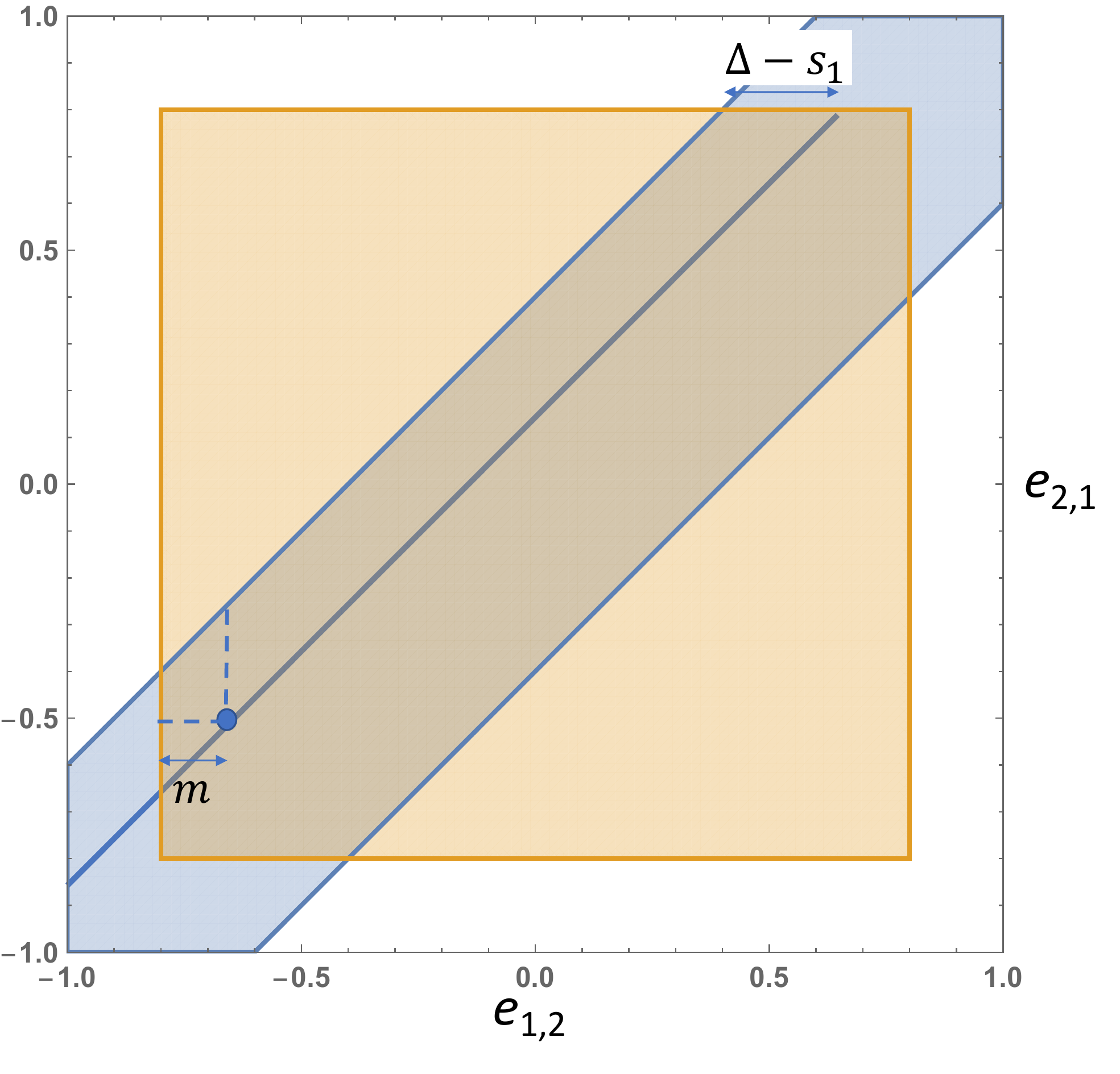}
\par\end{centering}
\caption{\label{fig:NCNT2}Illustration for Theorem \ref{thm:NCNT2}, $n=2$,
a detailed analog of Figs. \ref{fig:points}B and \ref{fig:points}C.
Upper panel: the case $\Delta-s_{1}\left(\mathbf{\mathbf{\mathbf{e_{b}^{*}}}}\right)\protect\leq m\left(\mathbf{\mathbf{e_{b}^{*}}}\right)$.
Lower panel: the case $\Delta-s_{1}\left(\mathbf{e_{b}^{*}}\right)>m\left(\mathbf{e_{b}^{*}}\right)$.}
\end{figure}
\par\end{center}

By geometric considerations, $m\left(\mathcal{\mathbf{\mathbf{e_{b}}}}\right)$
is also an $L_{p}$-distance between $\mathbb{R}_{\mathbf{b}}$ and
$\mathbf{\mathbf{\mathbf{e_{b}^{*}}}}$, for any $p\geq1$. Because
of this, using the same reasoning as in the case of $\textnormal{CNT}_{2}^{\left(p\right)}$,
the $L_{p}$-distance from $\mathbf{\mathbf{\mathbf{e_{b}^{*}}}}$
to the surface of $\mathbb{E}_{\mathbf{b}}$ is
\begin{equation}
\textnormal{NCNT}_{2}^{\left(p\right)}=\frac{1}{4}\min\left(n^{\frac{1-p}{p}}\left(\Delta-s_{1}\left(\mathbf{\mathbf{\mathbf{e_{b}^{*}}}}\right)\right),m\left(\mathbf{\mathbf{e_{b}^{*}}}\right)\right).
\end{equation}
As we see, unlike in the case of $\textnormal{CNT}_{2}^{\left(p\right)}$,
this is not simply a scaled version of $\textnormal{NCNT}_{2}$, indicating
that replacing the latter with $\textnormal{NCNT}_{2}^{\left(p\right)}$
is not inconsequential for the theory.

\begin{figure}
\begin{centering}
\includegraphics[scale=0.25]{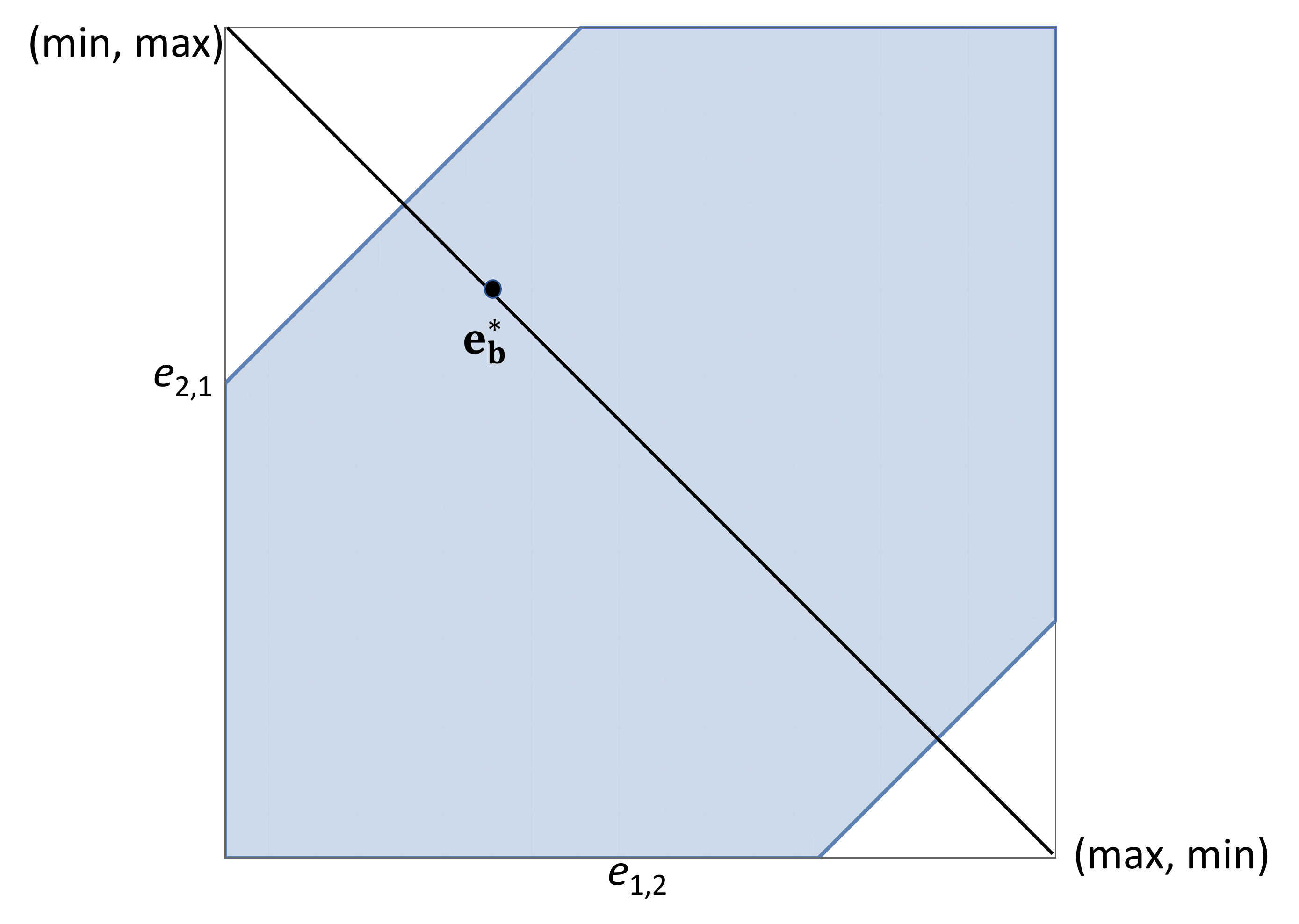}
\par\end{centering}
\begin{centering}
\includegraphics[scale=0.5]{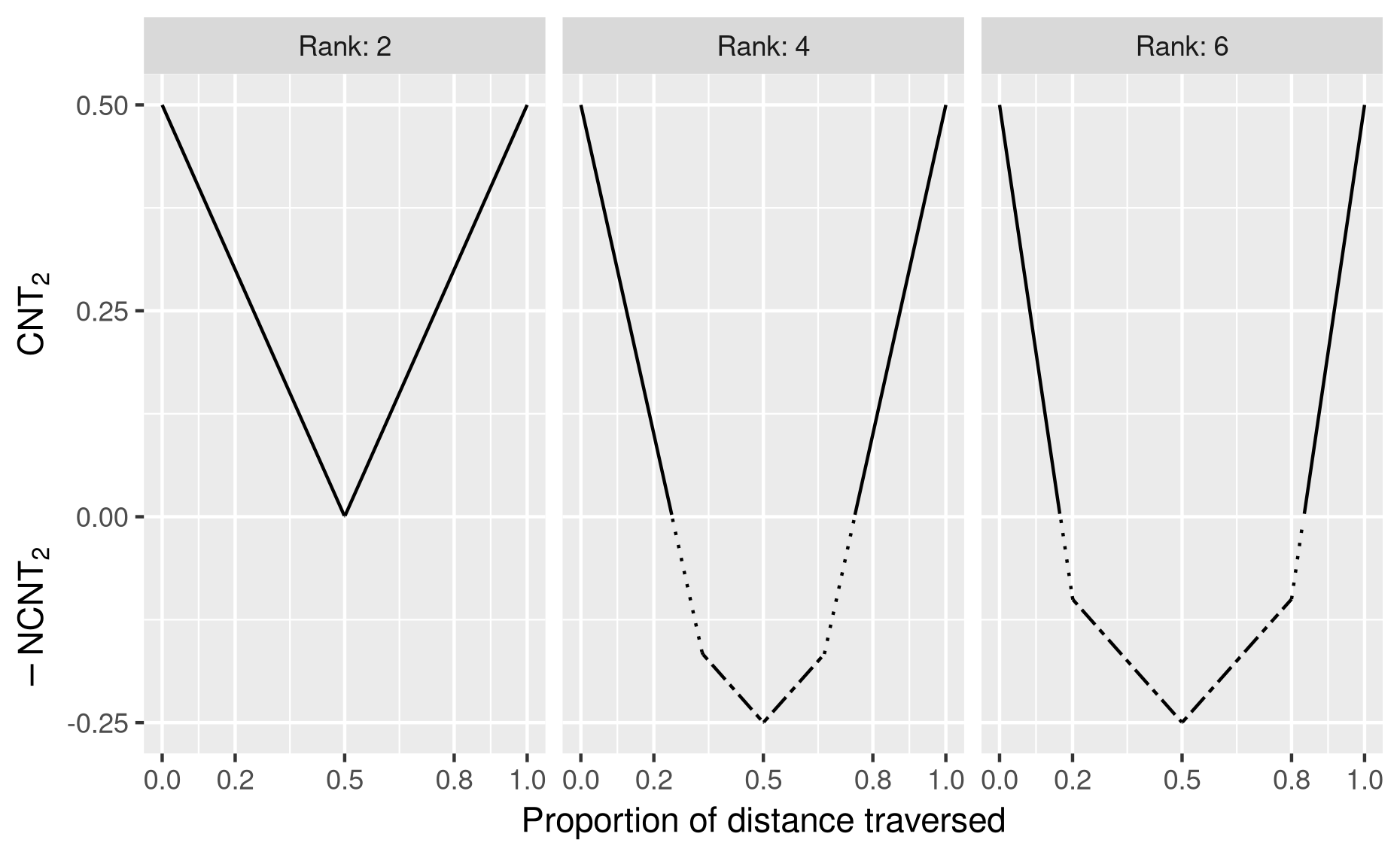}
\par\end{centering}
\begin{centering}
\includegraphics[scale=0.5]{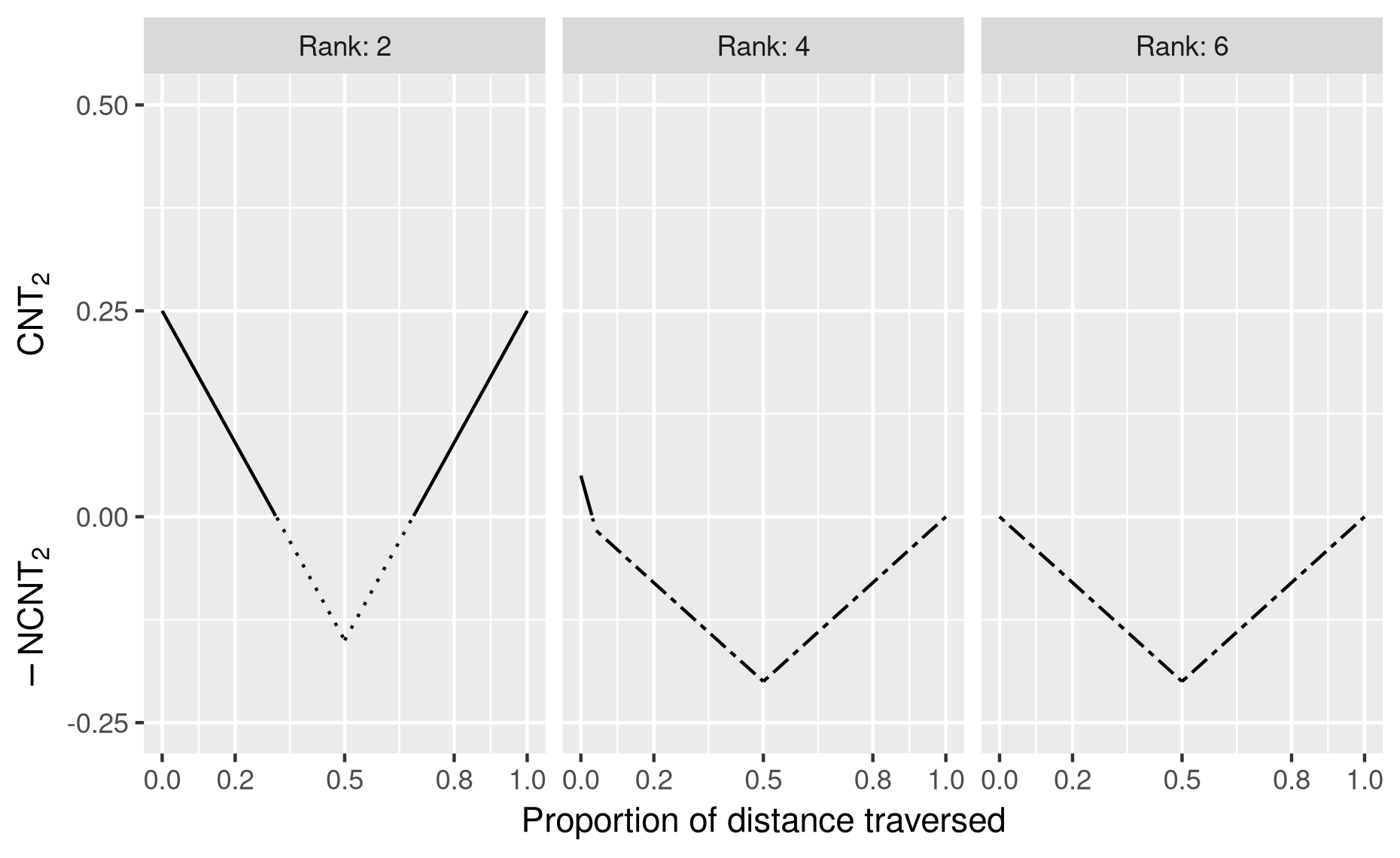}
\par\end{centering}
\caption{\label{fig:Dynamics}$\textnormal{\ensuremath{\textnormal{CNT}_{2}}}$
(solid lines, positive values) and $\textnormal{\ensuremath{-\textnormal{NCNT}_{2}}}$
(dotted or dashed lines, nonpositive values) as a function of the
position of $\mathbf{\mathbf{e_{b}^{*}}}$ on the diagonal connecting
two opposite vertices of $\mathbb{R}_{\mathbf{b}}$ for cyclic systems
of ranks 2, 4, 6. The dashed lines show the cases $\Delta-s_{1}\left(\mathbf{\mathbf{\mathbf{e_{b}^{*}}}}\right)>m\left(\mathbf{\mathbf{e_{b}^{*}}}\right)$,
when $\textnormal{\ensuremath{\textnormal{NCNT}_{2}}}=\frac{1}{4}m\left(\mathbf{\mathbf{e_{b}^{*}}}\right)$;
the dotted lines show the case $\Delta-s_{1}\left(\mathbf{\mathbf{\mathbf{e_{b}^{*}}}}\right)\protect\leq m\left(\mathbf{\mathbf{e_{b}^{*}}}\right)$,
when $\textnormal{\ensuremath{\textnormal{NCNT}_{2}}}=\frac{1}{4}\left(\Delta-s_{1}\left(\mathbf{\mathbf{\mathbf{e_{b}^{*}}}}\right)\right)$.
The lower set of graphs represents inconsistently connected systems,
with $\left\langle A_{i}^{i}\right\rangle =-0.2,\left\langle A_{i}^{i\ominus1}\right\rangle =0.1$,
for $i=1,\ldots,n$. The upper graphs represent consistently connected
systems, with the expectations of all random variables equal to zero.
As indicated in the illustration on the top, for the even-ranked systems,
it connects the vertex whose last coordinate is its single min-coordinate
to the vertex whose last coordinate is its single max-coordinate.}
\end{figure}

\begin{figure}
\begin{centering}
\includegraphics[scale=0.25]{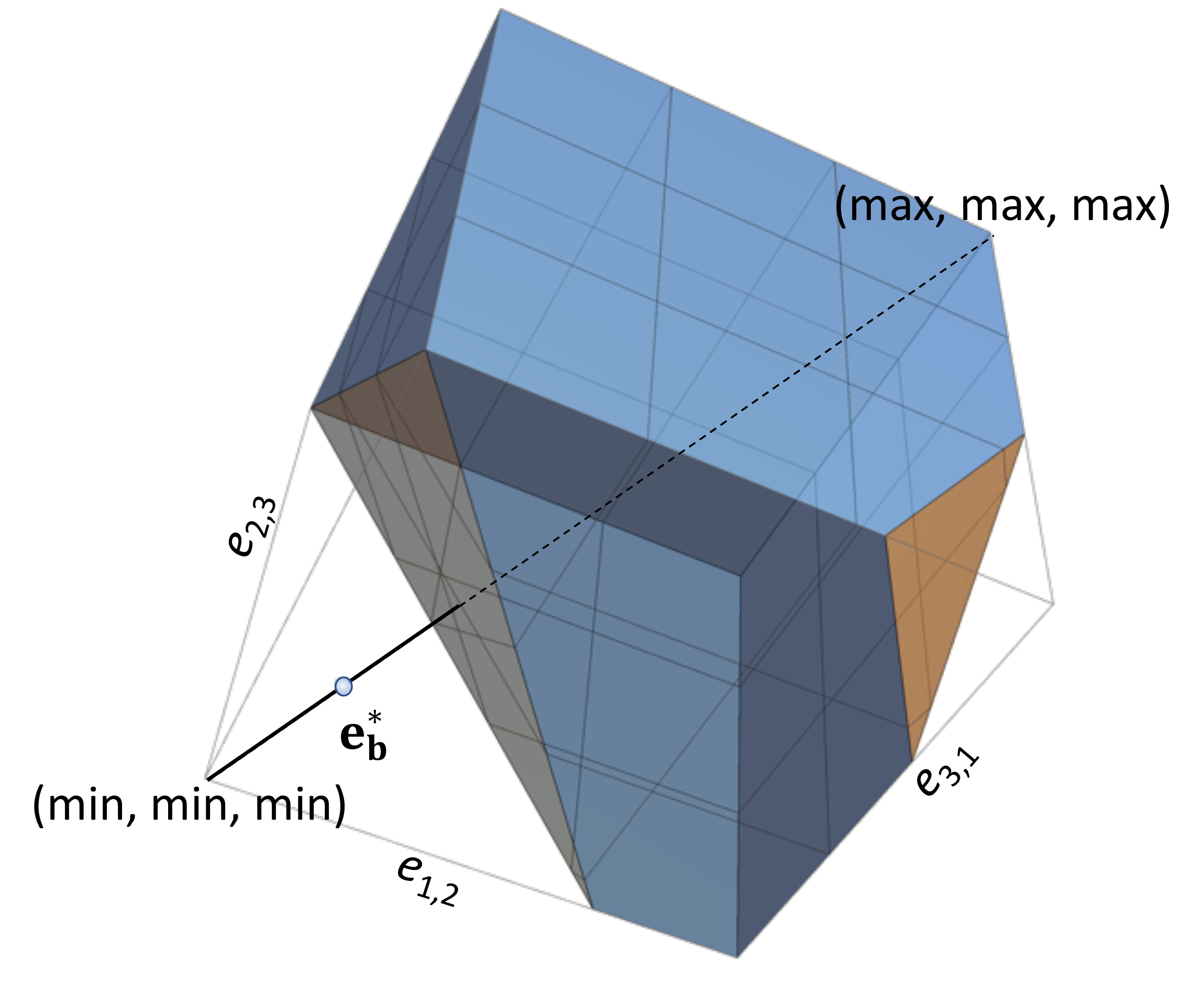}
\par\end{centering}
\begin{centering}
\includegraphics[scale=0.5]{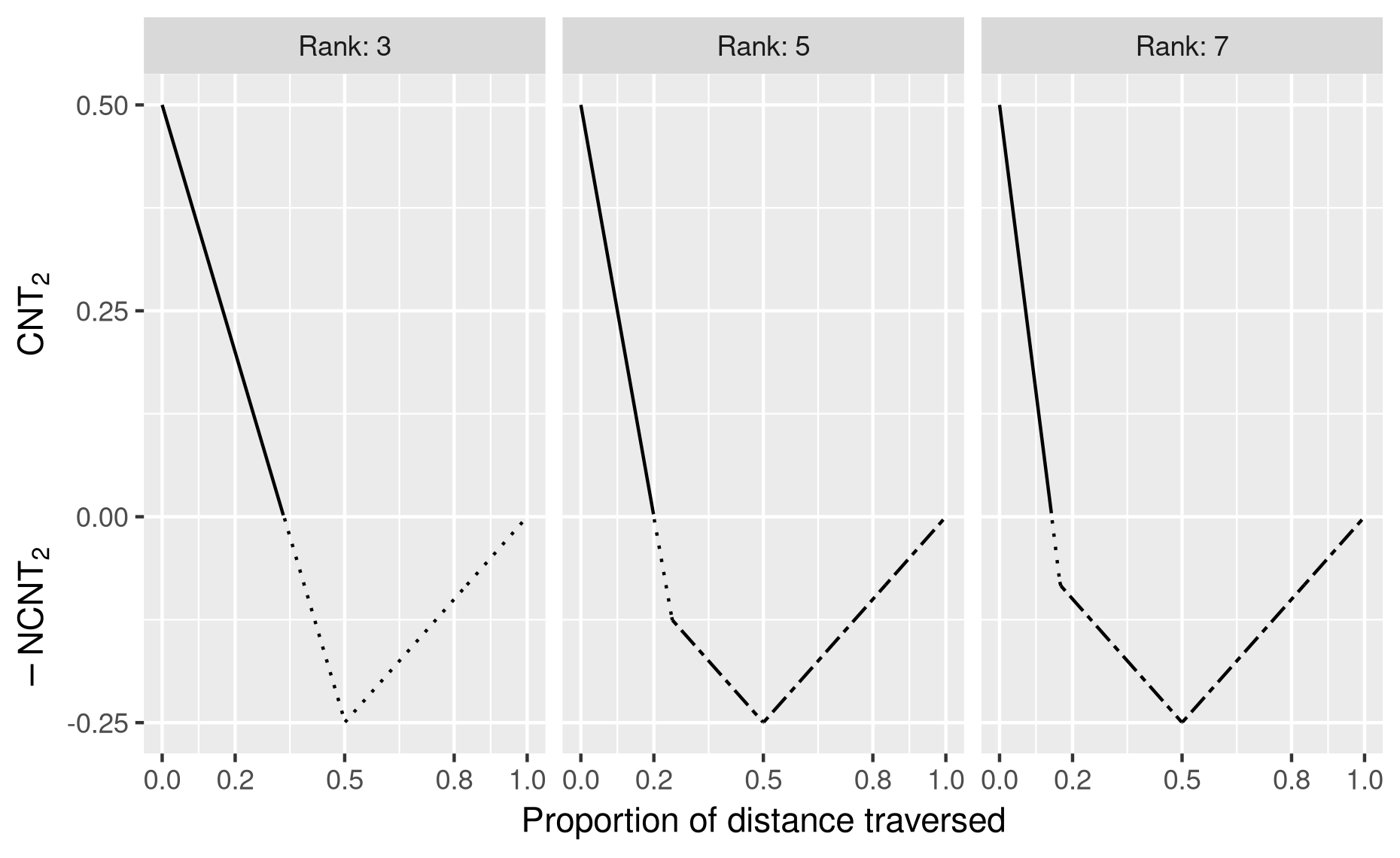}
\par\end{centering}
\begin{centering}
\includegraphics[scale=0.5]{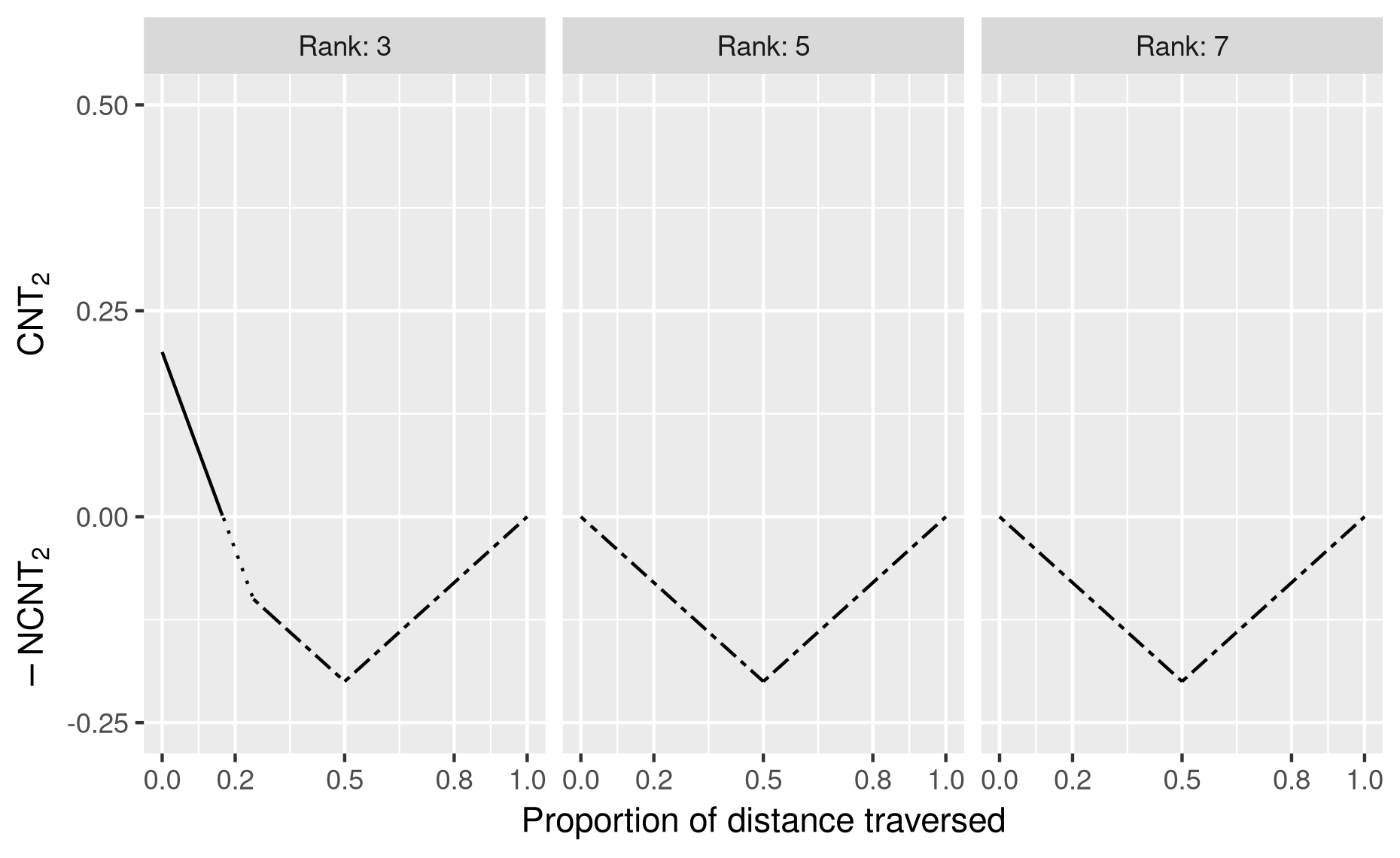}
\par\end{centering}
\caption{\label{fig:Dynamics-1}The same as Fig. \ref{fig:Dynamics}, but for
cyclic systems of ranks 3, 5, 7. As indicated in the illustration
on the top, for the odd-ranked systems, the diagonal connects the
vertex with all min-coordinates to the vertex with all max-coordinates.}
\end{figure}

Figures \ref{fig:Dynamics} and \ref{fig:Dynamics-1} illustrate the
dynamics of $\textnormal{\ensuremath{\textnormal{CNT}_{2}}}$ and
$\textnormal{\ensuremath{\textnormal{NCNT}_{2}}}$ as point $\mathbf{\mathbf{\mathbf{e_{b}^{*}}}}$
moves along the diagonal connecting two opposite vertices of $\mathbb{R}_{\mathbf{b}}$
for cyclic systems of several ranks. To emphasize that $\textnormal{\ensuremath{\textnormal{NCNT}_{2}}}$
is an extension of $\textnormal{\ensuremath{\textnormal{CNT}_{2}}}$
(and vice versa), we plot $\textnormal{\ensuremath{\textnormal{NCNT}_{2}}}$
with minus sign: as $\mathbf{\mathbf{\mathbf{e_{b}^{*}}}}$ moves
closer to the surface of $\mathbb{E}_{\mathbf{b}}$, $\textnormal{\ensuremath{\textnormal{CNT}_{2}}}$
decreases from a positive value to zero, the system becomes noncontextual,
and as the point continues to move inside the polytope, the value
of $-\ensuremath{\textnormal{NCNT}_{2}}$ proceeds to decrease continuously.

\section{\label{sec:Polytope-of-all}Polytope of all possible couplings}

We now need to gain insight into why $\textnormal{CNT}_{1}$ and $\textnormal{CNT}_{2}$
are the same for cyclic systems. Is it a peculiar coincidence? Does
$\textnormal{CNT}_{1}$, if interpreted geometrically, have the same
``nice'' properties as $\textnormal{CNT}_{2}$? The answer to the
first question turns out to be negative, and to second one affirmative.

In Sec. \ref{sec:Measures-of-contextuality} we introduced in (\ref{all-couplings})
the polytope $\mathbb{P}$ of all possible couplings for a system
with the low-marginals $\mathbf{p_{l}^{*}}$. As in the cases of $\mathbb{P}_{\mathbf{b}}$
and $\mathbb{P}_{\mathbf{c}}$, we redefine this polytope in terms
of $\pm1$-variables,

\begin{equation}
\mathbb{E}=\phi\left(\mathbb{P}\right),
\end{equation}
and use it to define a measure of contextuality 
\begin{equation}
\textnormal{CNT}_{0}=L_{1}\left(\mathbf{\mathbf{\left(\begin{array}{c}
\mathbf{p_{b}^{*}}\\
\mathbf{p_{c}^{*}}
\end{array}\right)}},\mathbb{P}\right)=\frac{1}{4}L_{1}\left(\mathbf{\mathbf{\left(\begin{array}{c}
\mathbf{e_{b}^{*}}\\
\mathbf{e_{c}^{*}}
\end{array}\right)}},\mathbb{E}\right).
\end{equation}
To investigate the properties of $\mathbb{E}$ and $\textnormal{CNT}_{0}$
we use the following result:
\begin{thm}[Kujala-Dzhafarov-Larsson \citep{KujDzhLar2015}]
 A system represented by $\left(\mathbf{e_{b}^{*}},\mathbf{e_{c}^{*}}\right)^{\intercal}$
is noncontextual if and only if 
\[
s_{1}\left(\mathbf{e_{b}^{*}},\mathbf{e_{c}^{*}}\right)\leq2n-2.
\]
\end{thm}

This can be understood as a special case of Theorem \ref{thm:KD2015}
if one uses the procedure of treating connections as if they were
additional contexts, rendering thereby any system consistently connected
\citep{KujDzhMeasures,AmaralDuarteOliveira2018}. Here and in the
following we write $s_{1}\left(\mathbf{e_{b}^{*}},\mathbf{e_{c}^{*}}\right)$
instead of the more correct $s_{1}\left(\left(\mathbf{e_{b}^{*}},\mathbf{e_{c}^{*}}\right)^{\intercal}\right)$.

It is evident now that the entire development in Secs. \ref{sec:Additional-terminology-and}
and \ref{sec:Measures-of-contextuality} can be repeated with $\mathbb{E}$
replacing $\mathbb{E}_{\mathbf{b}}$, except that the ambient cube
$\mathbb{C}$, extended noncontextuality polytope $\mathbb{N}$, and
the box $\mathbb{R}$ circumscribing $\mathbb{E}$ (replacing, respectively,
$\mathbb{C}_{\mathbf{b}}$, $\mathbb{N}_{\mathbf{b}}$, and $\mathbb{R}_{\mathbf{b}}$)
are $2n$-dimensional rather than $n$-dimensional, and the value
of $\Delta$ that defines the polytope is $2n-2$. In particular,
the shape of the polytope $\mathbb{N}$ is always a $2n$-demicube,
the convex hull of the $2^{2n-1}$ even vertices of $\mathbb{C}$,
similar to the $n$-demicubes shown in Fig. \ref{fig:pockets-1},
except that the minimal meaningful number of dimensions has to be
4 (representing a cyclic system of rank 2). The following analog of
Theorem \ref{thm:Main contextual} then holds.
\begin{thm}
The $L_{1}$-distance between $\mathbb{E}$ and a point $\mathbf{\left(\mathbf{\mathbf{e_{b}^{*}},\mathbf{e_{c}^{*}}}\right)^{\intercal}}$
representing a contextual system\emph{ }\textup{\emph{is a single-coordinate
distance, equal to $s_{1}\left(\mathbf{\mathbf{\mathbf{e_{b}^{*}},\mathbf{e_{c}^{*}}}}\right)-\left(2n-2\right)$
for all coordinates. This is the value of }}$\textnormal{\ensuremath{\textnormal{CNT}_{0}}}\cdot4$.
\end{thm}

It is easy to see now that
\begin{equation}
\textnormal{\ensuremath{\textnormal{CNT}_{0}}}=\textnormal{\ensuremath{\textnormal{CNT}_{1}}}=\textnormal{\ensuremath{\textnormal{CNT}_{2}}}.
\end{equation}
Indeed, the single-coordinate $L_{1}$-distance mentioned in the theorem
can be taken along an $\mathbf{e_{b}}$-coordinate or along an $\mathbf{e_{c}}$-coordinate,
and with all other coordinates being fixed at appropriate values,
this will be a single-coordinate $L_{1}$-distance from, respectively,
$\mathbb{E}_{\mathbf{b}}$ or $\mathbb{E}_{\mathbf{c}}$. Since we
know that 
\begin{equation}
\textnormal{\ensuremath{\textnormal{CNT}_{1}}}=\textnormal{\ensuremath{\textnormal{CNT}_{2}}}=\frac{1}{4}\left(s_{1}\left(\mathbf{\mathbf{e_{b}^{*}}}\right)-\Delta\right),
\end{equation}
and that
\begin{equation}
\textnormal{\ensuremath{\textnormal{CNT}_{0}}}=\frac{1}{4}\left(s_{1}\left(\mathbf{\mathbf{\mathbf{e_{b}^{*}},\mathbf{e_{c}^{*}}}}\right)-\left(2n-2\right)\right),
\end{equation}
we have an indirect proof that when $s_{1}\left(\mathbf{\mathbf{\mathbf{e_{b}^{*}},\mathbf{e_{c}^{*}}}}\right)>\left(2n-2\right)$
(i.e., the system is contextual),
\begin{equation}
s_{1}\left(\mathbf{\mathbf{e_{b}^{*}},\mathbf{e_{c}^{*}}}\right)=s_{1}\left(\mathbf{\mathbf{e_{b}^{*}}}\right)+n-\delta.
\end{equation}

Note that $\textnormal{\ensuremath{\textnormal{CNT}_{0}}}$, like
$\textnormal{CNT}_{1}$ and unlike $\textnormal{CNT}_{2}$, cannot
be naturally extended to a noncontextuality measure. Because $\mathbf{e_{c}^{*}}$
consists of the maximal possible values of $e^{i,i\oplus1}$ ($i=1,\ldots,n$),
any point $\mathbf{\left(\mathbf{\mathbf{e_{b}^{*}},\mathbf{e_{c}^{*}}}\right)^{\intercal}}$
representing a noncontextual system should lie on the surface of the
polytope $\mathbb{E}$, yielding 
\begin{equation}
s_{1}\left(\mathbf{\mathbf{\mathbf{e_{b}^{*}},\mathbf{e_{c}^{*}}}}\right)-\left(2n-2\right)=0.
\end{equation}
The argument leading to this conclusion was presented in Ref. \citep{KujDzhMeasures}
for $\mathbb{E}_{\mathbf{c}}$. When applied to $\mathbb{E}$, it
goes as follows: if $\mathbf{\mathbf{\mathbf{\left(\mathbf{\mathbf{e_{b}^{*}},\mathbf{e_{c}^{*}}}\right)^{\intercal}}}}$
were an interior point of $\mathbb{E}$, it would be surrounded by
a $2n$-ball entirely within $\mathbb{E}$, and one would be able
to increase any component of $\mathbf{e_{c}^{*}}$ while remaining
within this ball, which is not possible. $\textnormal{CNT}_{2}$ remains
the only one of the contextuality measures considered in the literature
that can be naturally extended into a noncontextuality measure.

\section{Conclusion with a glimpse into noncyclic systems}

Most of the regularities established in this paper do not generalize
to noncyclic systems. In particular, $\textnormal{CNT}_{1}$ and $\textnormal{CNT}_{2}$
do not generally coincide, nor is one of them any function of the
other \citep{newnote}. This can be seen in Fig. \ref{fig: New demonstration}
that presents the values of $\textnormal{CNT}_{1}$ and $\textnormal{CNT}_{2}$
for several systems described by

\begin{equation}
\begin{array}{|c|c|c|c||c}
\hline R_{1}^{1} & R_{2}^{1} &  &  & c_{1}\\
\hline  & R_{2}^{2} & R_{3}^{2} &  & c_{2}\\
\hline  &  & R_{3}^{3} & R_{4}^{3} & c_{3}\\
\hline R_{1}^{4} &  &  & R_{4}^{4} & c_{4}\\
\hline R_{1}^{5} & R_{2}^{5} & R_{3}^{5} & R_{4}^{5} & c_{5}\\
\hline\hline q_{1} & q_{2} & q_{3} & q_{4}
\end{array}\:.
\end{equation}
Here, all $R_{i}^{k}$ are uniformly distributed random variables,
and in each of the first four rows the two variables are always equal
to each other. Different systems in Fig. \ref{fig: New demonstration}
are obtained by varying the joint distribution of the four variables
in context $c_{5}$. We see that there is no functional relation between
$\textnormal{CNT}_{1}$ and $\textnormal{CNT}_{2}$: for either of
the measures, there are pairs of systems with different values of
this measure at a fixed value of the other.
\begin{center}
\begin{figure}[h]
\begin{centering}
\includegraphics[scale=0.5]{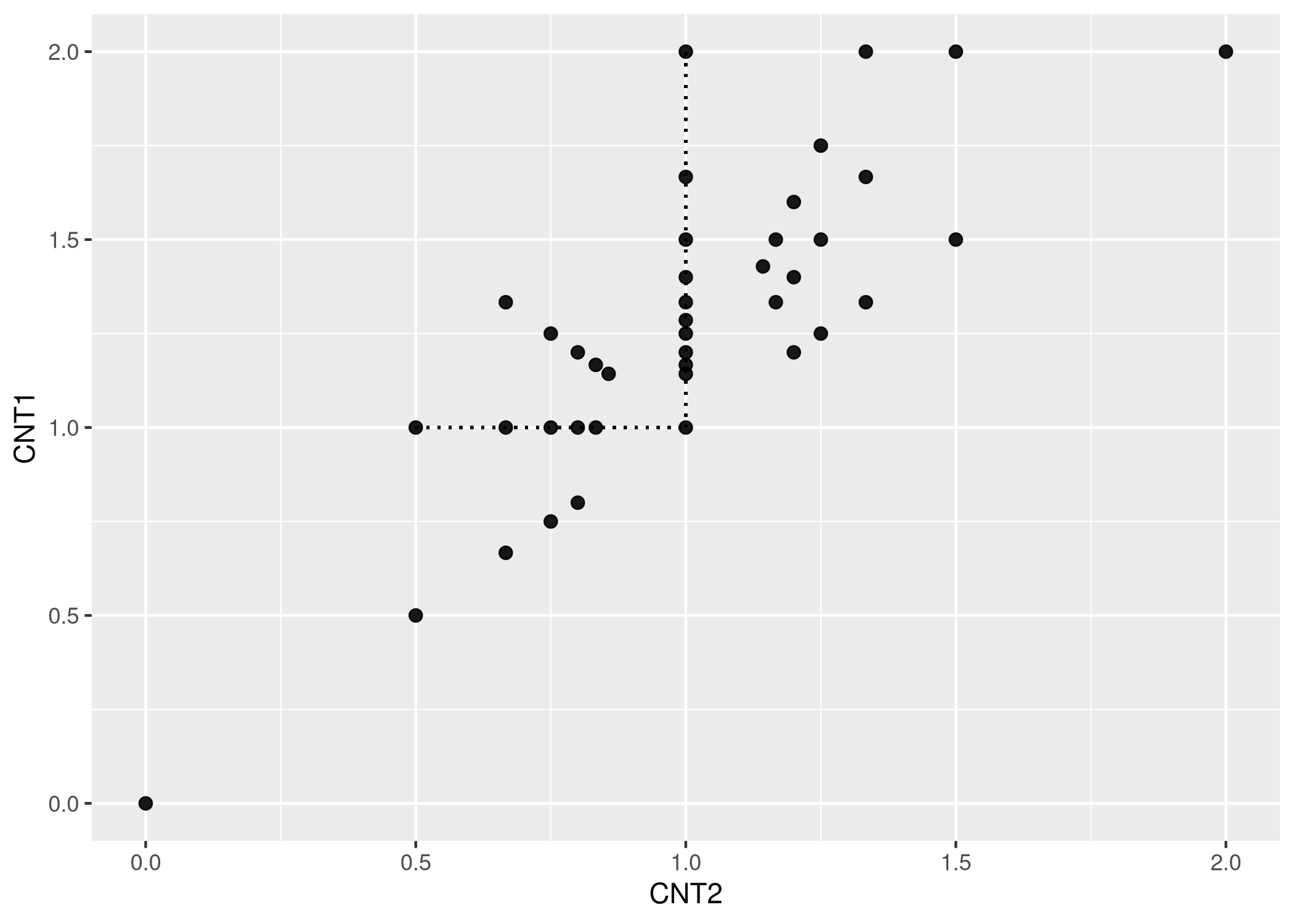}
\par\end{centering}
\caption{\label{fig: New demonstration}$\textnormal{CNT}_{1}$ vs $\textnormal{CNT}_{2}$
for system (66), with $\Pr\left[R_{j}^{i}=1\right]=1/2$ for all $i,j$
in the system. Each symbol corresponds to a specific choice of the
joint distribution of $R_{1}^{5},R_{2}^{5},R_{3}^{5},R_{4}^{5}$,
while keeping $R_{1}^{1}=R_{2}^{1}$, $R_{2}^{2}=R_{3}^{2}$, $R_{3}^{3}=R_{4}^{3}$,
and $R_{4}^{4}=R_{1}^{4}$. Neither of $\textnormal{CNT}_{1}$ and
$\textnormal{CNT}_{2}$ is a function of the other, as indicated by
the horizontally and vertically aligned points.}
\end{figure}
\par\end{center}

It might be tempting to think that cyclic systems could help one in
at least detecting if not measuring (non)contextuality of a system.
Clearly, if a system contains a contextual cyclic subsystem, then
it is contextual. This is not surprising, however, because this is
true for any contextual subsystem, cyclic or not \citep{DzhCerKuj2017,DzhKujFoundations2017}.
Could it be, one might wonder, that a system is always noncontextual
if it does not contain a contextual cyclic subsystem? The answer is
negative, as we see from the following counterexample. Let a system
of dichotomous random variables be
\begin{equation}
\begin{array}{|c|c|c|c||c}
\hline R_{1}^{1} & R_{2}^{1} & R_{3}^{1} &  & c_{1}\\
\hline  & R_{2}^{2} & R_{3}^{2} & R_{4}^{2} & c_{2}\\
\hline R_{1}^{3} &  & R_{3}^{3} & R_{4}^{3} & c_{3}\\
\hline\hline q_{1} & q_{2} & q_{3} & q_{4}
\end{array},
\end{equation}
with four contents measured in three contexts. Let the joint distributions
of the three bunches be
\begin{equation}
\begin{array}{|c|c|c|c|}
\hline R_{1}^{1} & R_{2}^{1} & R_{3}^{1} & \\
\hline -1 & -1 & +1 & \nicefrac{1}{4}\\
\hline -1 & +1 & -1 & \nicefrac{1}{4}\\
\hline +1 & -1 & -1 & \nicefrac{1}{4}\\
\hline +1 & +1 & +1 & \nicefrac{1}{4}
\\\hline \end{array},\;\begin{array}{|c|c|c|c|}
\hline R_{2}^{2} & R_{3}^{2} & R_{4}^{2} & \\
\hline -1 & -1 & +1 & \nicefrac{1}{4}\\
\hline -1 & +1 & -1 & \nicefrac{1}{4}\\
\hline +1 & -1 & -1 & \nicefrac{1}{4}\\
\hline +1 & +1 & +1 & \nicefrac{1}{4}
\\\hline \end{array},\;\begin{array}{|c|c|c|c|}
\hline R_{1}^{3} & R_{3}^{3} & R_{4}^{3} & \\
\hline +1 & +1 & -1 & \nicefrac{1}{4}\\
\hline +1 & -1 & +1 & \nicefrac{1}{4}\\
\hline -1 & +1 & +1 & \nicefrac{1}{4}\\
\hline -1 & -1 & -1 & \nicefrac{1}{4}
\\\hline \end{array}.
\end{equation}
Here, the probabilities of the triples of values in each bunch are
shown in the rightmost columns, with all remaining triples having
probability zero. One can check that all random variables are distributed
uniformly, 
\begin{equation}
\Pr\left[R_{i}^{k}=-1\right]=\Pr\left[R_{i}^{k}=+1\right]=\frac{1}{2},
\end{equation}
so the system is consistently connected. All pairs $\left(R_{i}^{k},R_{j}^{k}\right)$
are also uniformly distributed, 
\begin{equation}
\begin{array}{|c|c|c|}
\hline R_{i}^{k} & R_{j}^{k} & \\
\hline -1 & -1 & \nicefrac{1}{4}\\
\hline -1 & +1 & \nicefrac{1}{4}\\
\hline +1 & -1 & \nicefrac{1}{4}\\
\hline +1 & +1 & \nicefrac{1}{4}
\\\hline \end{array}.
\end{equation}
This means that the system is \emph{strongly consistently connected}:
whenever a set of contents is measured in two contexts, their joint
(here, pairwise) distributions coincide. Because the variables in
each bunch are pairwise independent, any cyclic subsystem of this
system is noncontextual. The entire system, however is contextual.
Indeed, in the hypothetical coupling satisfying the definition of
noncontextuality, if $\left(S_{1}^{1},S_{2}^{1},S_{3}^{1}\right)=\left(-1,-1,1\right)$,
then $\left(S_{2}^{2},S_{3}^{2},S_{4}^{2}\right)$ can only be $\left(-1,1,-1\right)$,
and $\left(S_{1}^{3},S_{3}^{3},S_{4}^{3}\right)$ can only be $\left(-1,1,1\right)$.
The reason for this is that in this hypothetical coupling we should
have $\left(S_{1}^{1},S_{3}^{1}\right)=\left(S_{1}^{3},S_{3}^{3}\right)$,
and $\left(S_{2}^{1},S_{3}^{1}\right)=\left(S_{2}^{2},S_{3}^{2}\right)$.
However it should also be true that $\left(S_{3}^{2},S_{4}^{2}\right)=\left(S_{3}^{3},S_{4}^{3}\right)$,
and this is not the case in the above triples: $\left(S_{3}^{2},S_{4}^{2}\right)=\left(1,-1\right)$
while $\left(S_{3}^{3},S_{4}^{3}\right)=\left(1,1\right)$. This completes
the counterexample.

To summarize, we know now that the regular way in which the noncontextuality
polytope $\mathbb{P}_{\mathbf{b}}$ (or $\mathbb{E}_{\mathbf{b}}$)
and the polytope of all possible couplings $\mathbb{P}$ (or $\mathbb{E}$)
create pockets at the vertices of the circumscribing boxes makes $\textnormal{CNT}_{1}$
and $\textnormal{CNT}_{2}$ single-coordinate distances that are equal
to each other. Both of them are proportional to the degree of violation
of the generalized Bell criterion derived in Ref. \citep{KujDzhProof2016},
$s_{1}\left(\mathbf{e_{b}^{*}}\right)-\Delta$. We have known from
Ref. \citep{KujDzhMeasures}, that $\textnormal{CNT}_{2}$, unlike
$\textnormal{CNT}_{1}$, naturally extends to a measure of noncontextuality,
$\textnormal{NCNT}_{2}$, and this can be taken as a reason for preferring
$\textnormal{CNT}_{2}$ to $\textnormal{CNT}_{1}$. $\textnormal{NCNT}_{2}$
is a single-coordinate distance, and in the case of cyclic systems,
the properties of the noncontextuality polytope make $\textnormal{NCNT}_{2}$
proportional, with the same proportionality coefficient as for $\textnormal{CNT}_{2}$
and $\textnormal{CNT}_{1}$, to the smaller of two quantities: the
degree of compliance with the generalized Bell inequality, $\Delta-s_{1}\left(\mathbf{e_{b}^{*}}\right)$,
and the distance $m\left(\mathbf{e_{b}^{*}}\right)$ of $\mathbf{e_{b}^{*}}$
from the surface of the circumscribing box $\mathbb{R}_{\mathbf{b}}$.
We also know that none of these regularities extend beyond the class
of cyclic systems, so the general theory of the relationship between
the measures considered in this paper has much left to develop.

\section*{Appendix: Proofs of the formal statements}

\paragraph*{Proof of Lemma \textup{\emph{\ref{lem:1 pocket}}}.}

Verify that, for any $\Delta$, each of the $n$ points \emph{
\[
\mathbf{x}_{k}=\left\{ \lambda_{1},\ldots,\lambda_{k}\left(1-n+\Delta\right),\ldots,\lambda_{n}\right\} ,k=1,\ldots,n,
\]
}satisfies
\[
\sum\lambda_{i}x_{i,i\oplus1}=n-1+\lambda_{k}^{2}\left(1-n+\Delta\right)=\Delta,
\]
whence so does the hyperplane passing through these points. Since
$n-2\leq\Delta\leq n$, the distance $n-\Delta$ is between 0 and
2, so that the hyperplane does cut each of the edges joined at the
vertex.

\paragraph{Proof of Lemma \ref{lem:pockets}.}

Two odd vertices have nonoverlapping sets of edges emanating from
them, and each of the two hyperplanes cuts its own set. The only case
when an axis from one set is cut at the same point as an axis from
another set is when the cuts are at the ends of the emanating edges,
and this means that $\Delta=n-2$.

\paragraph*{Proof of Lemma \ref{lem:point outside}.}

We need to show that for any other odd vertex $V'=\left\{ \lambda'_{i}:i=1,\ldots,n\right\} $
of $\mathbb{C}_{\mathbf{b}}$, $\sum\lambda'x_{i,i\oplus1}\leq\Delta_{\mathbf{x}}$.
This is indeed the case because for any value $\Delta'\geq n-2$,
the hyperplane segment $\sum\lambda'e_{i,i\oplus1}=\Delta'$ does
not cut any of the edges emanating from vertex $V$, except, possibly,
at their other ends (if $\Delta'=n-2$). Consequently, $\sum\lambda'x_{i,i\oplus1}<n-2\leq\Delta<\Delta_{\mathbf{x}}$.

\paragraph*{Proof of Lemma \ref{lem:even vertices}.}

By induction. For $n=2$ we have to show that the even vertices $\left(1-\left|e_{1}^{1}-e_{2}^{1}\right|,1-\left|e_{1}^{2}-e_{2}^{2}\right|\right)$
and $\left(\left|e_{1}^{1}+e_{2}^{1}\right|-1,\left|e_{1}^{2}+e_{2}^{2}\right|-1\right)$
are within $\mathbb{E}_{\mathbf{b}}$. For the former vertex this
means that
\[
\left|\left(1-\left|e_{1}^{1}-e_{2}^{1}\right|\right)-\left(1-\left|e_{1}^{2}-e_{2}^{2}\right|\right)\right|\leq\left|e_{1}^{1}-e_{1}^{2}\right|+\left|e_{2}^{1}-e_{2}^{2}\right|.
\]
Without loss of generality, let the left-hand side be $\left(1-\left|e_{1}^{1}-e_{2}^{1}\right|\right)-\left(1-\left|e_{1}^{2}-e_{2}^{2}\right|\right)$.
The inequality then is equivalent to
\[
\left|e_{1}^{2}-e_{2}^{2}\right|\leq\left|e_{1}^{2}-e_{1}^{1}\right|+\left|e_{1}^{1}-e_{2}^{1}\right|+\left|e_{2}^{1}-e_{2}^{2}\right|,
\]
which is true by the triangle inequality. For the second even vertex
we have to show that 
\[
\left|\left(\left|e_{1}^{1}+e_{2}^{1}\right|-1\right)-\left(\left|e_{1}^{2}+e_{2}^{2}\right|-1\right)\right|\leq\left|e_{1}^{1}-e_{1}^{2}\right|+\left|e_{2}^{1}-e_{2}^{2}\right|.
\]
Again, without loss of generality, let the left-hand side be $\left(\left|e_{1}^{1}+e_{2}^{1}\right|-1\right)-\left(\left|e_{1}^{2}+e_{2}^{2}\right|-1\right)$.

(1) If $e_{1}^{1}+e_{2}^{1}\geq0$, $e_{1}^{2}+e_{2}^{2}\geq0$, the
inequality acquires the form $\left(e_{1}^{1}-e_{1}^{2}\right)+\left(e_{2}^{1}-e_{2}^{2}\right)\leq\left|e_{1}^{1}-e_{1}^{2}\right|+\left|e_{2}^{1}-e_{2}^{2}\right|$,
which is true.

(2) If $e_{1}^{1}+e_{2}^{1}<0$, $e_{1}^{2}+e_{2}^{2}<0$, the inequality
acquires the form $\left(e_{1}^{2}-e_{1}^{1}\right)+\left(e_{2}^{2}-e_{2}^{1}\right)\leq\left|e_{1}^{1}-e_{1}^{2}\right|+\left|e_{2}^{1}-e_{2}^{2}\right|$,
which is true.

(3) If $e_{1}^{1}+e_{2}^{1}\geq0$, $e_{1}^{2}+e_{2}^{2}<0$, we have
\[
\begin{array}{r}
\left(e_{1}^{1}+e_{1}^{2}\right)+\left(e_{2}^{1}+e_{2}^{2}\right)=\left(e_{1}^{1}-e_{1}^{2}\right)+\left(e_{2}^{1}-e_{2}^{2}\right)+2\left(e_{1}^{2}+e_{2}^{2}\right)\\
\leq\left(e_{1}^{1}-e_{1}^{2}\right)+\left(e_{2}^{1}-e_{2}^{2}\right)\leq\left|e_{1}^{1}-e_{1}^{2}\right|+\left|e_{2}^{1}-e_{2}^{2}\right|,
\end{array}
\]
which is true. The fourth case is analogous.

Assume now that the statement of the theorem holds for all $2\leq k<n$.
We have to show that
\[
s_{1}\left(\mathbf{x}\right)\leq n-2+\delta^{\left(n\right)}
\]
for any even vertex of $\mathbb{R}_{\mathbf{b}}$. Without changing
the values of $s_{1}$ and any of the summands in $\delta^{\left(n\right)}$,
we can put the inequality in the canonical form (Lemma \ref{lem:Canonical form}),
\[
\sum_{i=1}^{n-1}x_{i,i\oplus1}-x_{n1}\leq n-2+\delta^{\left(n\right)}.
\]
Consider two cases.

(Case 1) At least one of the coordinates $x_{i,i\oplus1}$ ($i=1,\ldots,n-1$)
is a max-coordinate. Let it be $x_{12}=1-\left|x_{1}^{1}-x_{2}^{1}\right|$.
We can rewrite the inequality as
\[
\begin{array}{r}
\left(\sum_{i=2}^{n-1}x_{i,i\oplus1}-x_{n1}\right)+1-\left|x_{1}^{1}-x_{2}^{1}\right|\leq\left(n-3+\delta^{\left(n-1\right)}\right)\\
+1+\left|x_{1}^{1}-x_{1}^{n}\right|+\left|x_{2}^{1}-x_{2}^{2}\right|-\left|x_{2}^{2}-x_{1}^{n}\right|
\end{array}\tag{*}.
\]
The value of $\sum_{i=2}^{n-1}x_{i,i\oplus1}-x_{n1}$ is equal to
the $s_{1}$ of some system of rank $n-1$, and since the vector $\left\{ x_{23},\ldots,x_{n-1,n},x_{n1}\right\} $
contains an even number of min-coordinates,
\[
\left(\sum_{i=2}^{n-1}x_{i,i\oplus1}-x_{n1}\right)\leq n-3+\delta^{\left(n-1\right)}
\]
holds by the induction hypothesis. At the same time, obviously,
\[
1-\left|x_{1}^{1}-x_{2}^{1}\right|\leq1+\left|x_{1}^{1}-x_{1}^{n}\right|+\left|x_{2}^{1}-x_{2}^{2}\right|-\left|x_{2}^{2}-x_{1}^{n}\right|,
\]
and this establishes ({*}) for this case.

(Case 2) All coordinates $x_{i,i\oplus1}$ ($i=1,\ldots,n-1$) are
min-coordinates. Let us then replace two of them (which is possible
since $n-1\geq2$) with the corresponding max-coordinates --- this
will leave the number of the min-coordinates even. The left-hand side
of ({*}) can only increase, but we can use the argument of the previous
case to show that it is still less than the (unchanged) right-hand
side of ({*}).

This completes the proof.


\begin{thebibliography}{99}
\bibitem{Araujoetal2013}M. Araújo, M. T. Quintino, C. Budroni, M.
T. Cunha, and A. Cabello, \emph{All noncontextuality inequalities
for the n-cycle scenario}, Phys. Rev. A 88, 022118 (2013).

\bibitem{KujDzhLar2015}J. V. Kujala, E. N. Dzhafarov, and J.-Å. Larsson,
\emph{Necessary and sufficient conditions for extended noncontextuality
in a broad class of quantum mechanical systems}, Phys. Rev. Lett.
115, 150401 (2015).

\bibitem{KCBS2008}A. A. Klyachko, M. A. Can, S. Binicio\u{g}lu, and
A. S. Shumovsky, \emph{A simple test for hidden variables in spin-1
system}, Phys. Rev. Lett. 101, 020403 (2008).

\bibitem{Lapkiewicz2015}R. Lapkiewicz, P. Li, C. Schaeff, N. K. Langford,
S. Ramelow, M. Wie\'{s}niak, and A. Zeilinger, \emph{Experimental
non-classicality of an indivisible quantum system}, Nature 474, 490
(2011).

\bibitem{Bell1964}J. Bell, \emph{On the Einstein-Podolsky-Rosen paradox},
Phys. 1, 195 (1964).

\bibitem{Bell1966}J. Bell, \emph{On the problem of hidden variables
in quantum mechanics}, Rev. Modern Phys. 38, 447 (1966).

\bibitem{CHSH1969}J. F. Clauser, M. A. Horne, A. Shimony, and R.
A. Holt, \emph{Proposed experiment to test local hidden-variable theories},
Phys. Rev. Lett. 23, 880 (1969).

\bibitem{Fine1982}A. Fine, \emph{Joint distributions, quantum correlations,
and commuting observables}, J. Math. Phys. 23, 1306 (1982).

\bibitem{Bacciagaluppi2015}G. Bacciagaluppi, \emph{Leggett-Garg inequalities,
pilot waves and contextuality}, Int. J. Quant. Found. 1, 1 (2015).

\bibitem{KoflerBrukner2013}J. Kofler and C. Brukner, \emph{Condition
for macroscopic realism beyond the Leggett-Garg inequalities}, Phys.
Rev. A 87, 052115 (2013).

\bibitem{LeggGarg1985}A. J. Leggett and A. Garg, \emph{Quantum mechanics
versus macroscopic realism: Is the flux there when nobody looks?}
Phys. Rev. Lett. 54, 857 (1985).

\bibitem{SuppesZanotti1981}P. Suppes and M. Zanotti, \emph{When are
probabilistic explanations possible? }Synth. 48, 191 (1981).

\bibitem{DzhZhaKuj2016}E. N. Dzhafarov, R. Zhang, and J. V. Kujala,
\emph{Is there contextuality in behavioral and social systems?} Phil.
Trans. Roy. Soc. A 374, 20150099 (2016).

\bibitem{Wang}Z. Wang, T. Solloway, R. M. Shiffrin, and J. R. Busemeyer,
\emph{Context effects produced by question orders reveal quantum nature
of human judgments}, Proc. Natl. Acad. Sci. 111, 9431 (2014).

\bibitem{DzhCerKuj2017}E. N. Dzhafarov, V. H. Cervantes, and J. V.
Kujala, \emph{Contextuality in canonical systems of random variables},
Phil. Trans. Roy. Soc. A 375, 20160389 (2017).

\bibitem{DzhKujFoundations2017}E. N. Dzhafarov and J. V. Kujala,
\emph{Probabilistic foundations of contextuality}, Fortsch. Phys.
- Prog. Phys. 65, 1600040 (1-11) (2017).

\bibitem{KujDzhMeasures}J. V. Kujala and E. N. Dzhafarov, \emph{Measures
of contextuality and noncontextuality}, Phil. Trans. Roy. Soc. A 377,
20190149 (2019).

\bibitem{DzhKujLar2015}E. N. Dzhafarov, J. V. Kujala, and J.-Å. Larsson,
\emph{Contextuality in three types of quantum-mechanical systems},
Found. Phys. 45, 762 (2015).

\bibitem{KujDzhProof2016}J. V. Kujala and E. N. Dzhafarov, \emph{Proof
of a conjecture on contextuality in cyclic systems with binary variables},
Found. Phys. 46, 282 (2016).

\bibitem{Malinowskietal.}M. Malinowski, C. Zhang, F. M. Leupold,
J. Alonso, J. P. Home, and A. Cabello, A, \emph{Probing the limits
of correlations in an indivisible quantum system}, arXiv:1712.06494v2
(2018).

\bibitem{Ariasetal.2015}M. Arias, G. Canas, E. S. Gomez, J. B. Barra,
G. B. Xavier, G. Lima, V. D\textquoteright Ambrosio, F. Baccari, F.
Sciarrino, and A. Cabello, \emph{Testing noncontextuality inequalities
that are building blocks of quantum correlations,} Phys. Rev. A 92,
032126 (2015)

\bibitem{Crespietal.2017}A. Crespi, M. Bentivegna, I. Pitsios, D.
Rusc, D. Poderini, G. Carvacho, V. D\textquoteright Ambrosio, A. Cabello,
F. Sciarrino, and R., Osellame, \emph{Single-photon quantum contextuality
on a chip,} ACS Photonics 4, 2807-2812. (2017).

\bibitem{Fluhmannetal2018}C. Flühmann, V. Negnevitsky, M. Marinelli,
and J. P. Home, \emph{Sequential modular position and momentum measurements
of a trapped ion mechanical oscillator,} Phys. Rev. X 8, 021001 (2018).

\bibitem{Zhanetal.2017}X. Zhan, P. Kurzy\'{n}ski, D. Kaszlikowski,
K. Wang, Z. Bian, Y. Zhang, and P. Xue, \emph{Experimental detection
of information deficit in a photonic contextuality scenario,} Phys.
Rev. Lett. 119, 220403 (2017).

\bibitem{AbramBarbMans2017}S. Abramsky, R. S. Barbosa, and S. Mansfield,
\emph{The contextual fraction as a measure of contextuality,} Phys.
Rev. Lett. 119, 050504 (2017).

\bibitem{Brunneretal2014}N. Brunner, D. Cavalcanti, S. Pironio, V.
Scarani, and S. Wehner,\emph{ Bell nonlocality}, Rev. Mod. Phys. 86,
419 (2014).

\bibitem{AmaralCunhaCabello2015}B. Amaral, M. T. Cunha, and A. Cabello,
\emph{Quantum theory allows for absolute maximal contextuality,} Phys.
Rev. A 92, 062125 (2015).

\bibitem{Grudkaetal2014}A. Grudka, K. Horodecki, M. Horodecki, P.
Horodecki, R. Horodecki, P. Joshi, W. K\l obus, and A. Wójcik, \emph{Quantifying
Contextuality}, Phys. Rev. Lett. 112, 120401 (2014).

\bibitem{Kleinmanetal2011}M. Kleinmann, O. Gühne, J\'{.}R. Portillo,
J.-Å. Larsson, and A. Cabello, \emph{Memory cost of quantum contextuality},
New J. Phys. 13, 113011 (2011).

\bibitem{Bermejoetal2017}J. Bermejo-Vega, N. Delfosse, D. E. Browne,
C. Okay, and R. Raussendorf, \emph{Contextuality as a resource for
models of quantum computation with qubits,} Phys. Rev. Lett. 119,
120505 (2017).

\bibitem{Howardetal.2014}M. Howard, J. Wallman, V. Veitch, and J.
Emerson, \emph{Contextuality supplies the \textquoteleft magic\textquoteright{}
for quantum computation,} Nature 510, 351--355 (2014).

\bibitem{Brukneretal.2004}C. Brukner, M. Zukowski, J.-W. Pan, and
A. Zeilinger, \emph{Bell\textquoteright s inequalities and quantum
communication complexity,} Phys. Rev. Lett. 92, 127901 (2004).

\bibitem{footnote2}As Ref. \citep{KujDzhProof2016} plays an important
role in the present paper (see Theorem \ref{thm:KD2015} and Lemmas
\ref{lem:variants}-\ref{lem:Canonical form}), we should mention
that we have noticed two unfortunate typos in the introduction to
that paper (in Eq. 7 and Sec. 1.4). They are corrected in the arXiv'ed
version of the paper, arXiv:1503.02181, and on the authors' websites.

\bibitem{Tuenter2006}H. J. H. Tuenter, \emph{Minimum L1-distance
projection onto the boundary of a convex set: Simple characterization},
J. Optim. Theor. Appl., 112, 441 (2002).

\bibitem{Dzh2019}E. N. Dzhafarov, \emph{On joint distributions, counterfactual
values, and hidden variables in understanding contextuality,} Phil.
Trans. Roy. Soc. A 377, 20190144 (2019).

\bibitem{footnote3}In all our previous publications $\Delta$ was
simply $n-2+\delta$. However, at $\Delta\geq n$ any cyclic system
of rank $n$ is noncontextual. This will become apparent in Sec. \ref{sec:Properties-of-the}:
at $\Delta=n$ the noncontextuality polytope fills the entire hypercube
of possible values of $\mathbf{e_{b}}$, and a further increase in
$\Delta$ does not change this.

\bibitem{Dzh2017Nothing}E. N. Dzhafarov, \emph{Replacing nothing
with something special: Contextuality-by-Default and dummy measurements},
in \emph{Quantum Foundations, Probability and Information.} (Springer,
Berlin, 2017), pp 39-44.

\bibitem{AmaralDuarteOliveira2018}B. Amaral, C. Duarte, and R. I.
Oliveira, \emph{Necessary conditions for extended noncontextuality
in general sets of random variables,} J. Math. Phys. 59, 072202 (2018).

\bibitem{newnote}The rest of this paragraph and Fig. \ref{fig: New demonstration}
are modified with respect to the published version in accordance with
\emph{Erratum: Contextuality and noncontextuality measures and generalized
Bell inequalities for cyclic systems {[}Phys. Rev. A 101, 042119 (2020){]}},
published in Phys. Rev. A.
\end{thebibliography}
\end{document}